\documentclass[preprint,3p,12pt,sort&compress]{elsarticle}
\usepackage{mathrsfs}
\usepackage{amsmath}
\usepackage{stmaryrd}
\usepackage{bbding}
\usepackage{dcolumn}
\usepackage{graphicx}
\usepackage{amsfonts}
\usepackage{amssymb}
\usepackage{psfrag}
\usepackage{wrapfig}
\usepackage{subfigure}
\usepackage{makeidx}
\usepackage{bm}
\usepackage{epsf}
\usepackage{epsfig}
\usepackage{setspace}
\usepackage{graphicx}
\usepackage{epstopdf}
\usepackage{psfrag}
\usepackage{subfigure}
\usepackage{color}
\usepackage{enumerate}
\usepackage{verbatim} 

\usepackage{tikz}
\usetikzlibrary{positioning,calc}
\usetikzlibrary{arrows,shapes}
\tikzstyle{block} = [rectangle,rounded corners,thin,align=center,fill=green!20,draw=black!20]
\tikzstyle{line} = [-latex]

\begin{document}

\title{A stochastic kinetic scheme for multi-scale flow transport with uncertainty quantification}

\author[KIT]{Tianbai Xiao\corref{cor}}
\ead{tianbaixiao@gmail.com}

\author[KIT]{Martin Frank}
\ead{martin.frank@kit.edu}

\address[KIT]{Karlsruhe Institute of Technology, Karlsruhe, Germany}

\cortext[cor]{Corresponding author}

\begin{abstract}

Gaseous flows show a diverse set of behaviors on different characteristic scales. Given the coarse-grained modeling in theories of fluids, considerable uncertainties may exist between the flow-field solutions and the real physics. To study the emergence, propagation and evolution of uncertainties from molecular to hydrodynamic level poses great opportunities and challenges to develop both sound theories and reliable multi-scale numerical algorithms. In this paper, a new stochastic kinetic scheme will be developed that includes uncertainties via a hybridization of stochastic Galerkin and collocation methods. Based on the Boltzmann-BGK model equation, a scale-dependent evolving solution is employed in the scheme to construct governing equations in the discretized temporal-spatial domain.
Therefore typical flow physics can be recovered with respect to different physical characteristic scales and numerical resolutions in a self-adaptive manner. We prove that the scheme is formally asymptotic-preserving in different flow regimes with the inclusion of random variables, so that it can be used for the study of multi-scale non-equilibrium gas dynamics under the effect of uncertainties.

Several numerical experiments are shown to validate the scheme. We make new physical observations, such as the wave-propagation patterns of uncertainties from continuum to rarefied regimes. These phenomena will be presented and analyzed quantitatively. The current method provides a novel tool to quantify the uncertainties within multi-scale flow evolutions.
\end{abstract}


\maketitle

\section{Introduction}
Hilbert's 6th problem \cite{hilbert1902mathematical} has served as an intriguing beginning of trying to describe the behavior of interacting many-particle systems, including the gas dynamic equations, across different scales. It has been shown since then that some hydrodynamic equations can be derived from the asymptotic limits of kinetic solutions \cite{chapman1970mathematical, grad1949kinetic,eu1980modified,levermore1996moment,sone2012kinetic,torrilhon2016modeling}.

Multi-scale kinetic algorithms aim at a discretized Hilbert's passage between scales. Instead of coupling physical laws at different scales, asymptotic-preserving (AP) methods are based on solving kinetic equations uniformly, with connection to their hydrodynamic limits. When the mesoscopic structure cannot be resolved by the current numerical resolution, the scheme mimics the collective behaviors of kinetic solutions at hydrodynamic level in a self-adaptive manner.
This scale-bridging property has been validated to be feasible in various AP schemes \cite{jin2010asymptotic,degond2007mach,filbet2010class,berthon2011asymptotic,lemou2008new,liu2010analysis,crispel2007asymptotic,degond2017asymptotic}, among them unified gas-kinetic schemes (UGKS) \cite{xu2010unified,liu2016unified,xiao2017well,xiao2019unified}, and high-order/low-order (HOLO) algorithms \cite{chacon2017multiscale}.

So far most kinetic schemes have been constructed for deterministic solutions. Given the coarse-grained approximation in fluid theories and errors from numerical simulations, considerable uncertainties may be introduced inevitably. A typical example is the collision kernel employed in the kinetic equations, which measures the strength of particle collisions in different directions. Even if scattering theory provides a one-to-one correspondence between the intermolecular potential law and its collision kernel, the differential cross sections become too complicated except for simple Maxwell and hard-sphere molecules. As a result, phenomenological models, e.g. the Lennard-Jones molecules \cite{lennard1924determination}, have to be constructed to reproduce the correct coefficients of viscosity, conductivity and diffusivity.
The adjustable model parameters need to be calibrated by experiments, which introduce errors into the simulations that ought to be deterministic. How predictive are the simulation results from the idealized models? How can one explicitly assess the effects of uncertainties on the quality of model predictions? To answer such questions lies at the core of uncertainty quantification (UQ).

Although the UQ field has undergone rapid development over the past few years, its applications on computational fluid dynamics mainly focus on macroscopic fluid dynamic equations with standard stochastic settings. Limited work has been conducted either on the Boltzmann equation at kinetic scale or on the evolutionary process of uncertainty in multi-scale physics \cite{hu2016stochastic,shu2017stochastic,hu2019stochastic}. 
Given the nonlinear system including intermolecular collisions, initial inputs, fluid-surface interactions and geometric complexities, uncertainties may emerge from molecular-level nature, develop upwards, affect macroscopic collective behaviors, and vice versa.
To study the emergence, propagation and evolution of uncertainty poses great opportunities and challenges to develop both sound theories and reliable multi-scale numerical algorithms.

Generally, the methods for UQ study can be classified into intrusive and non-intrusive ones, depending on the methodology to treat random variables. Monte-Carlo sampling (MCS) is the simplest non-intrusive method, in which many realizations of random inputs are generated based on the prescribed probability distribution.
For each realization we solve a deterministic problem, and then  post-processing is employed to estimate uncertainties. 
MCS is intuitive and straightforward to implement, but a large number of realizations are needed due to the slow convergence with respect to sampling size. This remains true for other variants of MCS like quasi or multi-level Monte-Carlo, which differ in the nodes and weights that are used in the postprocessing.

On the other hand, intrusive methods work in a way such that we reformulate the original deterministic system. One commonly used intrusive strategy is the stochastic Galerkin (SG) method \cite{xiu2010numerical}, in which the stochastic solutions are expressed into orthogonal polynomials of the input random parameters.
As a spectral method in random space, it promises spectral convergence when the solution depends smoothly on the random parameters \cite{canuto1982approximation,xiu2003modeling,gottlieb2008galerkin,jin2017uniform}. However, in the Galerkin system all the expansion coefficients are nearly always coupled, which becomes cumbersome in complicated systems with strong nonlinearity. 

The stochastic collocation (SC) method  \cite{mathelin2005stochastic,xiu2005high,babuvska2007stochastic}, although a non-intrusive method, can be seen as a middle way. It combines the strengths of non-intrusive sampling and SG by evaluating the generalized polynomial chaos (gPC) expansions \cite{xiu2010numerical} on quadrature points in random space. As a result, a set of decoupled equations can be derived and solved with deterministic solvers on each quadrature point. Provided the solutions posses sufficient smoothness over random space, the SC methods maintain similar convergence as SG, but suffers from aliasing errors due to limited number of quadrature points.

The stochastic collocation (SC) and stochastic Galerkin (SG) methods can be combined when the integrals that are necessary for SG inside the algorithm are computed numerically using SC. 
Tracking the evolution of phase-space variables with quadrature rules is very similar in spirit to kinetic schemes to solve kinetic equations. This is the main idea of this paper: to solve an intrusive SG system for the Bhatnagar-Gross-Krook (BGK) equation \cite{bhatnagar1954model} by using SC, and by combining this with the integration that is necessary in particle velocity space to update macroscopic conservative flow quantities.
Similar to the unified gas-kinetic schemes (UGKS) \cite{xu2010unified,liu2016unified}, a scale-dependent interface flux function in the SG setting is constructed from the integral solution of the BGK equation, which considers the correlation between particle transport and collisions. We thus combine the advantages of SG and SC methods with the construction principle of kinetic schemes, and obtain an efficient and accurate scheme for multi-scale flow transport problems with uncertainties.


The rest of this paper is organized as follows.
Sec. \ref{sec:kinetic theory} is a brief introduction of gas kinetic theory and its stochastic formulation.
Sec. \ref{sec:solution algorithm} presents the numerical implementation of the current scheme and detailed solution algorithm.
Sec. \ref{sec:numerical experiment} includes numerical experiments to demonstrate the performance of the current scheme and analyze some new physical observations.
The last section is the conclusion.

\section{Kinetic theory and stochastic formulation}\label{sec:kinetic theory}

\subsection{Kinetic theory of gases}

The Boltzmann equation describes gas dynamics by tracking the temporal-spatial evolution of particle distribution function $f(t, \mathbf{x},\mathbf{u})$, where $\mathbf{x}\in \mathcal{R}^3$ is space variable and  $\mathbf{u}\in \mathcal{R}^3$ is particle velocity.
In the absence of an external force field, the deterministic Boltzmann equation for a monatomic dilute gas writes,
\begin{equation}
\frac{\partial f}{\partial t}+\mathbf{u}\cdot\nabla_{\mathbf x} f = Q(f,f) = \int_{\mathcal{R}^3} \int_{\mathcal{S}^2} \left[ f(\mathbf{u}')f(\mathbf{u}_1')-f(\mathbf{u})f(\mathbf{u}_1) \right] \mathcal B(\cos \theta,g) d\mathbf{\Omega} d\mathbf{u}_1,
\label{eqn:boltzmann equation}
\end{equation}
where $\mathbf{u},\mathbf{u_1}$ are the pre-collision velocities of two colliding particles, and $\mathbf{u}',\mathbf{u_1}'$ are the corresponding post-collision velocities.
The collision kernel $\mathcal B(\cos \theta,g)$ measures the strength of collisions in different directions, where $\theta$ is the deflection angle and $g=|\mathbf g|=|\mathbf{u}-\mathbf{u_1}|$ is the magnitude of relative pre-collision velocity, and $\mathbf{\Omega}$ is the unit vector along the relative post-collision velocity $\mathbf{u}'-\mathbf{u_1}'$,
and the deflection angle $\theta$ satisfies the relation $\cos \theta = \mathbf \Omega \cdot \mathbf g/g$.

Now let us consider the gas evolution with stochastic parameters, e.g.\ the collision kernel $\mathcal B(\cos \theta, g, \mathbf z)$ with random variable $\mathbf{z}\in \mathcal{R}^d$ of $d$ dimensions, then the Boltzmann equation becomes, 
\begin{equation}
\frac{D }{D t} f(t, \mathbf x, \mathbf u, \mathbf z) = Q(f,f)(t, \mathbf x, \mathbf u, \mathbf z),
\label{eqn:stochastic boltzmann equation}
\end{equation}
where $D/Dt$ denotes the material derivative terms on the left-hand side of Eq.(\ref{eqn:boltzmann equation}).
The macroscopic conservative flow variables are related to the moments of the particle distribution function over velocity space,
\begin{equation}
\mathbf{W} (t, \mathbf x, \mathbf z) =\left(
\begin{matrix}
\rho \\
\rho \mathbf U \\
\rho E
\end{matrix}
\right)=\int f\psi d\mathbf u,
\label{eqn:moment relation}
\end{equation}
where $\psi=\left(1,\mathbf u,\frac{1}{2} \mathbf u^2 \right)^T$ is a vector of collision invariants, and temperature is defined as
\begin{equation}
\frac{3}{2} k T (t, \mathbf x, \mathbf z) = \frac{1}{2n} \int (\mathbf u - \mathbf U)^2  f d\mathbf u,
\label{eqn:temperature relation}
\end{equation}
where $k$ is the Boltzmann constant and $n$ is number density of the gas.
Regardless of the value of the collision kernel in random space, the collision operator satisfies the compatibility condition,
\begin{equation}
\int Q(f,f)\psi d\mathbf u=0.
\label{eqn:compatibility condition}
\end{equation}

Substituting the $H$ function,
\begin{equation*}
	H(t, \mathbf x, \mathbf z) = - \int f \ln f d \mathbf{u},
\end{equation*}
into the Boltzmann equation we have
\begin{equation}
\frac{\partial H}{\partial t} =-\int(1+\ln f) \frac{\partial f}{\partial t} d \mathbf{u} =-\iiint(1+\ln f)\left(f^{\prime} f_{1}^{\prime}-f f_{1}\right) \mathcal B d \Omega d \mathbf{u} d \mathbf{u}_{1}.
\end{equation}
From the H-theorem \cite{cercignani1988boltzmann} we know that entropy is locally maximal when $f$ is a Maxwellian
\begin{equation}
\mathcal M(t, \mathbf x, \mathbf u, \mathbf z)=\rho \left( \frac{\lambda}{\pi} \right)^{\frac{3}{2}} e^{-\lambda(\mathbf u-\mathbf U)^2},
\label{eqn:maxwell distribution}
\end{equation}
where $\lambda=m/(2kT)$.
The macroscopic variables $\{\rho(t, \mathbf x, \mathbf z), \mathbf U(t, \mathbf x, \mathbf z), \lambda(t, \mathbf x, \mathbf z)\}$ vary in random space.

Due to the complicated fivefold integration in the Boltzmann collision operator, simplified kinetic model equations can been constructed, e.g.\ the Bhatnagar-Gross-Krook (BGK) model.
The BGK relaxation operator can be planted into the current stochastic system similarly, which writes,
\begin{equation}
f_t+\mathbf{u}\cdot\nabla_{\mathbf x} f = Q(f) = \nu(\mathcal M-f). \\
\label{eqn:bgk equation}
\end{equation}
Given a random collision kernel $\mathcal B(\cos \theta, g, \mathbf z)$, the collision frequency here is also a function of random variable $\nu(\mathbf z)$.
The BGK model simplifies the computation significantly, but still possesses some key properties of the original Boltzmann equation, e.g., the H-theorem.
In this paper, we will only conduct numerical simulations with the BGK relaxation term.

\subsection{Generalized polynomial chaos formulation of kinetic equation}
Consider the generalized polynomial chaos (gPC) expansion of particle distribution with degree $N$, i.e.,
\begin{equation}
f(t,\mathbf x,\mathbf u,\mathbf z) \simeq f_N = \sum_{|\mathbf i|=0}^N \hat f_\mathbf i (t,\mathbf x,\mathbf u) \Phi_\mathbf i (\mathbf z),
\label{eqn:polynomial chaos}
\end{equation}
where the $K$-dimensional index takes the form $\mathbf i=(i_1,i_2,\cdots,i_K)$ and $|\mathbf i|=i_1+i_2+\cdots+i_K$.
The $\hat f_i$ is the coefficient of $\mathbf i$-th polynomial chaos expansion,
and the basis functions used are orthogonal polynomials \{$\Phi_\mathbf i(\mathbf z)$\} satisfying the following constraints,
\begin{equation}
\mathbb{E}[\Phi_\mathbf j (\mathbf z) \Phi_\mathbf k (\mathbf z)] = \gamma_\mathbf k \delta_{\mathbf j \mathbf k}, \quad 0 \leq |\mathbf j|, |\mathbf k| \leq N,
\end{equation}
where 
\begin{equation}
\gamma_\mathbf k=\mathbb{E}[\Phi_\mathbf k^2 (\mathbf z)], \quad 0 \leq |\mathbf k| \leq N,
\end{equation}
are the normalization factors.
The expectation value defines a scalar product,
\begin{equation}
\mathbb{E}[\Phi_\mathbf j (\mathbf z) \Phi_\mathbf k (\mathbf z)] = \int_{I_{\mathbf z}} \Phi_{\mathbf j}(\mathbf z) \Phi_{\mathbf k}(\mathbf z) p(\mathbf z) d \mathbf z,
\label{eqn:continuous expectation value}
\end{equation}
for continuous distribution of $\mathbf z$ and
\begin{equation}
\mathbb{E}[\Phi_\mathbf j (\mathbf z) \Phi_\mathbf k (\mathbf z)] = \sum_i \Phi_{\mathbf j}(\mathbf z_i) \Phi_{\mathbf k}(\mathbf z_i) w(\mathbf z_i) ,
\label{eqn:discrete expectation value}
\end{equation}
for discrete distribution, where $p(\mathbf z)$ is the probability density function, and $w(\mathbf z)$ is the corresponding quadrature weight function in random space.
In the following we use the notation $\langle \Phi_\mathbf j \Phi_\mathbf k \rangle$ to denote the integration formulas in Eq.(\ref{eqn:continuous expectation value}) and (\ref{eqn:discrete expectation value}) uniformly.

Given the correspondence between macroscopic and mesoscopic variables, from Eq.(\ref{eqn:moment relation}) we can derive,
\begin{equation}
\begin{aligned}
\mathbf W & \simeq \int f_N \psi d\mathbf u = \int \sum_{\mathbf i}^N \hat f_\mathbf i (t,\mathbf x,\mathbf u) \Phi_\mathbf i (\mathbf z) \psi d\mathbf u = \sum_{\mathbf i} \left( \int \hat f_\mathbf i \psi d\mathbf u \right) \Phi_\mathbf i\\
& \simeq \mathbf W_N = \sum_{\mathbf i}^N \hat w_\mathbf i \Phi_\mathbf i ,
\end{aligned}
\end{equation}
and the compatibility condition is satisfied
\begin{equation}
\int Q(f_N)\psi d\mathbf u=0.
\end{equation}

After substituting the Eq.(\ref{eqn:polynomial chaos}) into the kinetic equation (\ref{eqn:boltzmann equation}) and (\ref{eqn:bgk equation}), and performing a Galerkin projection, we then obtain
\begin{equation}
\frac{\partial \hat f_{\mathbf{i}}}{\partial t}+\mathbf{u} \cdot \nabla_{\mathbf{x}} \hat f_{\mathbf{i}}= \hat Q_{\mathbf{i}}\left(f_N\right),
\label{eqn:stochastic bgk equation}
\end{equation}
where $Q_{\mathbf{i}}$ is the $\mathbf i$-th projection of the collision operator onto the basis polynomials. We assume the same gPC expansion for the collision frequency,
\begin{equation}
\nu_N = \sum_{|\mathbf i|=0}^N \hat \nu_\mathbf i \Phi_\mathbf i,
\end{equation}
and thus the collision term becomes,
\begin{equation}
\hat Q_{\mathbf{i}} (f_N) = \frac{\sum_\mathbf j^N \sum_\mathbf k^N \hat \nu_\mathbf j \hat m_\mathbf k \langle \Phi_\mathbf j \Phi_\mathbf k \Phi_\mathbf i \rangle - \sum_\mathbf j^N \sum_\mathbf k^N \hat \nu_\mathbf j \hat f_\mathbf k \langle \Phi_\mathbf j \Phi_\mathbf k \Phi_\mathbf i \rangle }{\langle \Phi_\mathbf k^2 \rangle},
\label{eqn:galerkin bgk term}
\end{equation}
with $\hat m_\mathbf k$ and $\hat \nu_\mathbf j$ being the coefficients of gPC expansions for the Maxwellian distribution and collision frequency.

\subsection{Maxwellian distribution in generalized polynomial chaos}\label{sec:maxwellian evaluation}

For a deterministic system, the evaluation of the Maxwellian distribution given in Eq.(\ref{eqn:maxwell distribution}) is straight-forward.
However, given a generalized polynomial chaos (gPC) system, the multiplication and division can't be operated directly on the stochastic moments without modifying the orthogonal basis.
Starting from a known particle distribution function in Eq.(\ref{eqn:polynomial chaos}), here we draw a brief outline to approximately evaluate the Maxwellian distribution function in the gPC expansion.

1. Derive the macroscopic conservative variables from particle distribution function with gPC expansion,
\begin{equation}
\mathbf{W}_N  =\left(
\begin{matrix}
\rho_N \\
(\rho \mathbf U)_N \\
(\rho E)_N
\end{matrix}
\right)=\sum_{\mathbf i}^N \left( \int \hat f_\mathbf i \psi d\mathbf u \right) \Phi_\mathbf i ;
\end{equation}

2. Locate conservative variables on quadrature points $\mathbf z_j$ of random space and calculate primitive variables, e.g. flow velocity
\begin{equation}
\mathbf U(\mathbf z_j)
= \frac{(\rho \mathbf U)_N(\mathbf z_j)}{\rho_N (\mathbf z_j)},
\end{equation}
and
\begin{equation}
\lambda(\mathbf z_j) = \frac{3\rho_N(\mathbf z_j)}{4[(\rho E)_N(\mathbf z_j) - (\rho U)_N^2(\mathbf z_j)/2\rho_N(\mathbf z_j) ]};
\end{equation}

3. Calculate Maxwellian distribution on quadrature points
\begin{equation}
\mathcal M(\mathbf u, \mathbf z_j)=\rho_N(\mathbf z_j) \left( \frac{\lambda(\mathbf z_j)}{\pi} \right)^{\frac{3}{2}} e^{-\lambda(\mathbf z_j)(\mathbf u-\mathbf U(\mathbf z_j))^2},
\end{equation}
and decompose it into a gPC expansion
\begin{equation}
\mathcal M_N = \sum_{|\mathbf i|=0}^ N \hat m_\mathbf i \Phi_\mathbf i,
\end{equation}
with each coefficient in the expansion being given by a quadrature rule
\begin{equation}
\hat m_\mathbf i = \frac{\langle \mathcal M,\Phi_\mathbf i \rangle}{\langle \Phi_{\mathbf i}^2 \rangle} = \frac{\sum_j \mathcal M(\mathbf z_j) \Phi_{\mathbf i}(\mathbf z_j) p(\mathbf z_j)}{\int_{I_{\mathbf z}}  ( \Phi_{\mathbf i}(\mathbf z) )^2 p(\mathbf z) d \mathbf z}.
\end{equation}

Note that as a case of particle distribution function, the Maxwellian distribution certainly has one-to-one correspondence with macroscopic variables,
\begin{equation}
\mathbf{W}_N  =\left(
\begin{matrix}
\rho_N \\
(\rho \mathbf U)_N \\
(\rho E)_N
\end{matrix}
\right)=
\left(
\begin{matrix}
\sum_\mathbf i^N \hat \rho_\mathbf i \Phi_\mathbf i \\
\sum_\mathbf i^N (\hat{\rho \mathbf U})_\mathbf i \Phi_\mathbf i \\
\sum_\mathbf i^N (\hat{\rho E})_\mathbf i \Phi_i
\end{matrix}
\right) =
\sum_{|\mathbf i|=0}^N \left( \int \hat m_\mathbf i \psi d\mathbf u \right) \Phi_\mathbf i.
\end{equation}
Furthermore, the compatibility condition (\ref{eqn:compatibility condition}) still holds for the gPc-expanded collision term (\ref{eqn:galerkin bgk term}), i.e.
\begin{equation}
\begin{aligned}
\int \hat Q_{\mathbf{i}} (f_N) \psi d\mathbf u =& \frac{1}{\langle \Phi_\mathbf k^2 \rangle} \left(
\int \sum_\mathbf j^N \sum_\mathbf k^N \hat \nu_\mathbf j \hat m_\mathbf k \langle \Phi_\mathbf j \Phi_\mathbf k \Phi_\mathbf i \rangle \psi d\mathbf u \right. \\
& \left. - \int \sum_\mathbf j^N \sum_\mathbf k^N \hat \nu_\mathbf j \hat f_\mathbf k \langle \Phi_\mathbf j \Phi_\mathbf k \Phi_\mathbf i \rangle \psi d\mathbf u \right) \\
=&0.
\end{aligned}
\label{eqn:galerkin compatibility condition}
\end{equation}

\section{Solution algorithm} \label{sec:solution algorithm}

\subsection{Update algorithm}

The current numerical algorithm is constructed within the finite volume framework.
We adopt the notation of cell averaged macroscopic conservative variables and particle distribution function in a control volume,
\begin{equation*}
\begin{aligned}
&\mathbf W_{t^n,\mathbf x_i,\mathbf z_k}=\mathbf W_{i,k}^n=\frac{1}{\Omega_{i}(\mathbf x) \Omega_{k}(\mathbf z)} \int_{\Omega_{i}}\int_{\Omega_{k}} \mathbf W(t^n, \mathbf x, \mathbf z)d\mathbf x d\mathbf z,\\
&f_{t^n,\mathbf x_i,\mathbf u_j,\mathbf z_k}=f_{i,j,k}^n=\frac{1}{\Omega_{i}(\mathbf x)\Omega_{j}(\mathbf u)\Omega_{k}(\mathbf z)} \int_{\Omega_{i}} \int_{\Omega_{j}} \int_{\Omega_{k}} f(t^n,\mathbf x,\mathbf u,\mathbf z) d\mathbf xd\mathbf u d\mathbf z,
\end{aligned}
\end{equation*}
along with their $\mathbf m$-th coefficients in the gPC expansions,
\begin{equation*}
\begin{aligned}
&\hat {\mathbf W}_\mathbf m(t^n,\mathbf x_i)=\hat {\mathbf W}_{i, \mathbf m}^n=\frac{1}{\Omega_{i}(\mathbf x)} \int_{\Omega_{i}} \hat {\mathbf W}_\mathbf m(t^n, \mathbf x)d\mathbf x ,\\
&\hat f_\mathbf m(t^n,\mathbf x_i,\mathbf u_j)=\hat f_{i,j,\mathbf m}^n=\frac{1}{\Omega_{i}(\mathbf x)\Omega_{j}(\mathbf u)} \int_{\Omega_{i}} \int_{\Omega_{j}}\hat f_\mathbf m(t^n,\mathbf x,\mathbf u) d\mathbf xd\mathbf u,
\end{aligned}
\end{equation*}
where $\Omega_{i}$, $\Omega_{j}$ and $\Omega_{k}$ are the cell area in the discretized physical, velocity and random space.

The update of the macroscopic variables and the distribution function at the $k$-th collocation point can be formulated as
\begin{equation}
\mathbf{W}_{i,k}^{n+1}=\mathbf{W}_{i,k}^n+\frac{1}{\Omega_{i}}\int_{t^n}^{t^{n+1}}\sum_{r} {\mathbf{F}}_r \cdot \Delta \mathbf S_r dt,
\label{eqn:macro update}
\end{equation}
\begin{equation}
f_{i,j,k}^{n+1}=f_{i,j,k}^n+\frac{1}{\Omega_{i}}\int_{t^n}^{t^{n+1}} \sum_{r} F_r \Delta S_r dt+ \int_{t^n}^{t^{n+1}} Q(f_{i,j,k}) dt,
\label{eqn:micro update}
\end{equation}
where $F_r$ is the time-dependent flux function of distribution function at cell interface, $\mathbf{F}_r$ is the flux of conservative variables, and $\Delta  S_r$ is the interface area.

For the update of the macroscopic variables and the distribution function, Eq.(\ref{eqn:macro update}) and (\ref{eqn:micro update}) can be solved in a coupled way.
Since there is no stiff source term in the macroscopic conservation laws, Eq.(\ref{eqn:macro update}) can be solved first, and then the updated variables at $n+1$ time step can be employed to evaluate the Maxwellian distribution in Eq.(\ref{eqn:micro update}) implicitly, which forms an implicit-explicit (IMEX) strategy.

At the same time, the update of the stochastic Galerkin coefficients for distribution function can be formulated as,
\begin{equation}
\hat f_{i,j,\mathbf m}^{n+1}=\hat f_{i,j,\mathbf m}^n+\frac{1}{\Omega_{i}}\int_{t^n}^{t^{n+1}} \sum_{r} \hat F_r \Delta S_r dt+ \int_{t^n}^{t^{n+1}} \hat Q_\mathbf m(f_{N i,j}) dt,
\label{eqn:galerkin micro update}
\end{equation}
where $\hat F_r$ is the $\mathbf m$-th coefficient in the gPC expansion of interface flux function. Taking the moments over velocity space, with the compatibility condition given in Eq.(\ref{eqn:galerkin compatibility condition}), the update for the moments of macroscopic conservative variables writes,
\begin{equation}
\hat {\mathbf{W}}_{i, \mathbf m}^{n+1}=\hat{\mathbf{W}}_{i,\mathbf m}^n+\frac{1}{\Omega_{i}}\int_{t^n}^{t^{n+1}}\sum_{r} \hat{\mathbf{F}}_r \Delta S_r dt,
\label{eqn:galerkin macro update}
\end{equation}
The update of Eq.(\ref{eqn:galerkin micro update}) and (\ref{eqn:galerkin macro update}) can be also treated in the IMEX way. However, now the implicit update of the collision term in Eq.(\ref{eqn:galerkin bgk term}) for the $\mathbf m$-th coefficient needs to take the contributions from all other orders into account, which forms a linear system for the source term.

\subsection{Multi-scale interface flux}\label{sec:flux evaluation}

Based on the finite volume framework, a scale-adaptive interface flux function is needed in multi-scale modeling and simulation.
Different from purely upwind flux, here we use an integral solution of the kinetic model equation to construct a multi-scale flux function.
This integral solution originates from Kogan's monograph on rarefied gas dynamics \cite{koganrarefied} and has been inherited by a series of gas-kinetic schemes \cite{xu2010unified,liu2016unified,xiao2017well,xiao2018investigation,xiao2019unified}.

We make the additional approximation that for the loss term the collision frequency $\nu$ at the cell interface can be regarded as a local constant in phase space $(\bf x, u, z)$ within each time step. In random space, we approximate $\nu$ by its expected value, $\nu \simeq \mathbb E(\nu_N) = \hat \nu_0$. This allows us to rewrite the stochastic BGK equation (\ref{eqn:stochastic bgk equation}) for the $\mathbf m$-th moment along the characteristics,
\begin{equation}
\frac{D \hat f_\mathbf m}{D t} + \hat \nu_0 \hat f_\mathbf m = \hat \nu_0 \hat {m}_\mathbf m,
\end{equation}
which holds the following integral solution,
\begin{equation}
\hat f_\mathbf m(\mathbf x,t,\mathbf u)=\hat \nu_0 \int_{t^0}^t \hat m_\mathbf m(\mathbf x',t',\mathbf u)e^{-\hat \nu_0(t-t')}dt' +e^{-\hat \nu_0(t-t^0)} \hat f_\mathbf m(\mathbf x^0,t^0,\mathbf u),
\label{eqn:integral solution}
\end{equation}
where $\mathbf x'=\mathbf x-\mathbf u(t-t')$ is the particle trajectory, and $\mathbf x^0=\mathbf x-\mathbf u(t-t^0)$ is the location at initial time $t=t^0$. The above solution indicates a self-conditioned mechanism for multi-scale gas dynamics.
For example, when the evolving time $t-t^0$ is much less than the mean collision time $\tau=1/\hat \nu_0$, the latter term in Eq.(\ref{eqn:integral solution}) dominates and describes the free transport of particles.
And if $t-t^0$ is much larger than $\tau$, the second term approaches to zero, and then the distribution function will be an accumulation of Maxwellian along the characteristic lines, which provides the underlying wave-interaction physics for the Euler and the Navier-Stokes solutions.
Based on the competition between particle transport and wave interaction, there is a continuous transition from rarefied gas dynamics to hydrodynamics.

In the following, we present a detailed strategy for the construction of the  numerical flux. For brevity, we use one-dimensional physical, velocity and random spaces to illustrate the principle of the solution algorithm, while its extension to multi-dimensional cases is straight-forward.
For each time step, the evolving solution at cell interface $x_{i+1/2}=0$ from initial time $t^n=0$ can be rewritten into the following form,
\begin{equation}
\hat f_m(0,t,u_j)=\hat \nu_0 \int_{0}^{t} \hat m_m(x',t',u_j)e^{-\hat \nu_0(t-t')}dt' +e^{-\hat \nu_0 t}\hat f_m(-u_j t,0,u_j),
\label{eqn:interface integral solution}
\end{equation}
where $\hat f_m(-u_j t,0,u_j)$ is the initial distribution at each time step.

In the numerical scheme,
the initial distribution function around the cell interface can be obtained through reconstruction, e.g.
\begin{equation}
\hat f_m(x,0,u_j)=\left\{
\begin{aligned}
&\hat f_{i+1/2,j,m}^L, \quad x\le 0, \\
&\hat f_{i+1/2,j,m}^R, \quad x> 0,
\end{aligned}
\right.
\label{eqn:f0 reconstruct 1st}
\end{equation}
with first-order accuracy and
\begin{equation}
\hat f_m(x,0,u_j)=\left\{
\begin{aligned}
&\hat f_{i+1/2,j,m}^L+\hat \sigma_{i,j,m}x, \quad x\le 0, \\
&\hat f_{i+1/2,j,m}^R+\hat \sigma_{i+1,j,m}x, \quad x> 0,
\end{aligned}
\right.
\label{eqn:f0 reconstruct 2nd}
\end{equation}
with second-order accuracy.
Here $\{\hat f_{i+1/2,j,m}^L$, $\hat f_{i+1/2,j,m}^R\}$ are the reconstructed initial distribution functions at the left and right hand sides of a cell interface, and $\hat \sigma$ is the corresponding slope along $x$ direction.

In the following, we use the superscript $0$ for the interface at  $\{x=0,t=0\}$. However, all formulas generalize to to arbitrary interfaces.
The macroscopic conservative variables in the gPC expansions at the initial interface $\{x=0,t=0\}$ can be evaluated by taking moments over velocity space,
\begin{equation*}
\mathbf W^0_N = \sum_{m=0}^N \hat{w}^0_m \Phi_m, \ \hat{w}^0_m=\sum_{u_j>0} \hat f_{i+1/2,j,m}^L\psi \Delta u_j +\sum_{u_j<0} \hat f_{i+1/2,j,m}^R\psi \Delta u_j.
\end{equation*}
The collision frequency $\nu$, which we approximated by its expected value, at the interface may be predetermined or can be evaluated from macroscopic variables,
\begin{equation}
\nu^0 \sim \hat \nu^0_0 = \frac{\hat p^0_0}{\hat \mu^0_0},
\end{equation}
where $\hat p^0_0$ is the pressure, and $\mu_0^0$ is the viscosity with respect to a specific molecule at the cell interface.

The equilibrium distribution at $\{x=0,t=0\}$ can be determined as illustrated in Sec.\ \ref{sec:maxwellian evaluation},
and the $m$-th coefficient of equilibrium distribution around a cell interface can be constructed as
\begin{equation}
\hat m_m(x,t)=\hat m_m^0,
\label{eqn:g0 reconstruct 1st}
\end{equation}
with first-order accuracy and
\begin{equation}
\hat m_m(x,t)=\hat m_m^0\left(1+ax+{A}t\right),
\label{eqn:g0 reconstruct 2nd}
\end{equation}
up to second order.
The coefficients $\{ a, A\}$ are the spatial and temporal derivatives of the equilibrium distribution, which can be expanded into series with respect to collision invariants $\psi$,
\begin{equation*}
\begin{aligned}
&a= a_1+ a_2u+ a_3\frac{1}{2}u^2= a_{i}\psi_i, \\
&A=A_1+ A_2u+ A_3\frac{1}{2}u^2= A_{i}\psi_i.
\end{aligned}
\end{equation*}
The spatial derivatives $a$ are related to the slopes of the conservative variables around the cell interface,
\begin{equation*}
\left(\frac{\partial {\hat w_m}}{\partial x}\right)\simeq \frac{\hat w_{i+1}-\hat w_{i}}{\Delta x}=\int a \hat m^0_m \psi du= M^0_{\alpha \beta} a_\beta, 
\end{equation*}
where $M^0_{\alpha \beta}=\int \hat m_m^0 \psi_\alpha \psi_\beta du$ is a known matrix and ${\mathbf a}=(a_1,a_2,a_3)^T$. Here $\Delta x=x_{i+1}-x_i$ is the distance between two cell centers. The time derivative $A$ is related to the temporal variation of conservative flow variables,
\begin{equation*}
\frac{\partial {\hat{w}_m}}{\partial t}=\int  A\hat m^0_m \psi du,
\end{equation*}
and it can be calculated via the time derivative of the compatibility condition
\begin{equation*}
\frac{d}{dt}\int (\hat m_m-\hat f_m)\psi du \mid_{x=0,t=0}=0.
\end{equation*}
With the help of the Euler equations, it gives
\begin{equation*}
-\int u\frac{\partial \hat m_m}{\partial x}\psi du=\frac{\partial\hat{w}_m}{\partial t}=\int A \hat m^0_m \psi du ,
\end{equation*}
and the spatial derivatives in the above equation have been obtained from the initial equilibrium reconstruction in  Eq.(\ref{eqn:g0 reconstruct 2nd}).
Therefore, we have
\begin{equation*}
\int A \hat m^0_m \psi du=-\int au \hat m_m^0\psi du,
\end{equation*}
from which $A=( {A}_1, {A}_2, {A}_3)^T$ is fully determined.

After all coefficients are obtained, the time-dependent interface distribution function becomes
\begin{equation}
\begin{aligned}
\hat f_m(0,t,u_j)=&\left(1-e^{-\hat \nu_0 t}\right) \hat m^{0}_{j,m} \\
&+\left[(-1+e^{-\hat \nu_0 t})/\hat \nu_0 +te^{-\hat \nu_0 t}\right] u a \hat m_{j,m}^{0} \\
&+\left[\left(\hat \nu_0 t-1+e^{-\hat \nu_0 t}\right)/\hat \nu_0 \right] {A} \hat m_{j,m}^{0} \\
&+e^{-\hat \nu_0 t}\left[\left(\hat f_{i+1/2,j,m}^L-u_j t\hat \sigma_{i,j,m}\right)H\left[u_j\right]  \right.\\
&\left. + \left(\hat f_{i+1/2,j,m}^R-u_j t\hat \sigma_{i+1,j,m}\right)(1-H\left[u_j\right])\right] \\
=&\widetilde {m}_{i+1/2,j,m}+\widetilde f_{i+1/2,j,m},
\end{aligned}
\label{eqn:interface distribution gPC 2nd}
\end{equation}
where $H(u)$ is the heaviside step function. The notation $\widetilde {m}_{i+1/2,j,m}$ denotes the contribution of equilibrium state integration and $\widetilde f_{i+1/2,j,m}$ is related to the initial distribution.
If we consider first-order interface flux in space and time, then it reduces to 
\begin{equation}
\begin{aligned}
\hat f_m(0,t,u_j)=&\left(1-e^{-\hat \nu_0 t}\right) \hat m^{0}_{j,m} \\
&+e^{-\hat \nu_0 t}\left[\hat f_{i+1/2,j,m}^L H\left[u_j\right] + \hat f_{i+1/2,j,m}^R (1-H\left[u_j\right])\right].
\end{aligned}
\label{eqn:interface distribution gPC 1st}
\end{equation}

With the variation of the ratio between evolving time $t$ (i.e., the time step in the computation) and collision time $\tau=1/\hat \nu_0$, the above interface distribution function provides a self-conditioned multiple scale solution across different flow regimes. After the coefficients of distribution function at all orders are determined, the corresponding gPC expansion can be expressed as,
\begin{equation}
f_N(0,t,u) = \sum_{m=0}^N \hat f_{m}(0,t,u) \Phi_m , 
\end{equation}
and the corresponding fluxes of particle distribution function and conservative flow variables can be evaluated via
\begin{equation}
\begin{aligned}
&{{F}}_{N}= u f_N(0,t,u,\xi) , \\
&{\mathbf{F}}_{N}=\int u f_N(0,t,u,\xi) \psi du \simeq \sum w_j u_j f_N(0,t,u_j,\xi) \psi_j  ,
\end{aligned}
\end{equation}
where $u_j$ denotes a discretized point in particle velocity space, and $w_j$ is its integral weight in velocity space.

\subsection{Collision term} \label{sec:collision evaluation}

Besides the construction of the interface flux, the collision term needs to be evaluated inside each control volume for the update of the particle distribution function within a time step. In the current numerical scheme, to overcome the stiffness of the kinetic equation in the continuum limit, the implicit-explicit (IMEX) technique is used to solve the collision operator.
For simplicity, here we only discuss a fully implicit treatment of collision term, while the trapezoidal and other high-order integration techniques can be implemented similarly. The solution algorithm can be implemented in the following two ways.

\textbf{1. Stochastic Galerkin method}

Let us consider the stochastic Galerkin system given by Eq.(\ref{eqn:galerkin macro update}) and (\ref{eqn:galerkin micro update}).
In the one-dimensional case, the update algorithm for the $m$-th coefficient of gPC expansion inside cell $\{x_i,u_j\}$ reduces to
\begin{equation}
\hat {\mathbf{W}}_{i,m}^{n+1}=\hat{\mathbf{W}}_{i,m}^n+\frac{1}{\Delta x_{i}} (\hat{\mathbf{F}}_{i-1/2,m} - \hat{\mathbf{F}}_{i-1/2,m} ),
\label{eqn:galerkin macro update 1d}
\end{equation}
\begin{equation}
\hat f_{i,j,m}^{n+1}=\hat f_{i,j,m}^n+\frac{1}{\Delta x_{i}} (\hat F_{i-1/2,j,m}-\hat F_{i+1/2,j,m}) + \Delta t \hat Q_m^{n+1}(f_{N i,j}),
\label{eqn:galerkin micro update 1d}
\end{equation}
where $\mathbf{\hat F}_{i\pm 1/2,m} = \int_{t^n}^{t^{n+1}} u \hat f_{i\pm 1/2,m} \psi dud\xi dt$ and $\hat F_{i\pm 1/2,j,m} = \int_{t^n}^{t^{n+1}} u_j \hat f_{i\pm 1/2,j,m} dt$ are the time-integral interface fluxes for the macroscopic and mesoscopic gPC expansion coefficients.
The source term for the distribution function at $t^{n+1}$ time step is,
\begin{equation}
\hat Q_{{m}} (f_N^{n+1}) = \frac{\sum_p^N \sum_q^N \hat \nu_p^{n+1} \hat m_q^{n+1} \langle \Phi_p \Phi_q \Phi_m \rangle - \sum_p^N \sum_q^N \hat \nu_p^{n+1} \hat f_q^{n+1} \langle \Phi_p \Phi_q \Phi_m \rangle}{\langle \Phi_q^2 \rangle}.
\label{eqn:galerkin bgk term 1d}
\end{equation}
In the numerical simulation, the macroscopic system (\ref{eqn:galerkin macro update 1d}) is solved first, and the updated quantities can be used to evaluate the Maxwellian distribution as described in Sec. \ref{sec:maxwellian evaluation}.
The collision frequency $\nu_N^{n+1}$ can be predetermined or evaluated from macroscopic variables via
\begin{equation}
\nu_N^{n+1} = \sum_{p=0}^N \hat \nu_p^{n+1} \Phi_p, \ 
\hat \nu_p^{n+1}=\frac{\langle p_N^{n+1}/\mu_N^{n+1},\Phi_p \rangle}{\langle \Phi_p^2 \rangle},
\end{equation}
where $\{ p_N, \mu_N\}$ are pressure and viscosity in the gPC expansions. Notice that we use the full gPC expansion for $\nu$ to discretize the collision term.

Notice also that Eq. (\ref{eqn:galerkin micro update 1d}) and (\ref{eqn:galerkin bgk term 1d}) form a linear system,
\begin{equation}
\begin{aligned}
&\hat f_{i,j,m}^{n+1} + \frac{\sum_p^N \sum_q^N \hat \nu_p^{n+1} \hat f_q^{n+1} \langle \Phi_p \Phi_q \Phi_m \rangle}{\langle \Phi_q^2 \rangle} \Delta t \\
&=\hat f_{i,j,m}^n+\frac{1}{\Delta x_{i}} (\hat F_{i-1/2,j,m}-\hat F_{i+1/2,j,m}) + \frac{\sum_p^N \sum_q^N \hat \nu_p^{n+1} \hat m_q^{n+1} \langle \Phi_p \Phi_q \Phi_m \rangle}{\langle \Phi_q^2 \rangle} \Delta t,
\end{aligned}
\label{eqn:galerkin linear system}
\end{equation}
which can be expressed as
\begin{equation}
\mathbf A\mathbf f^{n+1}=\mathbf B,
\end{equation}
where $\bf A$ is the coefficient matrix of solution vector $\mathbf f=(\hat f^{n+1}_1, \hat f^{n+1}_2, \cdots, \hat f^{n+1}_{m^d})^T$,
and $\bf B$ is the right-hand side of Eq.(\ref{eqn:galerkin linear system}).

\textbf{2. Hybrid Galerkin-Collocation method}

It is clear that the linear system in Eq.(\ref{eqn:galerkin linear system}) will bring considerable computational cost as the gPC order increases.
To overcome this, we take advantage of the original kinetic equation (\ref{eqn:micro update}) with quadrature points $z_k$ in random space.
In the one-dimensional case, this reduces to 
\begin{equation}
f_{i,j,k}^{n+1}=f_{i,j,k}^n+\frac{1}{\Delta x_{i}} (F_{i-1/2,j,k}-F_{i+1/2,j,k}) + \Delta t  Q(f_{i,j,k}^{n+1}),
\label{eqn:collocation micro update 1d}
\end{equation}
where $F_{i\pm 1/2,j,k} = \int_{t^n}^{t^{n+1}} u_j f_{i\pm 1/2,j,k} dt$ is the time-integral interface flux for distribution function.
To make use of it, in the numerical algorithm, we first update the gPC coefficients of macroscopic variables to $t^{n+1}$ time step, and of the distribution function in the intermediate step $t^*$,
\begin{equation}
\hat {\mathbf{W}}_{i,m}^{n+1}=\hat{\mathbf{W}}_{i,m}^n+\frac{1}{\Delta x_{i}} (\hat{\mathbf{F}}_{i-1/2,m} - \hat{\mathbf{F}}_{i-1/2,m} ),
\end{equation}
\begin{equation}
\hat f_{i,j,m}^{*}=\hat f_{i,j,m}^n+\frac{1}{\Delta x_{i}} (\hat F_{i-1/2,j,m}-\hat F_{i+1/2,j,m}),
\end{equation}
which is then evaluated on the quadrature points $z_k$,
\begin{equation}
f_{i,j,k}^*=f_{Ni,j}^*(z_k) = \sum_m^N \hat f_{i,j,m}^{*} (z_k) \Phi_m (z_k).
\end{equation}
Afterwards, the collision term is treated via
\begin{equation}
\begin{aligned}
f_{i,j,k}^{n+1}=&f_{i,j,k}^* + \Delta t \nu_{i,j,k}^{n+1} (\mathcal M_{i,j,k}^{n+1} - f_{i,j,k}^{n+1}) \\
=&( f_{i,j,k}^* + \Delta t \nu_{i,j,k}^{n+1} \mathcal M_{i,j,k}^{n+1} )/(1+\Delta t \nu_{i,j,k}^{n+1}),
\end{aligned}
\label{eqn:micro update 1d}
\end{equation}
where the Maxwellian distribution function at time step $t^{n+1}$ can be evaluated in the same way as described in Sec.\ \ref{sec:maxwellian evaluation}. The updated distribution function can be reabsorbed into the gPC expansion,
\begin{equation}
\hat f^{n+1}_{i,j,m} = \frac{\langle f^{n+1}_{i,j},\Phi_m \rangle}{\langle \Phi_{m}^2\rangle} = \frac{\sum_k f^{n+1}_{i,j}( z_k) \Phi_{m}(z_k) p(z_k)}{\int_{I_{z}}  ( \Phi_{m}(z) )^2 p( z) d \mathbf z},
\end{equation}
and the final solution in gPC expansion at $t^{n+1}$ is,
\begin{equation}
f_{Ni,j}^{n+1}=\sum_{m=0}^N \hat f^{n+1}_{i,j,m} \Phi_m.
\end{equation}
So far, we have illustrated the principle for two update algorithms.
In the Sec.\ \ref{sec:numerical experiment}, we will compare these two methods based on numerical experiments.

\subsection{Time step}

In the current scheme, the time step is determined by the Courant-Friedrichs-Lewy condition in phase space,
\begin{equation}
\Delta t=\mathcal{C}  \frac{\Delta x_{min}}{u_{max} + U_{max} } ,
\label{eqn:time step}
\end{equation}
where $\mathcal{C}$ is the CFL number, $\Delta x_{min}=\min (|\Delta x_i|)$ is the finest mesh size, $u_{max}=\max (|u_j|)$ is the largest discrete particle velocity, and $U_{max}=\max (\hat u_1, \hat u_2, \cdots, \hat u_N)$ is the largest stochastic coefficient in the gPC expansions of fluid velocity.

\subsection{AP property of the numerical scheme}

In this part, a brief numerical analysis will be presented on the asymptotic property of the current scheme.
For simplicity, the one-dimensional case is used for illustration, and the collision frequency $\nu$ is assumed to be a local constant.
The solution algorithm for the stochastic collocation method given in Eq.(\ref{eqn:collocation micro update 1d}) is equivalent to
\begin{equation}
\hat f_{i,j,m}^{n+1} = \hat f_{i,j,m}^{n} + \frac{1}{\Delta x} \int_{t^n}^{t^{n+1}} u_j (\hat f_{i-1/2,j,m}-\hat f_{i+1/2,j,m}) dt + \nu \Delta t (\hat m_{i,j,m}^{n+1}-\hat f_{i,j,m}^{n+1}).
\label{eqn:galerkin update 1D}
\end{equation}

Now let us consider limiting cases of numerical flow dynamics.
In the collisionless limit where $\nu$ approaches zero, the relation $\nu \Delta t \ll 1$ holds naturally, and the fully discretized interface distribution in Eq.(\ref{eqn:interface distribution gPC 2nd}) becomes 
\begin{equation}
\begin{aligned}
\hat f_m(x_{i+1/2},t,u_j)=\left(\hat f_{i+1/2,j,m}^L-u_jt\hat \sigma_{i,j,m}\right)H\left[u_j\right]  
+\left(\hat f_{i+1/2,j,m}^R-u_j t\hat \sigma_{i+1,j,m}\right)(1-H\left[u_j\right]) ,
\end{aligned}
\label{eqn:interface distribution 1D collisionless}
\end{equation}
and Eq.(\ref{eqn:galerkin update 1D}) reduces to
\begin{equation}
\begin{aligned} 
\hat f_{i,j,m}^{n+1}=\hat f_{i,j,m}^{n} 
& +\frac{1}{\Delta x}\left[ \left(\Delta t \hat f_{i-1 / 2, j,m}^{L}-\frac{1}{2} \Delta t^{2} u_{j} \hat \sigma_{i-1, j,m}\right) H\left[u_{j}\right] \right.\\
& \left. +\left(\Delta t \hat f_{i-1 / 2, j,m}^{R}-\frac{1}{2} \Delta t^{2} u_{j} \hat \sigma_{i, j,m}\right)\left(1-H\left[u_{j}\right]\right) \right.  \\
& -\left. \left(\Delta t \hat f_{i+1 / 2, j,m}^{L}-\frac{1}{2} \Delta t^{2} u_{j} \hat \sigma_{i, j,m}\right) H\left[u_{j}\right] \right.\\
& \left. -\left(\Delta t \hat f_{i+1 / 2, j,m}^{R}-\frac{1}{2} \Delta t^{2} u_{j} \hat \sigma_{i+1, j,m}\right)\left(1-H\left[u_{j}\right]\right) \right] ,
\end{aligned}
\label{eqn:distribution update collisionless}
\end{equation}
which is a second-order upwind scheme for free molecular flow.

On the other hand, in the Euler regime with $\nu \rightarrow \infty$, the particle distribution is close to equilibrium state. In this case we rewrite the solution algorithm in Eq.\ (\ref{eqn:galerkin update 1D}) and take the limit, which results
\begin{equation}
\begin{aligned}
\lim _{\nu \rightarrow \infty} \hat f_{i,j,m}^{n+1} &= \lim _{\nu \rightarrow \infty} \left( \hat m_{i,j,m}^{n+1} -  \frac{\hat f_{i,j,m}^{n+1} - \hat f_{i,j,m}^{n}}{\nu \Delta t} - \frac{\int_{t^n}^{t^{n+1}} u_j (\hat f_{i-1/2,j,m}-\hat f_{i+1/2,j,m}) dt}{\nu \Delta t \Delta x} \right) \\
&= \hat m_{i,j,m}^{n+1}.
\end{aligned}
\label{eqn:distribution update ns transform}
\end{equation}
If we consider a fully resolved case where there exist continuous distributions of flow variables and their derivatives over the domain,
then the reconstruction technique used in Sec. \ref{sec:flux evaluation} is equivalent to central interpolation, and the interface solution in Eq.(\ref{eqn:interface distribution gPC 2nd}) becomes
\begin{equation}
\begin{aligned}
\hat f_m(x_{i+1/2},t,u_j)=&\left(1-e^{-\nu t}\right) \hat m_{i+1/2,j,m}^n\\
&+\left[(-1+e^{-\nu t})/\nu+te^{-\nu t}\right] u_j a_m \hat m_{i+1/2,j,m}^n \\
&+\left[ \left(\nu t-1+e^{-\nu t}\right)/\nu \right] A_m \hat m_{i+1/2,j,m}^n \\
&+e^{-\nu t} \left(\hat f^n_{i+1/2,j,m}-u_jt\hat \sigma^n_{i+1/2,j,m}\right) .
\end{aligned}
\label{eqn:interface distribution galerkin 1D continuous}
\end{equation}
The initial distribution function can be obtained via
\begin{equation}
\begin{aligned} 
\hat f^n_{i+1/2,j,m} &=\hat f_{i, j,m}^{n}+\frac{f_{i+1, j,m}^{n}-f_{i,j,m}^{n}}{\Delta x} \frac{1}{2} \Delta x \\ 
&=\hat m_{i+1/2, j,m}^{n}+O\left(\Delta x^{2}\right) .
\end{aligned}
\end{equation}
We substitute the above initial distribution into Eq.(\ref{eqn:interface distribution galerkin 1D continuous}), and we get
\begin{equation}
\begin{aligned}
\hat f_m(x_{i+1/2},t,u_j) =
& (1+ t A_m) \hat m_{i+1/2,j,m}^n \\
&+\left[(-1+e^{-\nu t})/\nu+te^{-\nu t}\right] u_j a_m \hat m_{i+1/2,j,m}^n \\
&+\left[ \left(-1+e^{-\nu t}\right)/\nu \right] A_m \hat m_{i+1/2,j,m}^n -u_jt\hat \sigma^n_{i+1/2,j,m} e^{-\nu t} \\
=& \hat m_m(x_{i+1/2},t,u_j) + O\left(\Delta t^{2}, \Delta x^{2}\right)
\end{aligned}
\label{eqn:ugks interface solution ns}
\end{equation}
as $\nu \rightarrow \infty$.
The interface flux for the macroscopic variables can be obtained by taking conservative moments $\psi$ to Eq.(\ref{eqn:ugks interface solution ns}), which results in
\begin{equation}
\hat{\mathcal F}_{w}=\left(\begin{array}{c}{\hat{\mathcal F}_{\rho}} \\ 
\hat {\mathcal F}_m \\ 
{\hat{\mathcal F}_{e}}\end{array} \right) = \left(\begin{array}{c}{\sum w_j u_j \hat f_{i+1/2,j,m} } \\ 
{\sum w_j u^2_j \hat f_{i+1/2,j,m} } \\ 
{\sum w_j \frac{1}{2} u^3_j \hat f_{i+1/2,j,m} }\end{array}\right)+O\left(\Delta t^{2}, \Delta x^{2}\right),
\end{equation}
where $u_j$ is the discretized point in particle velocity space, and $w_j$ is its quadrature weight.
Thus, the Euler equations can be obtained up to errors of order $O(\Delta t^2, \Delta x^2)$, i.e.,
\begin{equation}
\frac{\partial}{\partial t} \left(\begin{array}{c}{\hat \rho_m} \\ 
{(\hat {\rho U})_m} \\ 
{(\hat {\rho E})_m}\end{array} \right) + \frac{\partial}{\partial x} \left(\begin{array}{c}{\hat{\mathcal F}_{\rho}} \\ 
{\hat{\mathcal F}_m} \\ 
{\hat{\mathcal F}_{e}}\end{array} \right) = O\left(\Delta t^{2}, \Delta x^{2}\right)
\end{equation}
The above numerical analysis demonstrates that our current scheme, including the stochastic collocation formulation, is formally asymptotic-preserving (AP).

\subsection{Summary of the algorithm}

The solution algorithm of our stochastic kinetic scheme can be summarized as follows: It updates both conservative variables and distribution function in Eq.(\ref{eqn:macro update}) and Eq.(\ref{eqn:micro update}).
The scale-dependent flux function is determined by the particle distribution function at the interface, which comes from the integral solutions of kinetic model equation and is given in Eq.(\ref{eqn:interface distribution gPC 1st}) and Eq.(\ref{eqn:interface distribution gPC 2nd}). As shown in the theoretical analysis, the asymptotic-preserving property is preserved by the numerical algorithm.

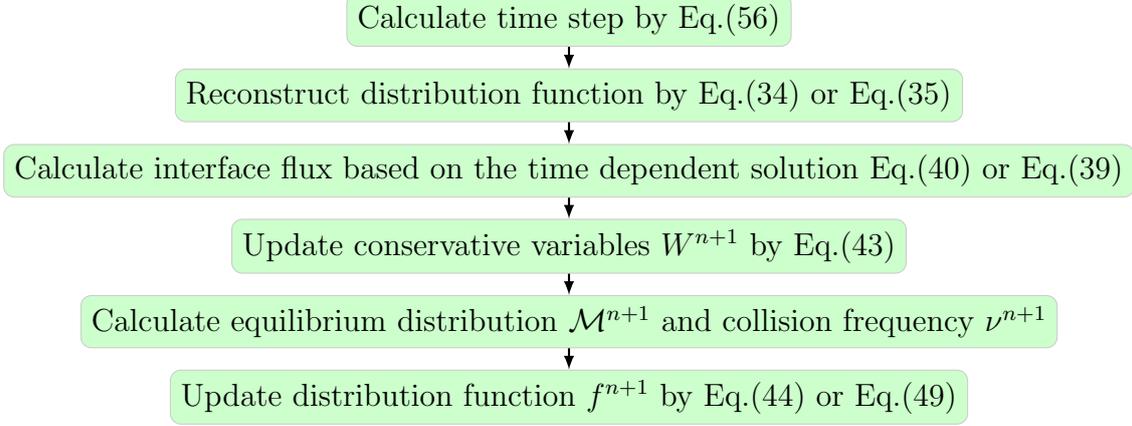
\begin{figure}[htb!]
	\centering
	\begin{tikzpicture}[thick]
	\path[line] node[block] (time) {Calculate time step by Eq.(\ref{eqn:time step})};
	\path[line] node[block,below of=time] (interp) {Reconstruct distribution function by Eq.(\ref{eqn:f0 reconstruct 1st}) or Eq.(\ref{eqn:f0 reconstruct 2nd})}
	(time) edge (interp);
	\path[line] node[block,below of=interp] (flux) {Calculate interface flux based on the time dependent solution Eq.(\ref{eqn:interface distribution gPC 1st}) or Eq.(\ref{eqn:interface distribution gPC 2nd})}
	(interp) edge (flux);
	\path[line] node[block,below of=flux] (new-w) {Update conservative variables $W^{n+1}$ by Eq.(\ref{eqn:galerkin macro update 1d})}
	(flux) edge(new-w);
	\path[line] node[block,below of=new-w] (new-g) {Calculate equilibrium distribution $\mathcal M^{n+1}$ and collision frequency $\nu^{n+1}$}
	(new-w) edge (new-g);
	\path[line] node[block,below of=new-g] (new-f) {Update distribution function $f^{n+1}$ by Eq.(\ref{eqn:galerkin micro update 1d}) or Eq.(\ref{eqn:collocation micro update 1d})}
	(new-g) edge (new-f);
	\end{tikzpicture}
	\caption{Flowchart of solution algorithm.}
	\label{pic:algorithm}
\end{figure}

\section{Numerical experiments} \label{sec:numerical experiment}

In this section, we will present some numerical results.
The goal of numerical experiments is not simply to validate the performance of the current scheme, but also to present and analyze new physical observations.
In order to demonstrate the multi-scale nature of the algorithm, simulations from Euler and Navier-Stokes to free molecule flow are presented. The following dimensionless flow variables are used in the calculations,
\begin{equation*}
\begin{aligned}
& \tilde{x}=\frac{x}{L_0}, \tilde{y}=\frac{y}{L_0}, \tilde{\rho}=\frac{\rho}{\rho_0}, \tilde{T}=\frac{T}{T_0}, \tilde{\mathbf u}=\frac{\mathbf u}{(2RT_0)^{1/2}}, \\ &\tilde{\mathbf U}=\frac{\mathbf U}{(2RT_0)^{1/2}}, \tilde{f}=\frac{f}{\rho_0 (2RT_0)^{3/2}}, \tilde{\mathbf P}=\frac{\mathbf P}{\rho_0 (2RT_0)}, \tilde{\mathbf q}=\frac{\mathbf q}{\rho_0 (2RT_0)^{3/2}},
\end{aligned}
\end{equation*}
where $R$ is the gas constant, $\mathbf u$ is the particle velocity, $\mathbf U$ is the macroscopic fluid velocity, $\mathbf P$ is the stress tensor, $\mathbf q$ is the heat flux. The subscript zero represents the reference state. For brevity, the tilde notation for dimensionless variables will be removed henceforth. In all simulations we consider one-dimensional monatomic gas, for which the corresponding gas constant is
$$
\gamma= (I+3) / (I+1)=3,
$$
with $I=0$ denoting the nonexistence of other molecular internal degrees of freedom, and the Maxwellian distribution function is
$$
\mathcal M=\rho \left( \frac{\lambda}{\pi} \right)^{\frac{1}{2}} e^{-\lambda(u- U)^2}.
$$

\subsection{Random collision kernel: homogeneous relaxation}

First let us consider a homogeneous relaxation problem.
The corresponding BGK equation is
\begin{equation}
\frac{\partial f}{\partial t} = \nu (\mathcal M - f),
\label{eqn:homegenous bgk 1d}
\end{equation}
and the initial condition of the particle distribution is
\begin{equation*}
f_0=u^2 e^{-u^2}.
\end{equation*}
The initial macroscopic variables are deterministic and fixed in time,
\begin{equation*}
\mathbf{W} =\left(
\begin{matrix}
\rho_0 \\
\rho_0 U_0 \\
\rho_0 E_0
\end{matrix}
\right)=\int f_0 \psi d u, \quad \mathcal M=\rho_0 \left( \frac{\lambda_0}{\pi} \right)^{\frac{1}{2}} e^{-\lambda_0(u- U_0)^2}.
\end{equation*}
In this case, the randomness comes from the collision kernel, which follows a Gaussian distribution in random space $\nu \sim (1, 0.2^2)$,
and can be written into a gPC expansion as
\begin{equation*}
\nu=1+0.2z,
\end{equation*}
where $z$ is a standard random variable with normal distribution $z \sim (0, 1)$, and serves as the first-order Hermite polynomial.

An integral solution of 
Eq.(\ref{eqn:homegenous bgk 1d}) can be constructed as,
\begin{equation}
f = f_0 e^{-\nu t} + (1-e^{-\nu t}) \mathcal M_0,
\end{equation}
which forms a combined log-normal distribution over random space.
Therefore, its mean and variance values can be derived theoretically, i.e.,
\begin{equation}
\begin{aligned}
&\mathbb E(f) = f_0 \exp(-t + (\sigma t)^2/2) + \mathcal M (1 - \exp(-t + (\sigma t)^2/2)), \\
&\mathbb S(f) = \left[ (f_0 - \mathcal M)^2 (\exp(\sigma^2 t^2) - 1) \exp(-2t + \sigma^2 t^2) \right]^{1/2}.
\end{aligned}
\label{eqn:homegenous mean and std}
\end{equation}
At the same time, we employ the stochastic Galerkin (SG) method given in Sec. \ref{sec:collision evaluation} with 4th-order Runge-Kutta method to conduct the numerical simulation.
The simulation is conducted within the time interval $t\in[0,10]$, with the time step fixed as $\Delta t=0.01$ here.
The particle velocity space is truncated as $[-6,6]$ with 200 uniform meshes, and the gPC expansion is employed up to 9th order.

\begin{figure}[htb!]
	\centering
	\subfigure[Expectation value]{
		\includegraphics[width=7.5cm]{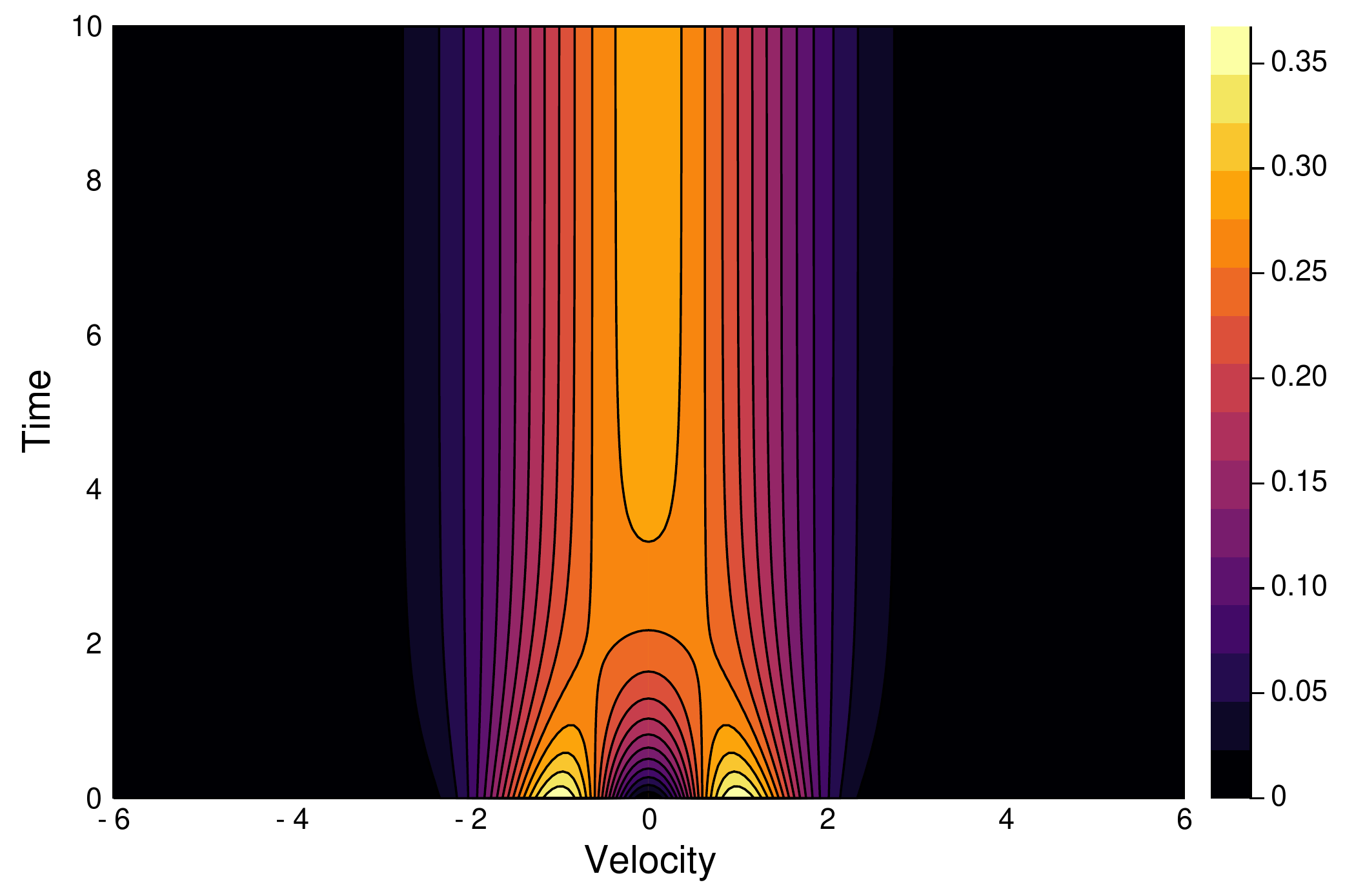}
	}
	\subfigure[Standard deviation]{
		\includegraphics[width=7.5cm]{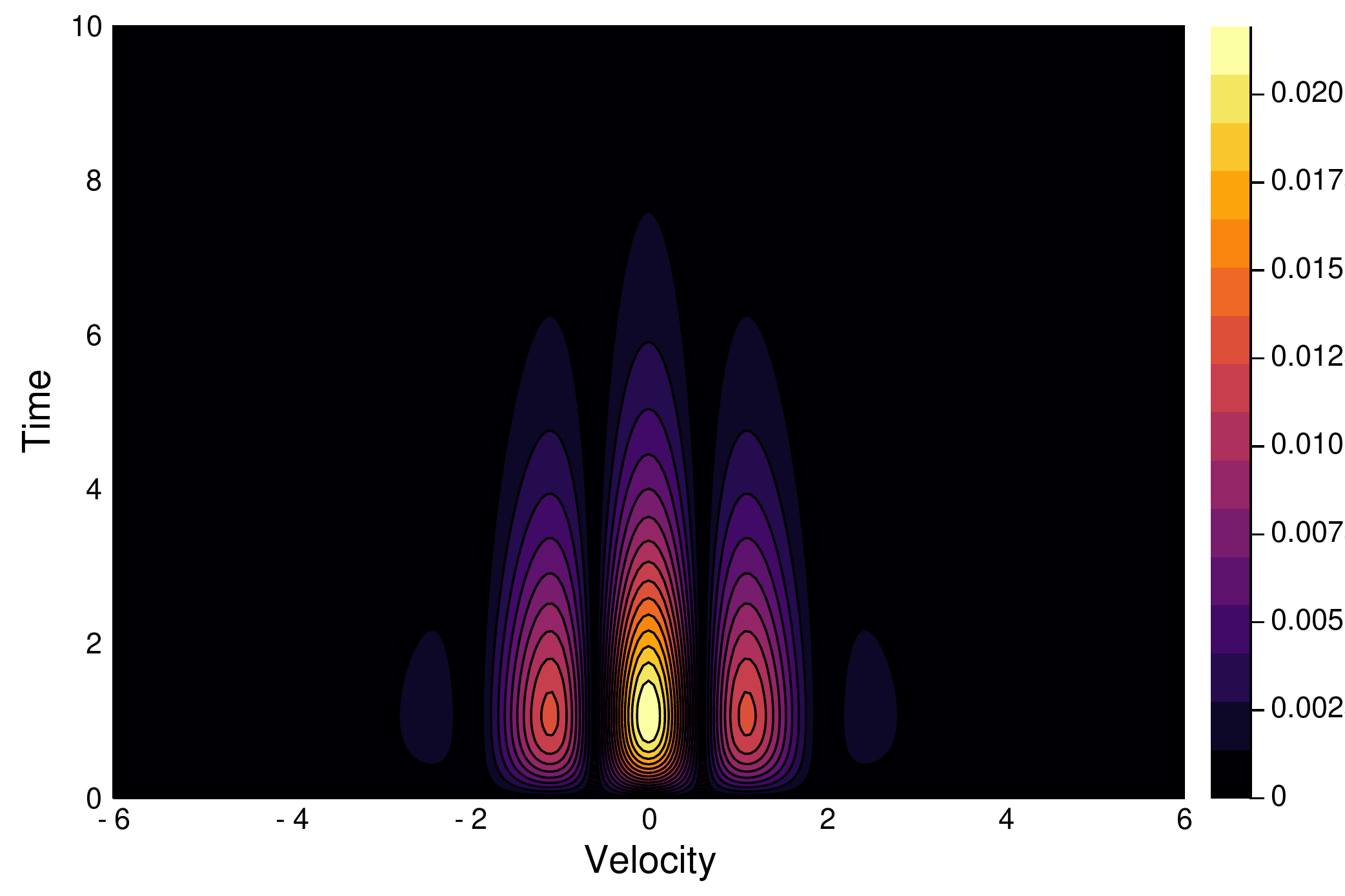}
	}
	\subfigure[Time derivative of mean value]{
		\includegraphics[width=7.5cm]{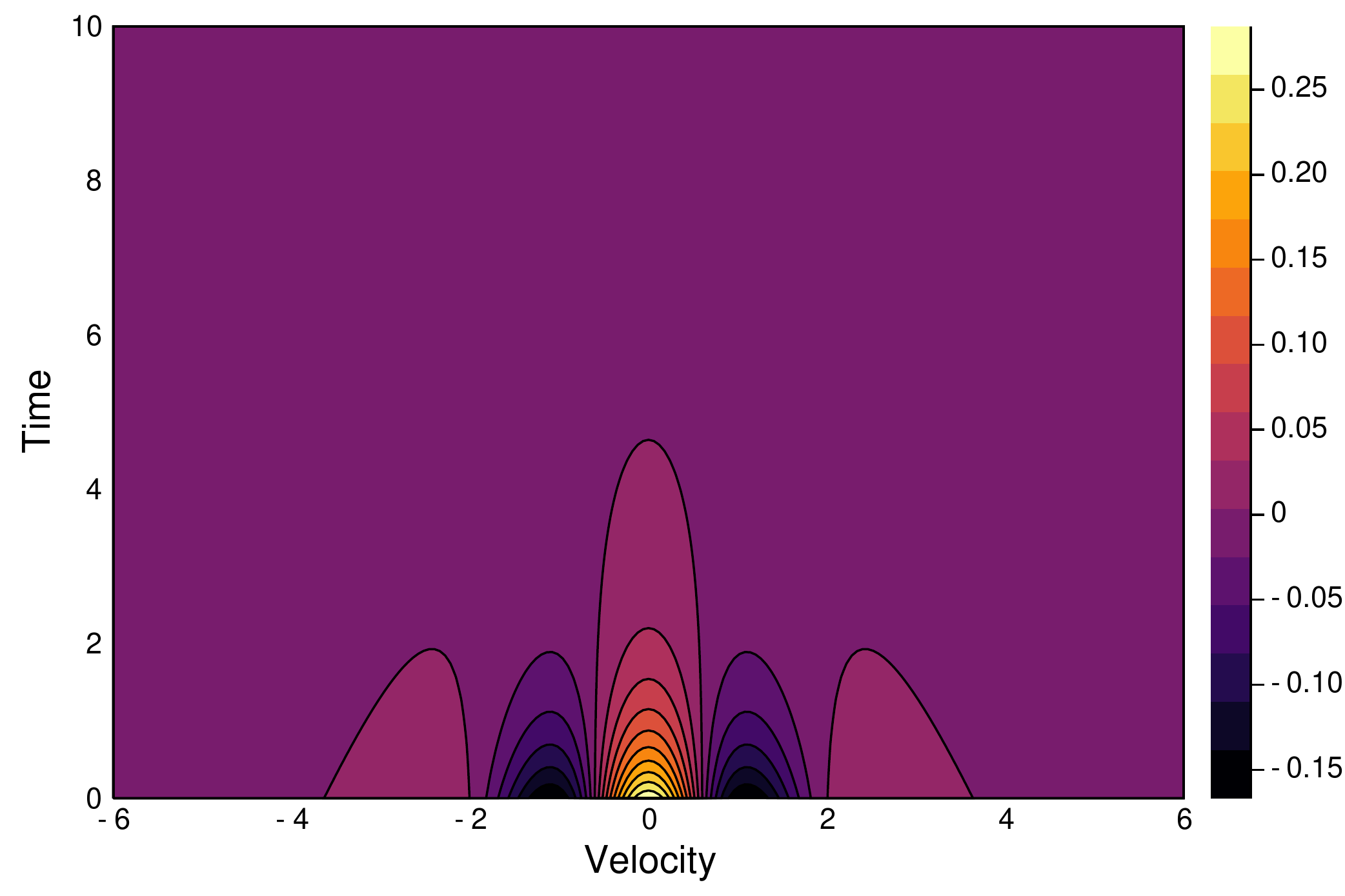}
	}
	\caption{Expectation value, its time derivatives, and standard deviation of particle distribution within $t,u\in [0,10]\times[-6,6]$ in the homogeneous relaxation problem.}
	\label{pic:relaxation micro evolution contour}
\end{figure}

\begin{figure}[htb!]
	\centering
	\subfigure[Expectation value]{
		\includegraphics[width=7.5cm]{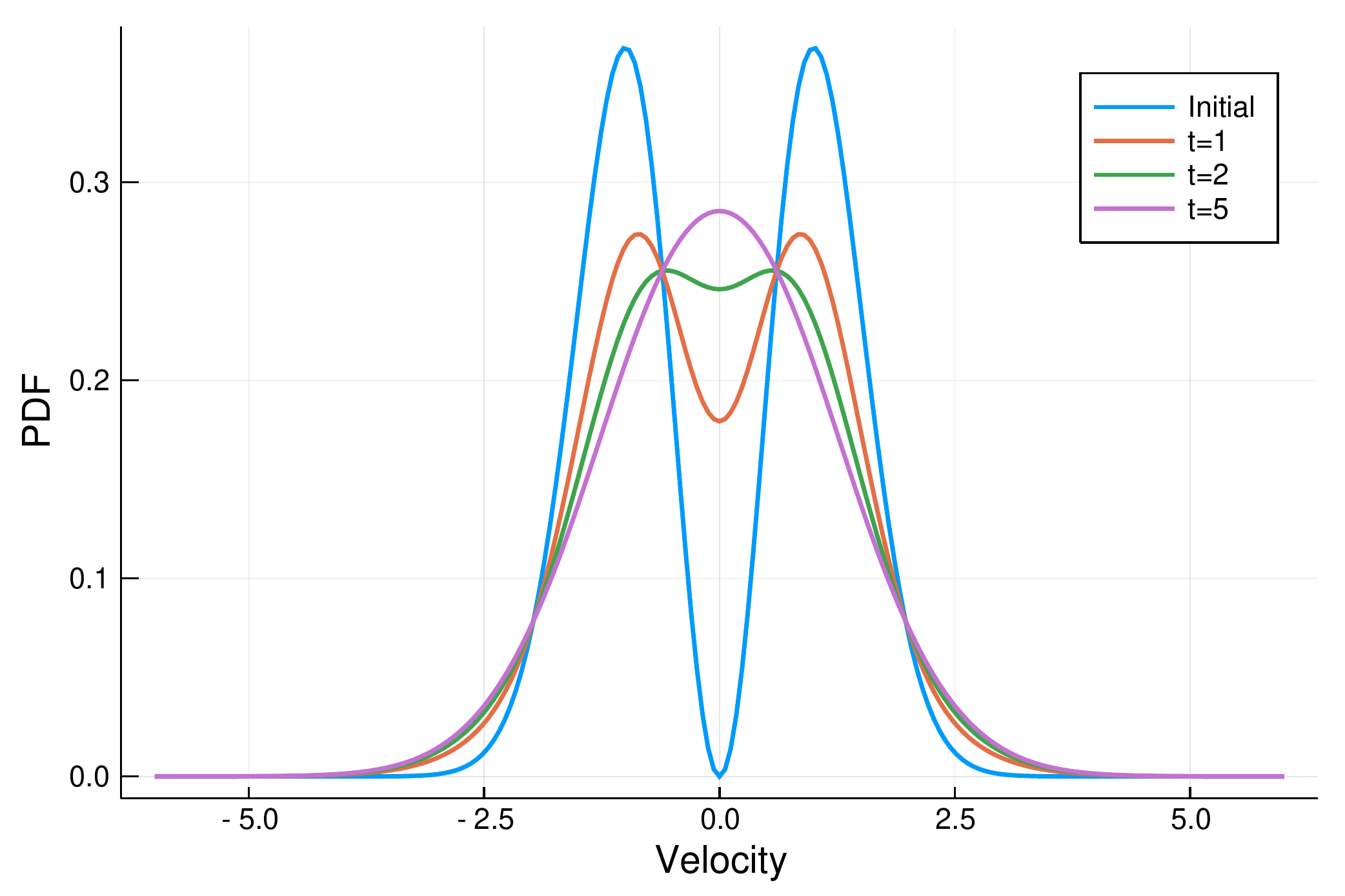}
	}
	\subfigure[Standard deviation]{
		\includegraphics[width=7.5cm]{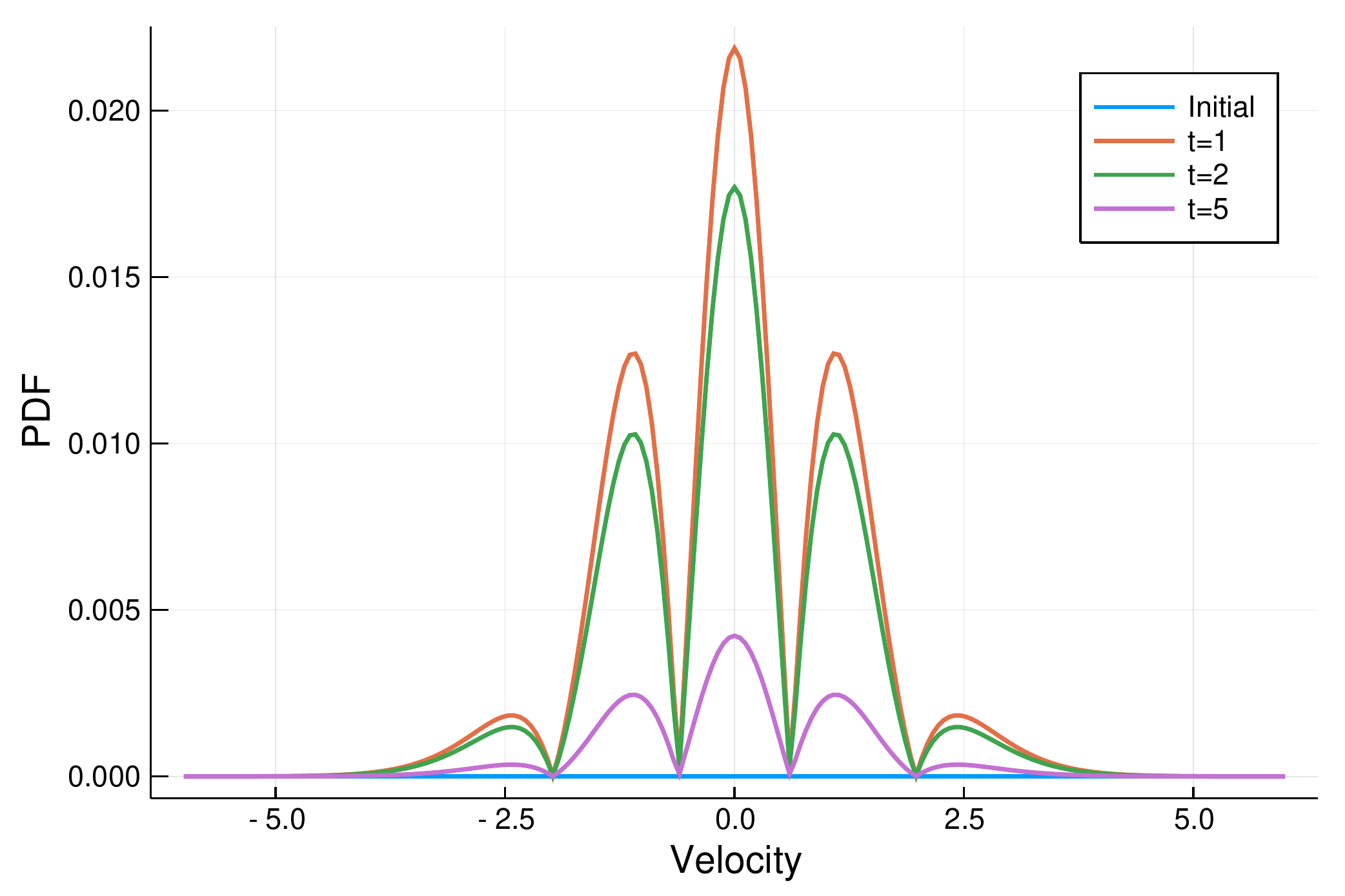}
	}
	\caption{Expectation value and standard deviation of particle distribution over velocity space at different output time in the homogeneous relaxation problem.}
	\label{pic:relaxation micro evolution curve}
\end{figure}

Fig.\ref{pic:relaxation micro evolution contour} presents the evolution of expectation and standard deviation of the particle distribution function over the entire phase space $\{ t\times u \}$,
and Fig.\ref{pic:relaxation micro evolution curve} picks up some curves over velocity space at typical output time.
As time goes, the initial bimodal particle distribution gradually approaches Maxwellian due to intermolecular collisions.
The maximum value of standard deviation emerges around $t=1$, which is a local maximum given in Eq.(\ref{eqn:homegenous mean and std}).
From a physical point of view, the random collision kernel results in more uncertainties where the distribution function is being reshaped by intermolecular interactions significantly. When $t>8$, with the distribution function being in a dynamical balance of a Maxwellian which is fully deterministic, the collision term has no explicit effects and the standard deviation approaches zero correspondingly.
Fig.\ref{pic:relaxation micro evolution contour}(b) and \ref{pic:relaxation micro evolution contour}(c) show the clear positive correlation between standard deviation and time derivative of expected value.

\begin{figure}[htb!]
	\centering
	\subfigure[Density]{
		\includegraphics[width=7.5cm]{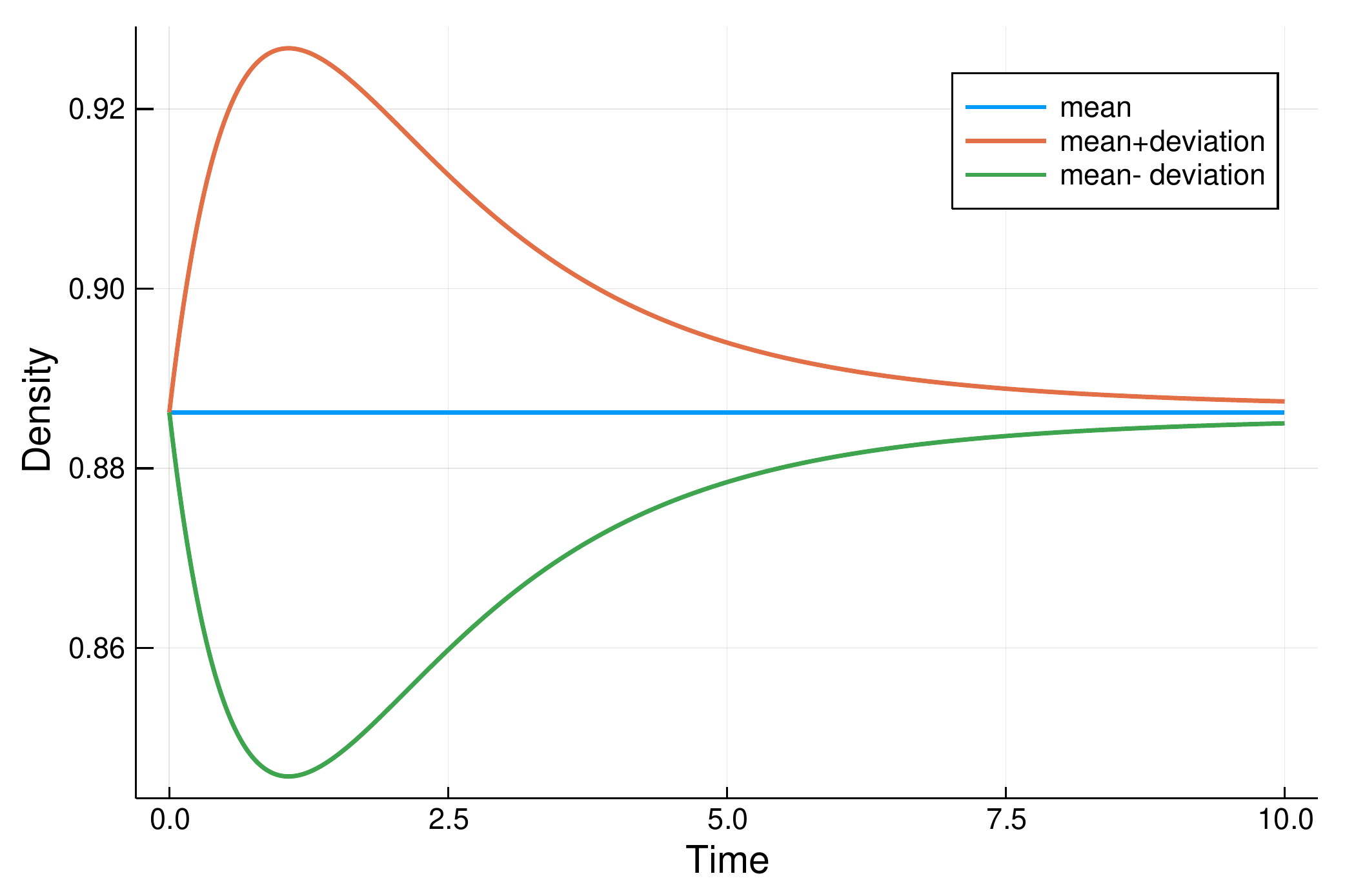}
	}
	\subfigure[Energy]{
		\includegraphics[width=7.5cm]{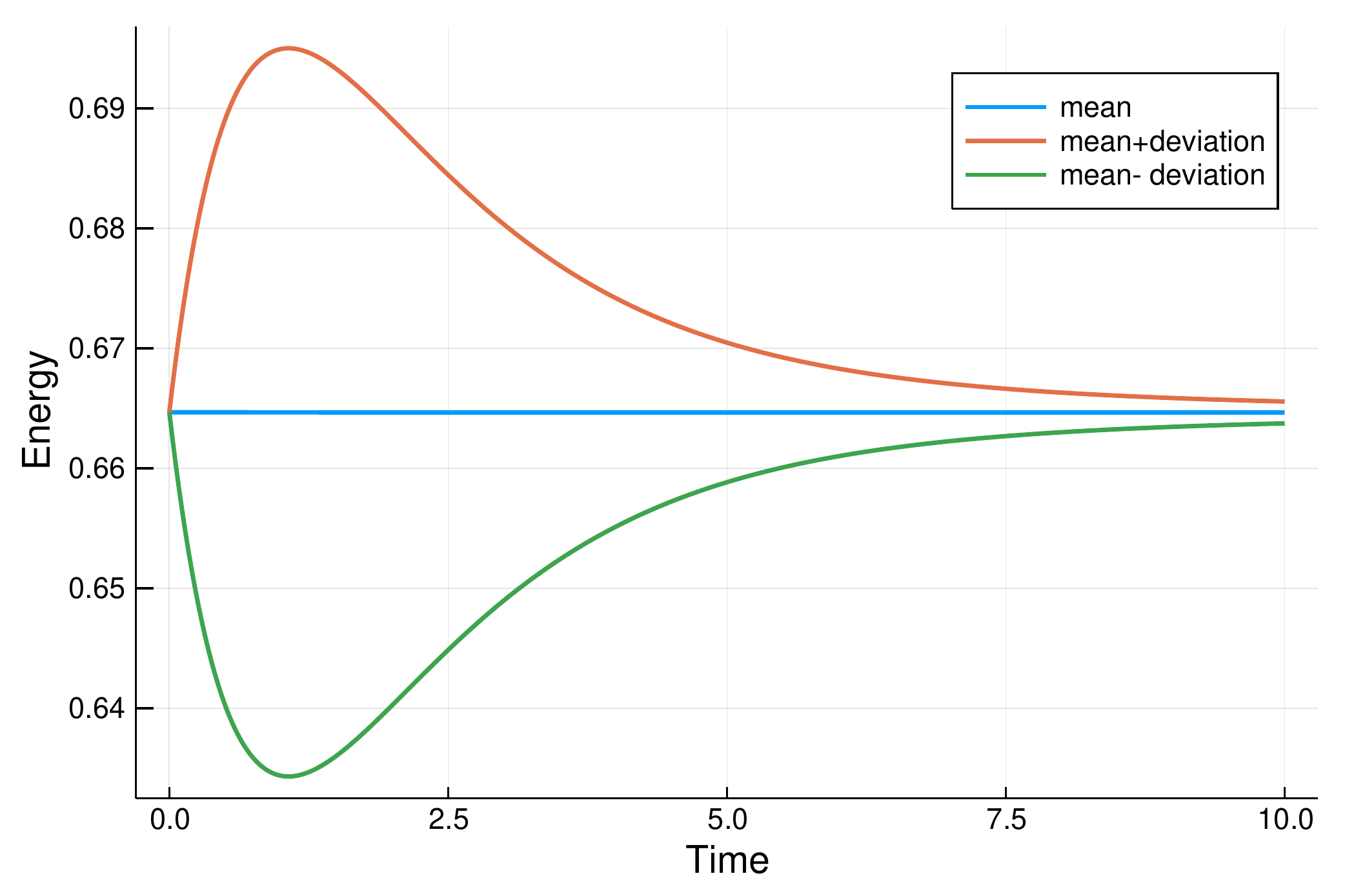}
	}
	\caption{Expectation values and standard deviations of macroscopic variables in the homogeneous relaxation problem.}
	\label{pic:relaxation macro evolution}
\end{figure}

Fig.\ref{pic:relaxation macro evolution} presents the time evolution of macroscopic density and total energy. Since there is no contribution of inhomogeneous transport, the total density and energy expectations are conserved. The pattern of standard deviation here coincides with that of particle distribution function, with a local maximum value emerging around $t=1$. Since the formulas given in Eq.(\ref{eqn:homegenous mean and std}) are always symmetric in velocity space, the macroscopic fluid velocity always equals to zero, and the random collision kernel only affects the evolution of density and energy under the current initial condition.

\begin{figure}[htb!]
	\centering
	\subfigure[$L_1$ error]{
		\includegraphics[width=7.5cm]{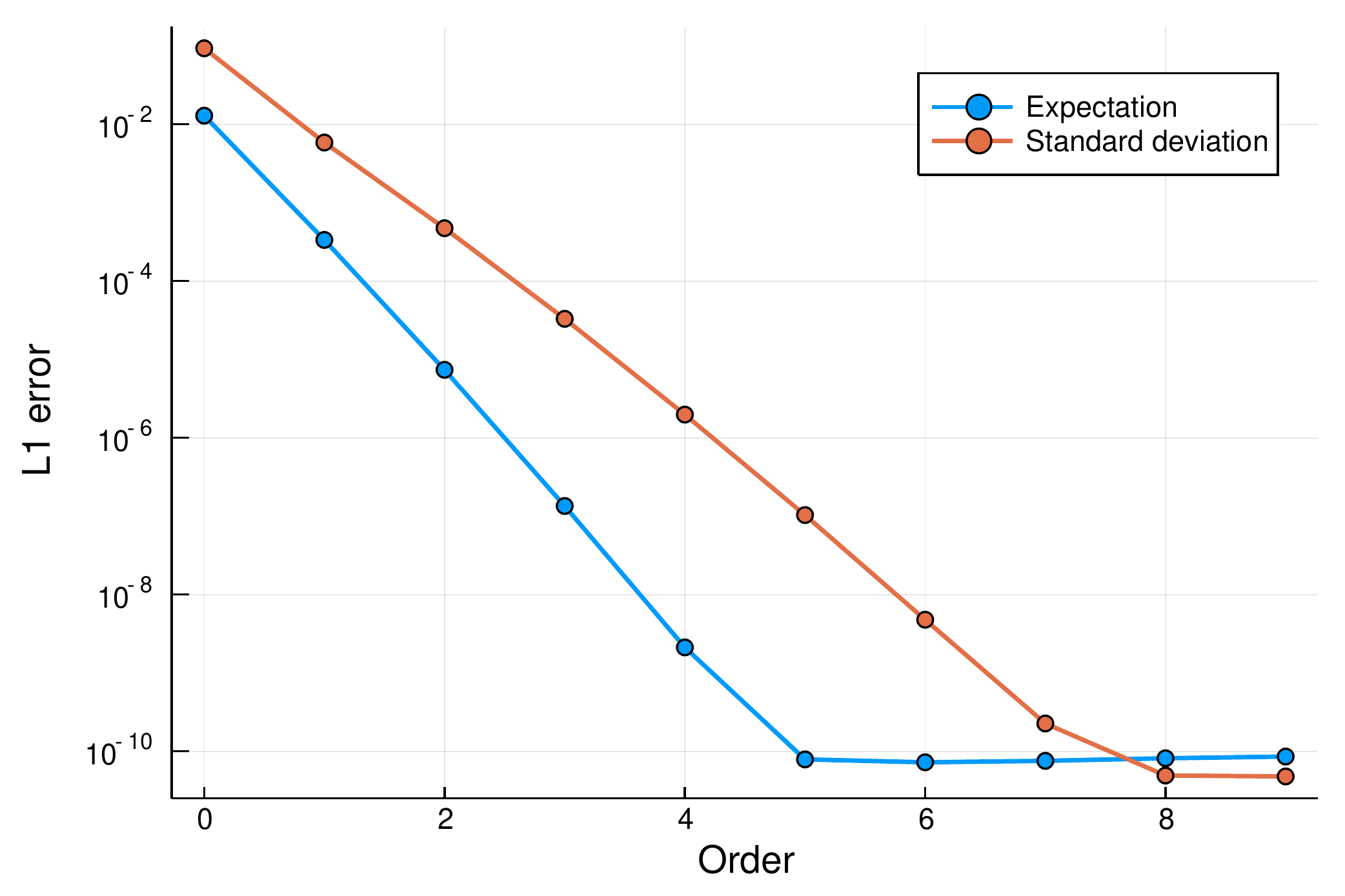}
	}
	\subfigure[$L_2$ error]{
		\includegraphics[width=7.5cm]{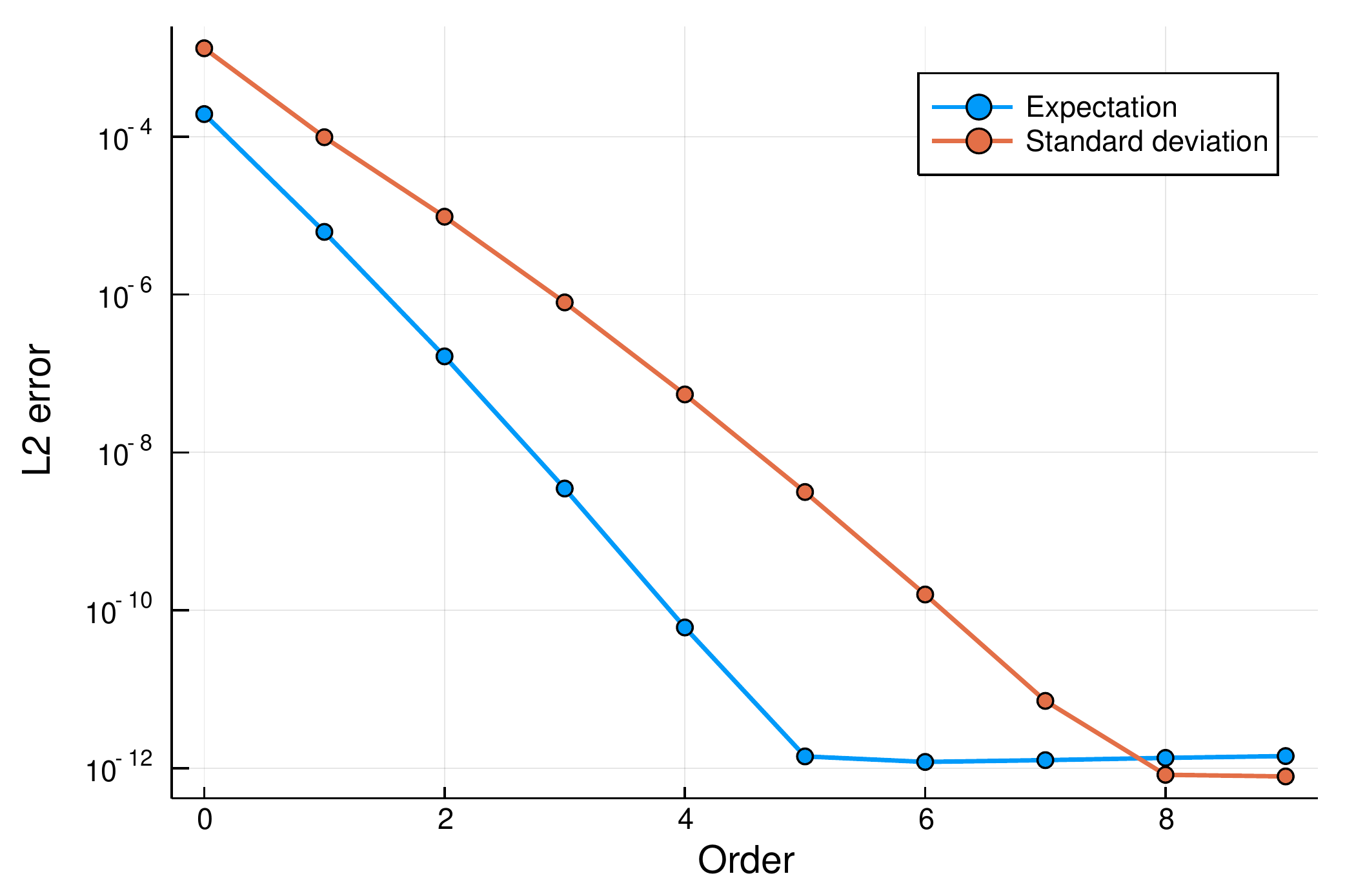}
	}
	\caption{Errors of expectation value and standard deviation of particle distribution function over the entire phase space $t,u\in [0,10]\times[-6,6]$ in the homogeneous relaxation problem.}
	\label{pic:relaxation error}
\end{figure}

To validate the current numerical scheme, Fig.\ref{pic:relaxation error} presents its convergence results compared to theoretical solutions under different gPC expansion orders. From the results, the spectral convergence of stochastic Galerkin method is clearly identified.

\subsection{Random collision kernel: normal shock structure}

From now on we turn to spatially inhomogeneous problems. The first problem is the normal shock structure \cite{muntz1969molecular}. Based on the reference frame of shock wave, the upstream and downstream gases are related with the well-known Rankine-Hugoniot relation, 
\begin{equation*}
\begin{aligned}
&\frac{\rho_+}{\rho_-}=\frac{(\gamma+1)\rm{Ma}^2}{(\gamma-1)\rm{Ma}^2+2},\\
&\frac{U_+}{U_-}=\frac{(\gamma-1)\rm{Ma}^2+2}{(\gamma+1)\rm{Ma}^2},\\
&\frac{T_+}{T_-}=\frac{ ((\gamma-1)\rm{Ma}^2+2) (2\gamma\rm{Ma}^2-\gamma+1) }{(\gamma+1)^2 \rm{Ma}^2},
\end{aligned}
\end{equation*}
where $\gamma$ is the ratio of specific heat, and the upstream macroscopic density, velocity and temperature are denoted with $\{ \rho_-, U_-, T_- \}$, and the downstream with $\{ \rho_+, U_+, T_+ \}$.
The upstream Mach number is defined as the ratio between fluid velocity and sound speed, i.e.,
\begin{equation*}
\rm{Ma}=\frac{U_-}{(\gamma/2\lambda_-)^{1/2}}.
\end{equation*}
Note that now the speed of sound $c=(\gamma/2\lambda_-)^{1/2}$ is larger than the most probable speed of molecule $(1/\lambda_-)^{1/2}$.
The randomness comes from the collision kernel, which follows Gaussian distribution in the random space,
\begin{equation*}
\nu=1+0.05z.
\end{equation*}

In this case, the upstream flow quantities are chosen as references in the nondimensionalization.
The physical domain is set as $x \in [-35, 35]$ with 100 uniform cells, where the reference length $L_0$ is the mean free path of upstreaming gas. 
The truncated particle velocity space is $u\in[-12,12]$, which is
discretized by 72 uniform quadrature points. 
The initial and boundary randomness in macroscopic variables and particle distribution function is set as zero, and the CFL number adopted is $0.95$. 
Both the stochastic Galerkin and hybrid Galerkin-collocation methods in Sec. \ref{sec:collision evaluation} are used in the simulation, with 5th order gPC expansion and 9 Gaussian quadrature points employed. The Monte-Carlo simulation with 10000 samplings is also conducted for reference.

\begin{figure}[htb!]
	\centering
	\subfigure[Density]{
		\includegraphics[width=7.5cm]{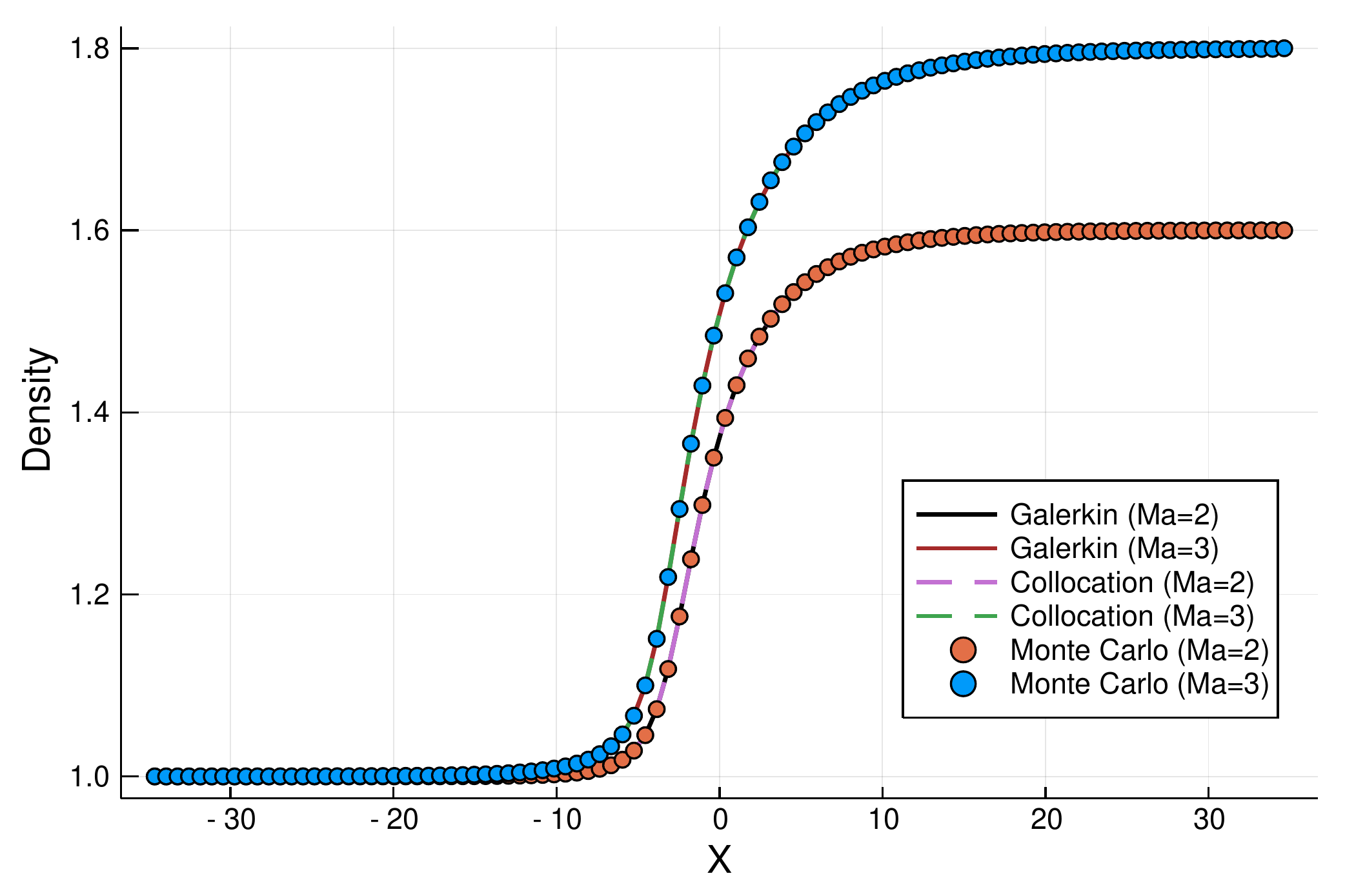}
	}
	\subfigure[Velocity]{
		\includegraphics[width=7.5cm]{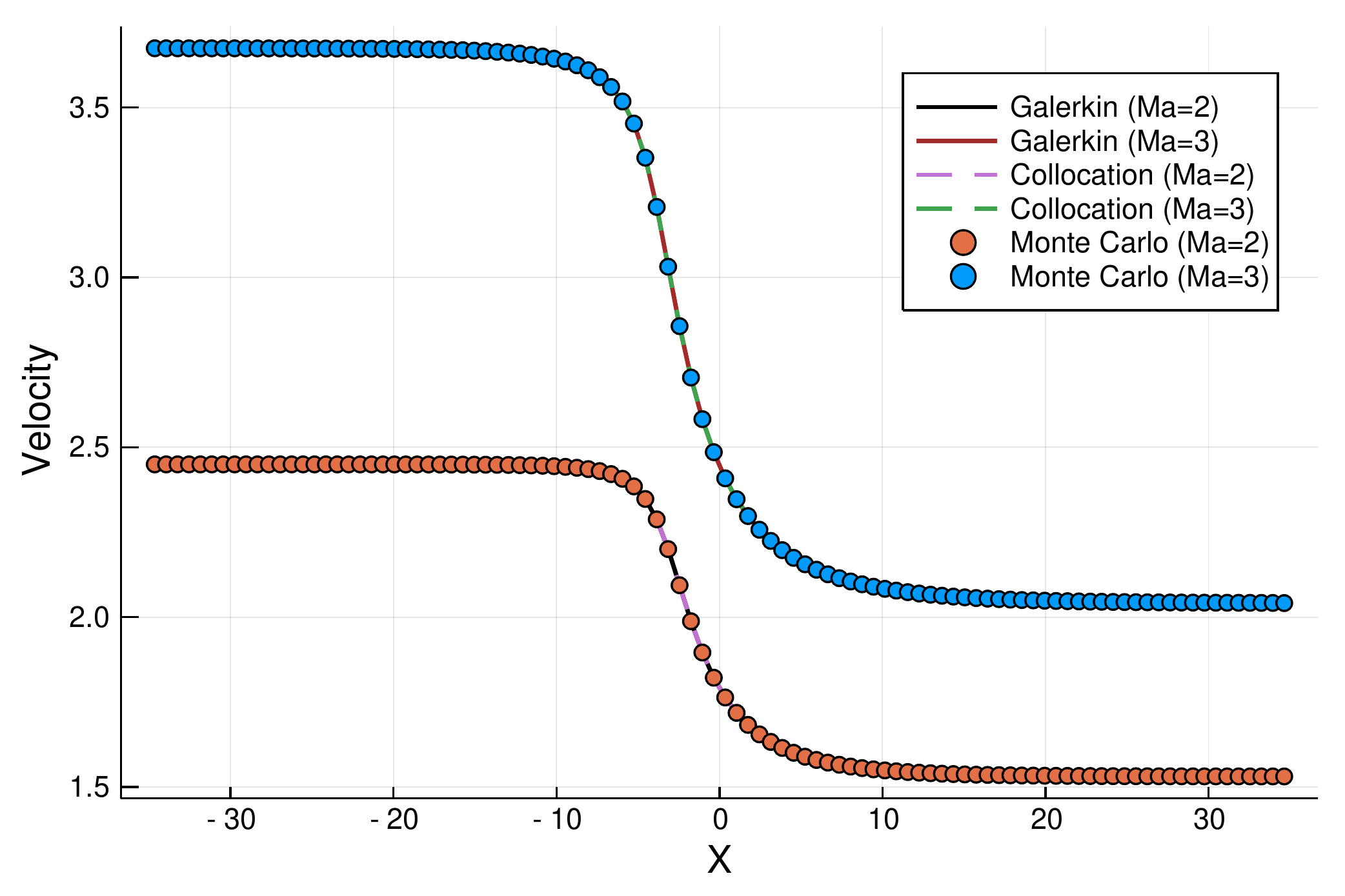}
	}
	\subfigure[Temperature]{
		\includegraphics[width=7.5cm]{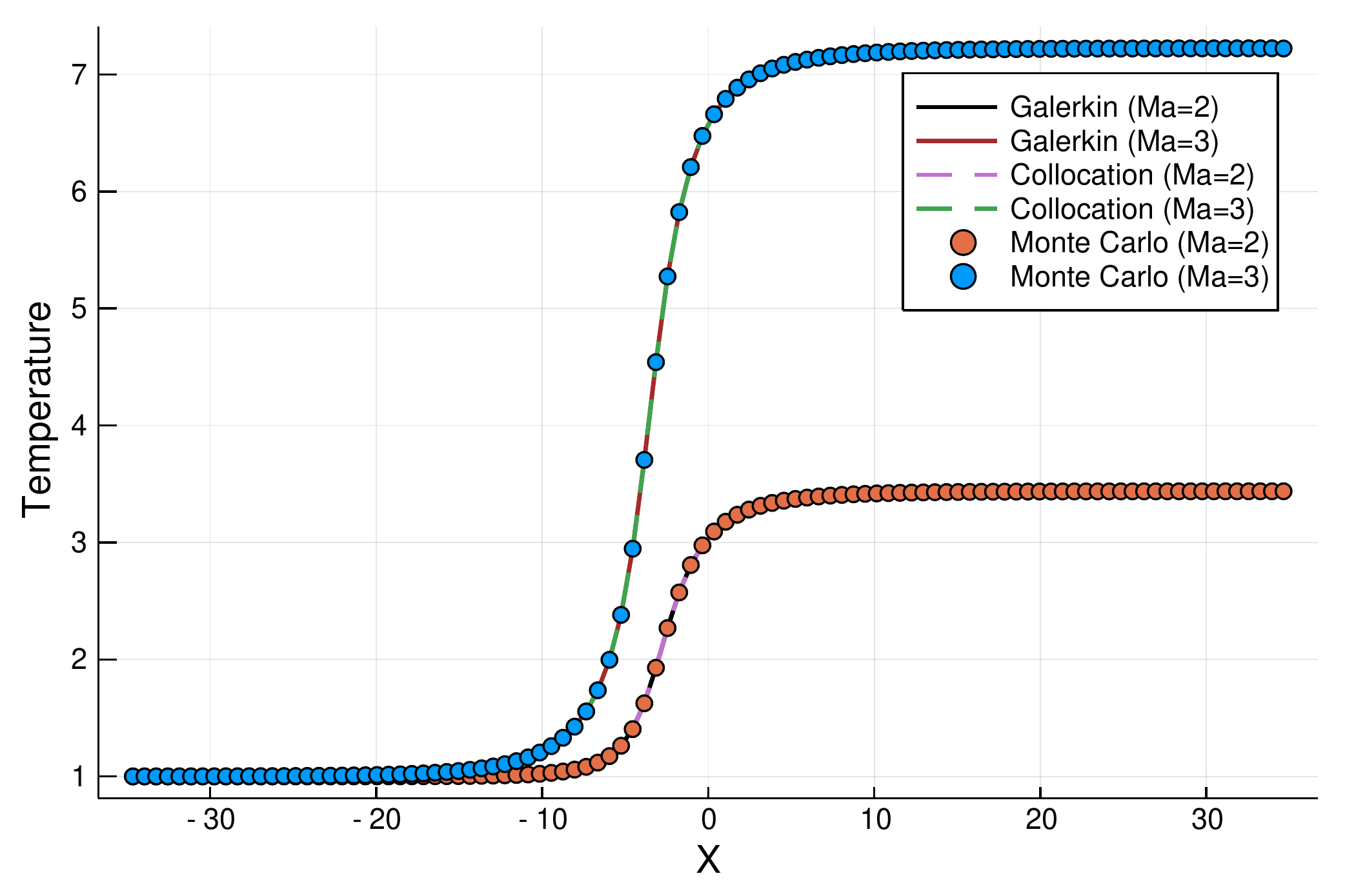}
	}
	\caption{Expectations of density, velocity and temperature around normal shock wave.}
	\label{pic:shock mean}
\end{figure}

\begin{figure}[htb!]
	\centering
	\subfigure[Density]{
		\includegraphics[width=7.5cm]{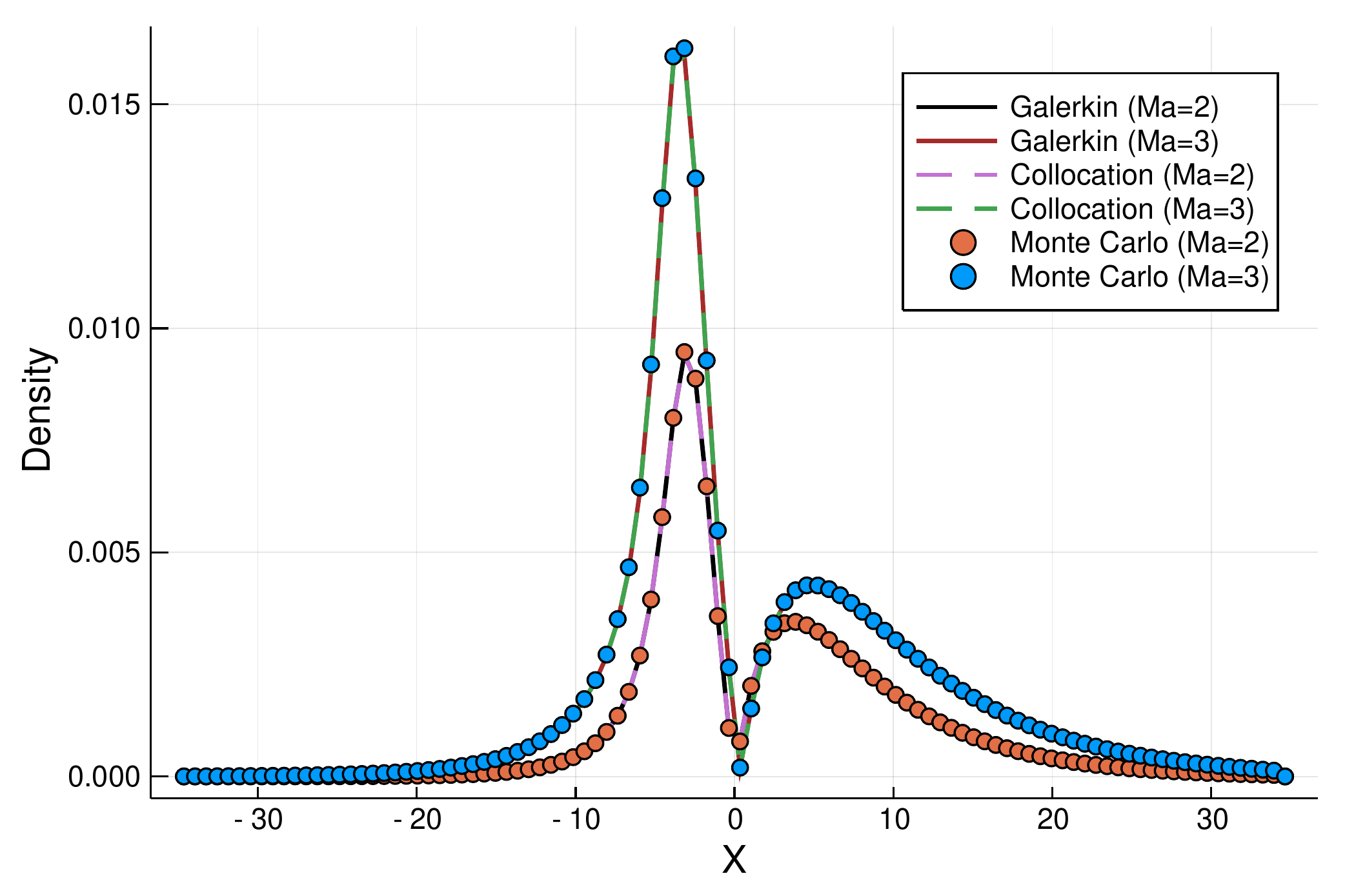}
	}
	\subfigure[Velocity]{
		\includegraphics[width=7.5cm]{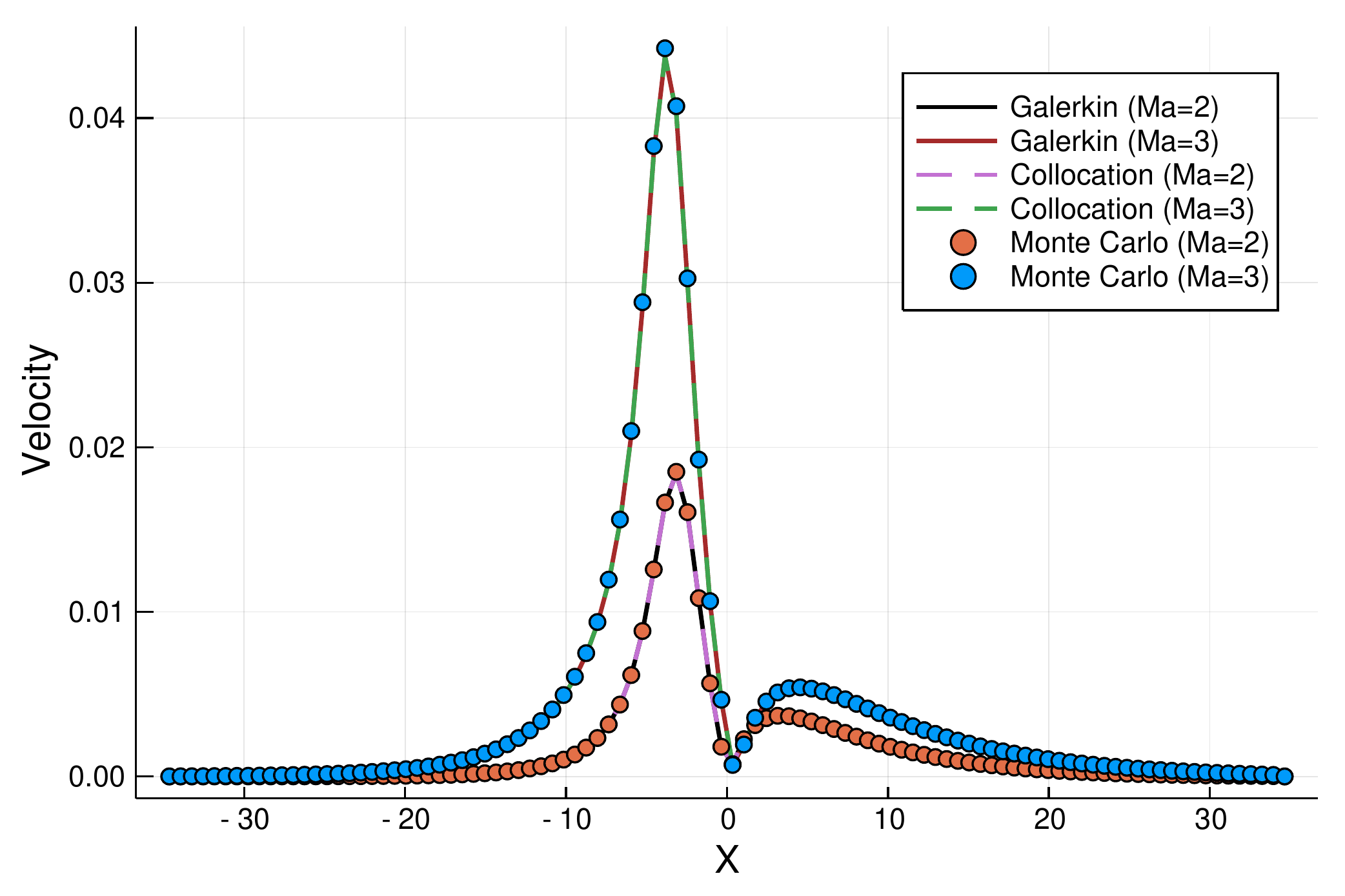}
	}
	\subfigure[Temperature]{
		\includegraphics[width=7.5cm]{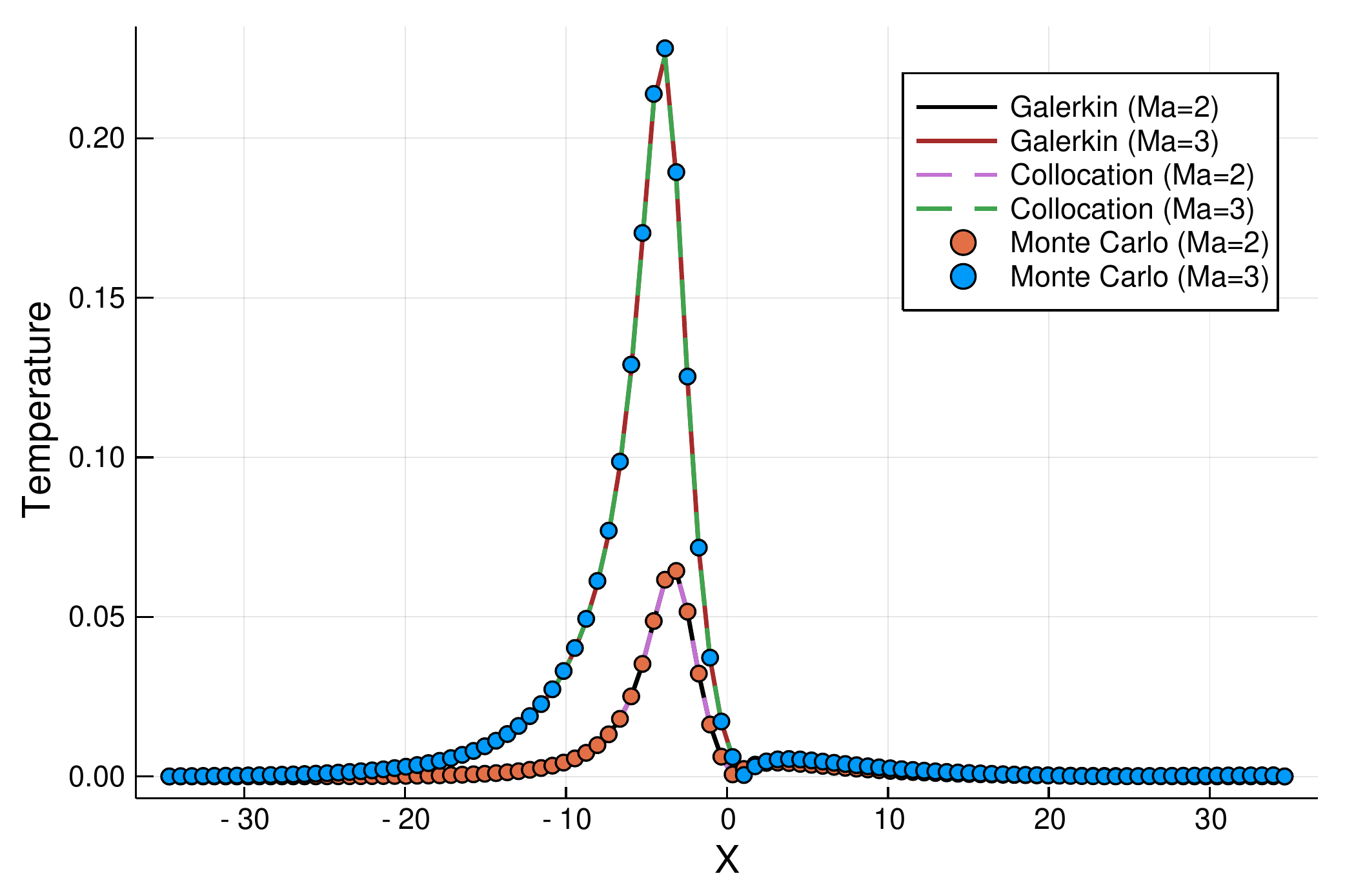}
	}
	\caption{Standard deviations of density, velocity and temperature around normal shock wave.}
	\label{pic:shock std}
\end{figure}

Fig.\ref{pic:shock mean} and \ref{pic:shock std} present the numerical solutions from the three methods at different upstream Mach numbers $\rm Ma=2$ and $3$, and
Table \ref{table:shock} shows their computational time costs. 
As shown, even with a a moderate number of samples, the Monte-Carlo method is much more time consuming than the intrusive stochastic methods.
Due to the nonlinearity held in the collision operator of kinetic equation, the proposed hybrid method is more than ten times faster than the standard SG method, but maintains the same equivalent accuracy.
At $\rm Ma=3$ the expected shock profile becomes wider than that of $\rm Ma=2$ due to the increasing momentum and energy transfers.
From Fig.\ref{pic:shock std}, it is clear that the shock wave serves as a main source for uncertainties with significant intermolecular interactions inside it.
Consistent with the behavior of expected flow quantities, the uncertainties at $\rm Ma=3$ are more significant and widely distributed than that of $\rm Ma=2$.
Besides, it is noticeable that the uncertainties of all flow variables present a bimodal pattern inside the shock profile.
Given the initial Rankine-Hugoniot jump relationship, the flow conditions at the center of shock $x=0$ are basically fixed, while the Mach number and collision kernel affect the shape and span of the shock profile.
As shown in Fig.\ref{pic:shock std}, the upstream half of the shock wave seems to be more sensitive to the random collision kernel, resulting in a steeper distribution of uncertainties.
After that, it approaches zero at the location of initial discontinuity and then arises again with a wider and moderate distribution at the downstream half.
Of all the three macroscopic flow variables, the density profile contains considerable magnitude of uncertainty in the downstream part, while the temperature randomness is nearly located in the upstream half.

\begin{table}
\centering
\begin{tabular}{cccc}
	\hline
	&Galerkin& Hybrid Galerkin-Collocation& Monte Carlo\\
	\hline
	Ma=2&5500& 488& 62575\\
	Ma=3&13344& 1180& 140430\\
	\hline
\end{tabular}
\caption{The computational time costs in seconds of different numerical methods in normal shock structure problem.}
\label{table:shock}
\end{table}

\subsection{Random initial input: multi-scale shock tube} \label{sec:sod}

The next case is Sod problem. 
The initial gas inside a one-dimensional tube $x\in[0,1]$ is set as,
\begin{equation*}
\begin{aligned}
&\rho_L,\quad U_L=0,\quad p_L=1.0, \quad x\le 0.5, \\
&\rho_R=0.125,\quad U_R=0,\quad p_R=0.1, \quad x> 0.5,
\end{aligned}
\end{equation*}
with the corresponding particle distribution function being Maxwellian everywhere.
We employ the variable hard-sphere (VHS) gas here, with its viscosity coefficient defined as,
\begin{equation*}
\mu=\mu_{ref} \left( \frac{T}{T_{ref}} \right)^\eta,
\end{equation*}
and the reference state is related with Knudsen number,
\begin{equation*}
\mu_{ref} =  \frac{5 (\alpha + 1) (\alpha + 2) \sqrt\pi}  {4 \alpha (5 - 2 \omega) (7 - 2 \omega)}   \mathrm{Kn}_{ref},
\end{equation*}
where the parameters $\{ \alpha, \omega, \eta \}$ take the value $\{1,0.5,0.72\}$.
The collision frequency is determined by
\begin{equation*}
\nu = \frac{p}{\mu},
\end{equation*}
where $p$ is the pressure.

In the simulation, the physical domain is divided into 150 uniform cells, and the particle velocity space $u\in[-5,5]$ is discretized into 72 uniform quadrature points to update the distribution function.
To test multi-scale performance of the current scheme, simulations are performed with different reference Knudsen numbers $\mathrm{Kn}_{ref} = 1.0\times 10^{-4}$, $1.0\times 10^{-2}$, and $1.0$, with respect to typical continuum, transition, and free molecular flow regimes.

In this case, the uncertainties get involved into the stochastic system through random initial inputs.
We consider two kinds of random distribution of the left-hand-side density, i.e. the normal and the uniform distributions over the random space,
\begin{equation*}
\begin{aligned}
&(1) \rho_L \sim \mathcal N[\mu,\sigma^2], \ \mu=1.0,\ \sigma=0.0289,\\
&(2) \rho_L \sim \mathcal U[a,b],\ a=0.95,\ b=1.05.
\end{aligned}
\end{equation*}
The parameters $\{\mu, \sigma, a, b\}$ are chosen in the way of keeping the same expectation and variance of initial density based on the probability theory.
The stochastic Galerkin method is employed with 6th-order gPC expansion, while the reference solutions are conducted by the collocation method with 800 uniform cells.

\begin{figure}[htb!]
	\centering
	\subfigure[Density]{
		\includegraphics[width=7.5cm]{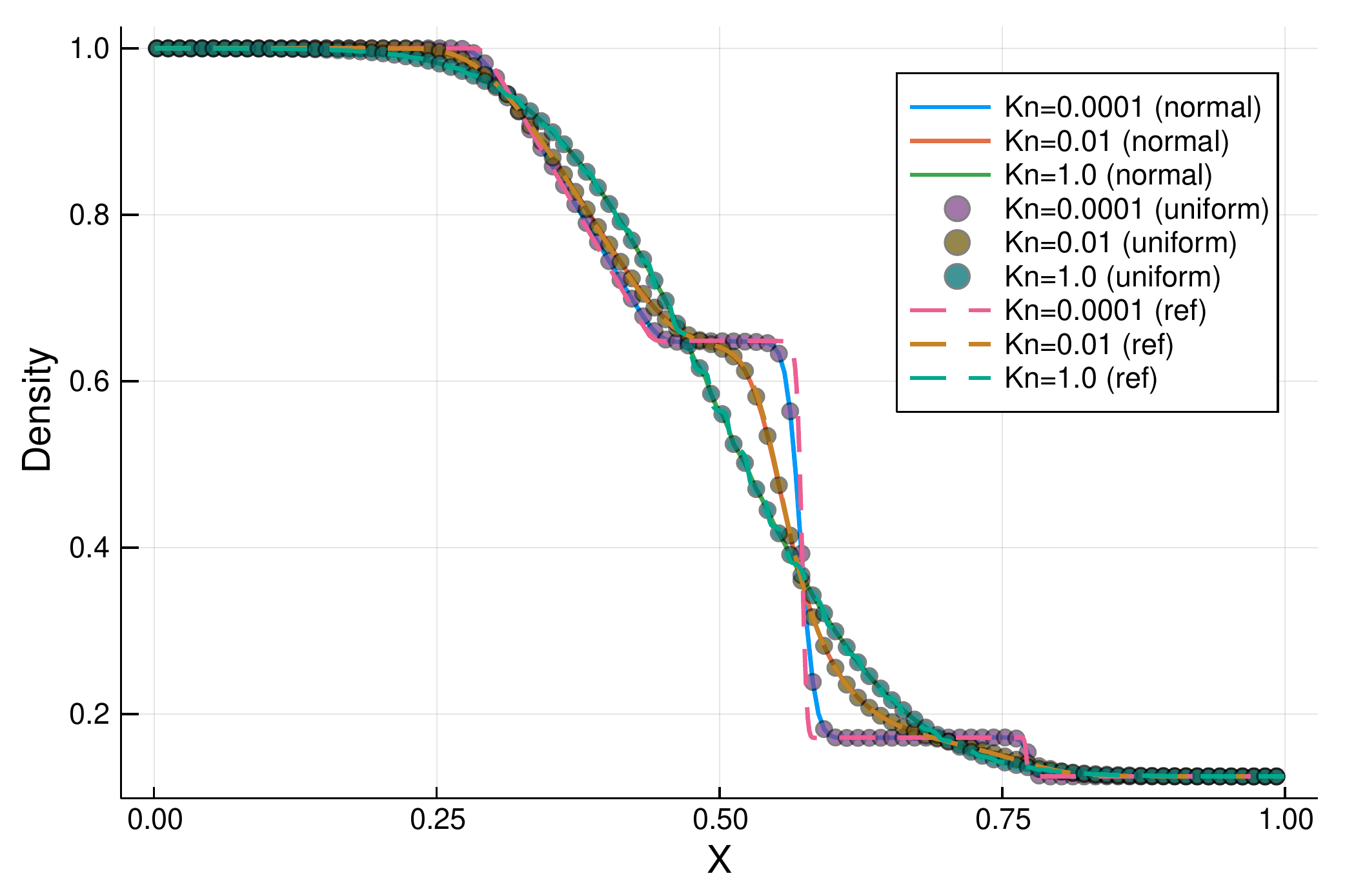}
	}
	\subfigure[Velocity]{
		\includegraphics[width=7.5cm]{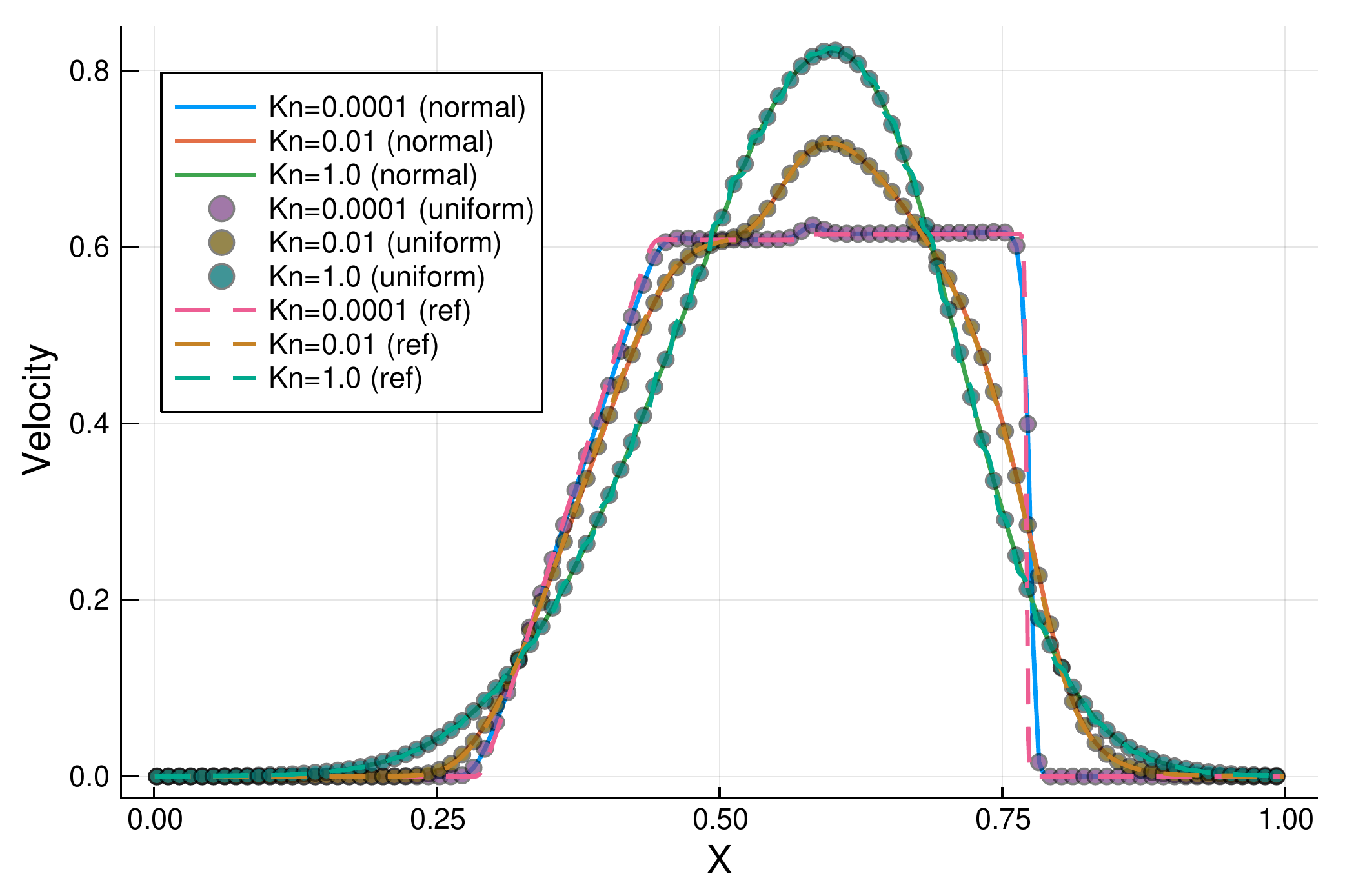}
	}
	\subfigure[Temperature]{
		\includegraphics[width=7.5cm]{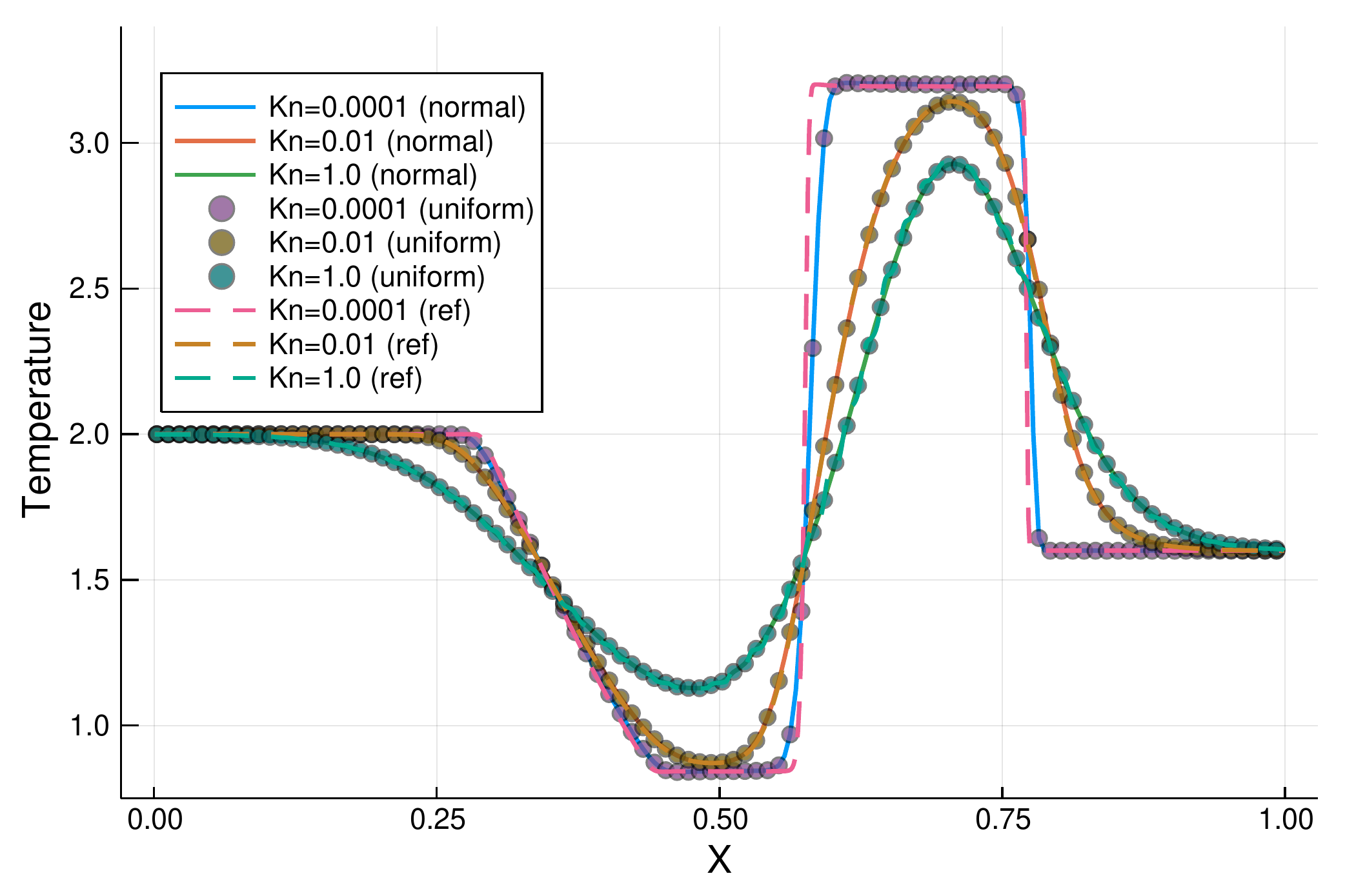}
	}
	\caption{Expectations of density, velocity and temperature inside shock tube at different Knudsen numbers.}
	\label{pic:sod mean}
\end{figure}

\begin{figure}[htb!]
	\centering
	\subfigure[Density]{
		\includegraphics[width=7.5cm]{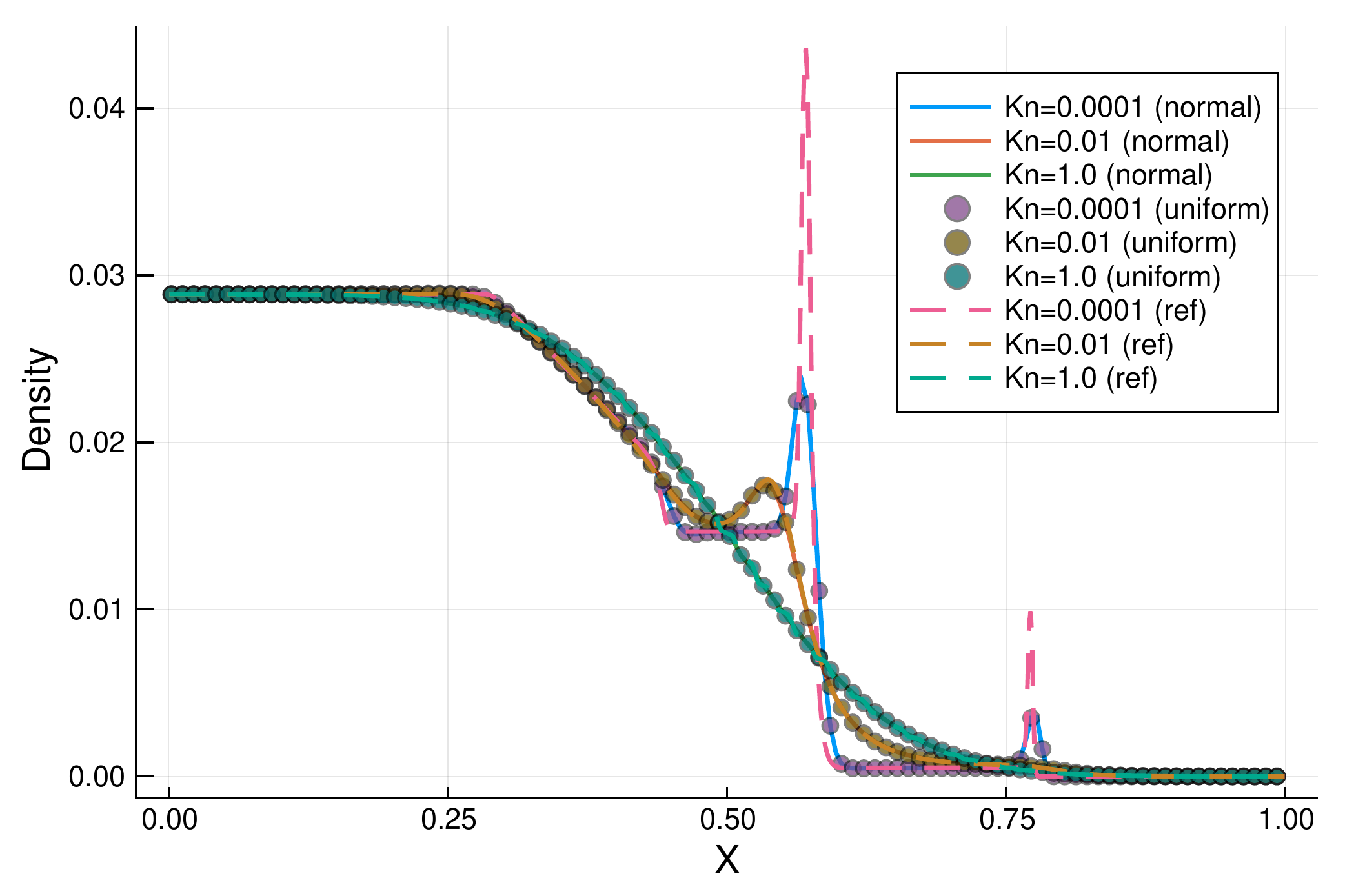}
	}
	\subfigure[Velocity]{
		\includegraphics[width=7.5cm]{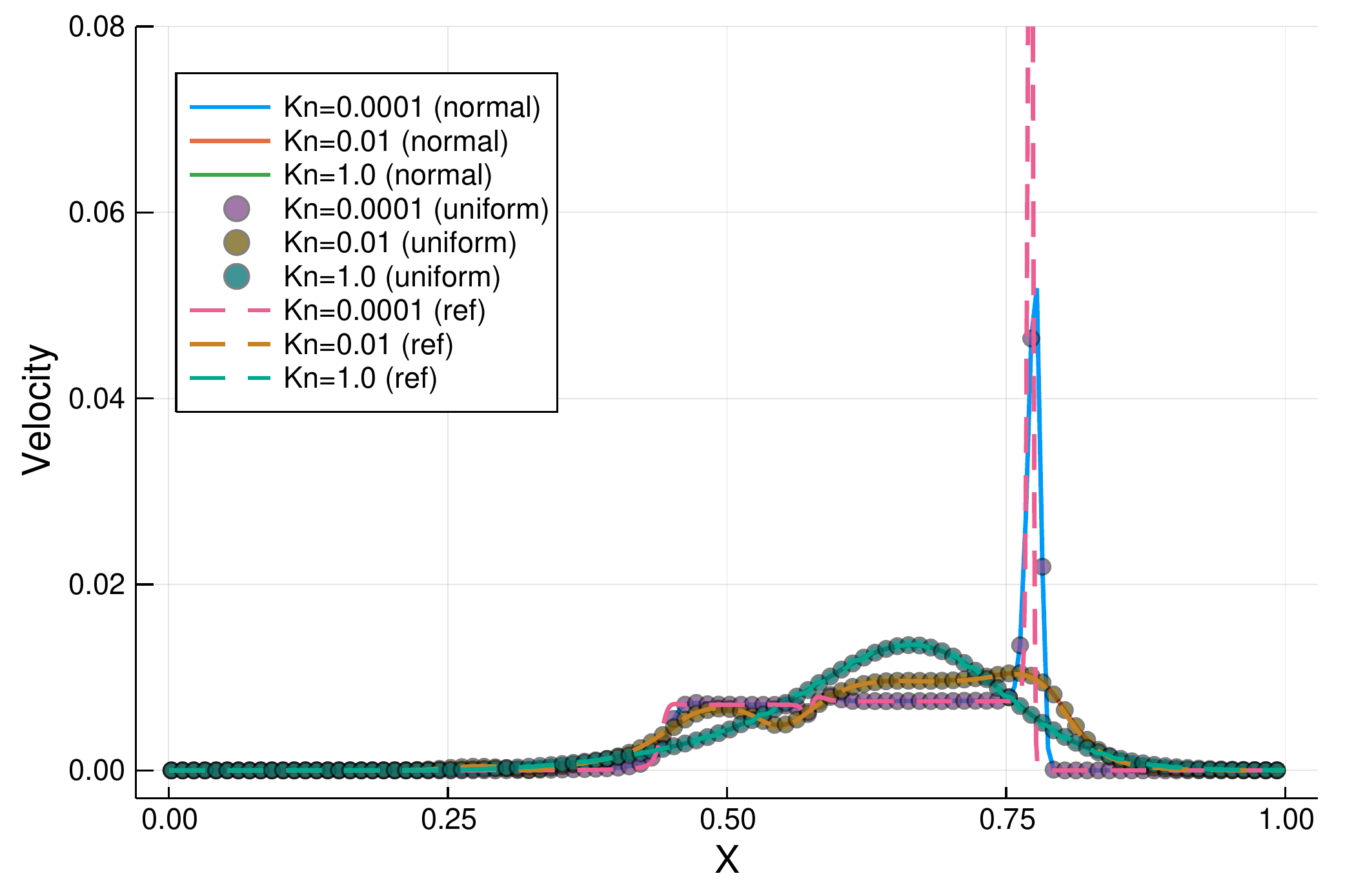}
	}
	\subfigure[Temperature]{
		\includegraphics[width=7.5cm]{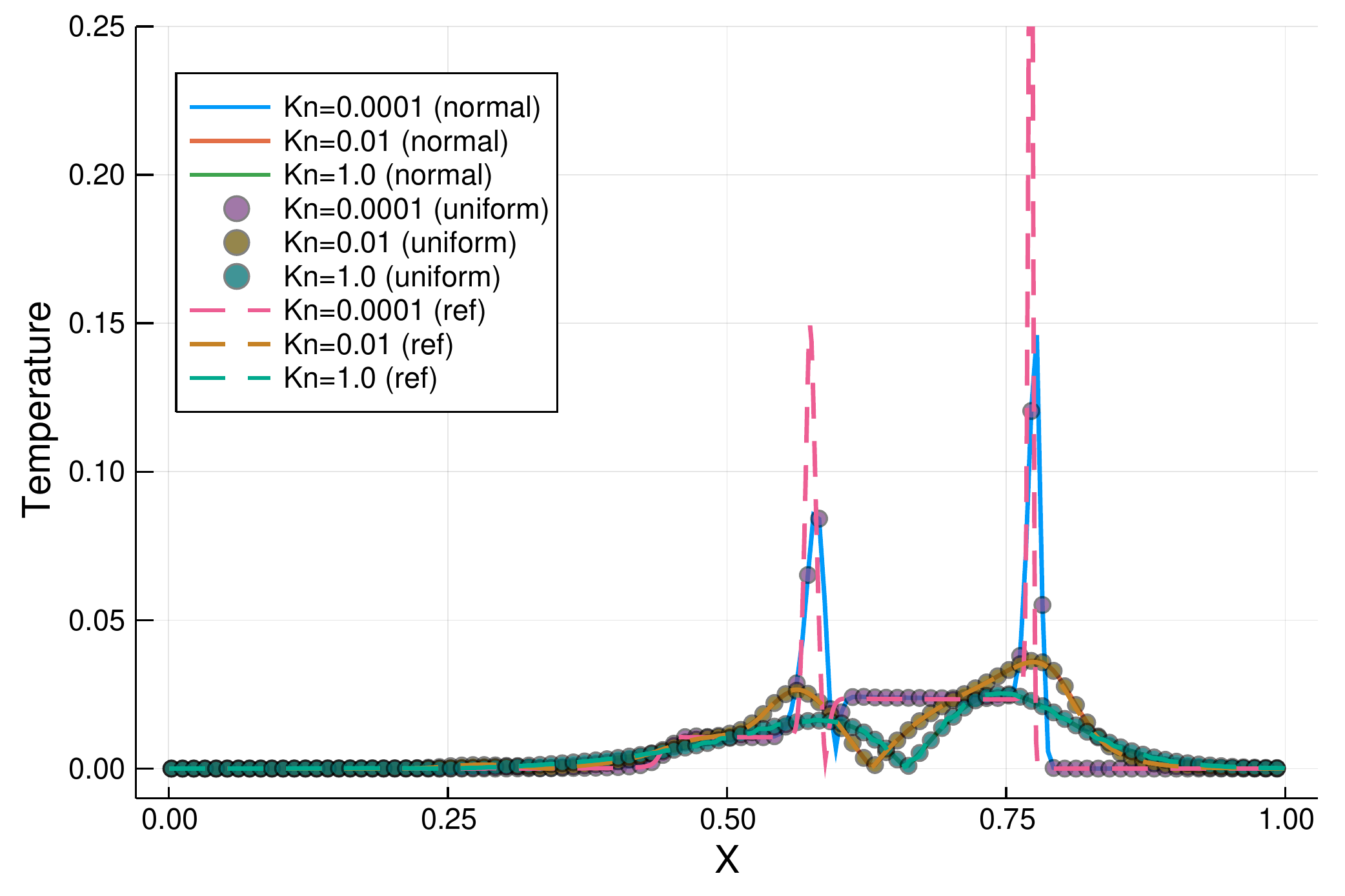}
	}
	\caption{Standard deviations of density, velocity and temperature inside shock tube at different Knudsen numbers.}
	\label{pic:sod std}
\end{figure}

The numerical expectation solutions of macroscopic variables at $t = 0.12$ are shown in Fig.\ref{pic:sod mean} and \ref{pic:sod std}. 
In the continuum regime with $\mathrm{Kn}_{ref} = 1.0\times 10^{-4}$, the molecular relaxation time is much smaller than the time step. 
As a result, the current scheme becomes a shock-capturing method due to limited resolution in space and time, and thus produces Euler solutions of the Riemann problem with wave-interaction structures.
With increasing reference Knudsen number and molecular mean fee path, the degrees of freedom for individual particle free transport increase, and the flow physics changes significantly along with the enhanced transport phenomena.
From $\mathrm{Kn}_{ref} = 1.0\times 10^{-4}$ to $\mathrm{Kn}_{ref} = 1.0$, a smooth transition is recovered from the Euler solutions of Riemann problem to collisionless Boltzmann solutions.

Fig.\ref{pic:sod std} presents the standard deviations at the same output instant.
Generally speaking, the uncertainties travel along with the wave structure of expectation values and present similar propagating patterns.
At $\mathrm{Kn}_{ref} = 1.0\times 10^{-4}$, structures such as rarefaction wave, contact discontinuity and shock also form inside the profiles of standard deviation.
Given the uncertainty from initial gas density at the left hand side of the tube, it can be seen that the wave structures serves as other sources where the local maximums of variance emerge.
Compared with the expectation value, it seems that the variance is more sensitive to physical discontinuities and holds finer-scale structures.
As a result, the overshoots near contact discontinuity and shock cannot be well resolved by the shock-capturing scheme due to the limited resolution and there exist deviations between numerical and reference solutions, but it is clear that all the key structures are preserved.
With increasing Knudsen numbers, the profiles of standard deviations get much smoother along with the wave-propagation structures inside the tube.
At $\mathrm{Kn}_{ref} = 1.0$, the density and velocity variance profiles show similar transition layers as their expectation values between upstream and downstream flow conditions.

\begin{figure}[htb!]
	\centering
	\subfigure[Density]{
		\includegraphics[width=5cm]{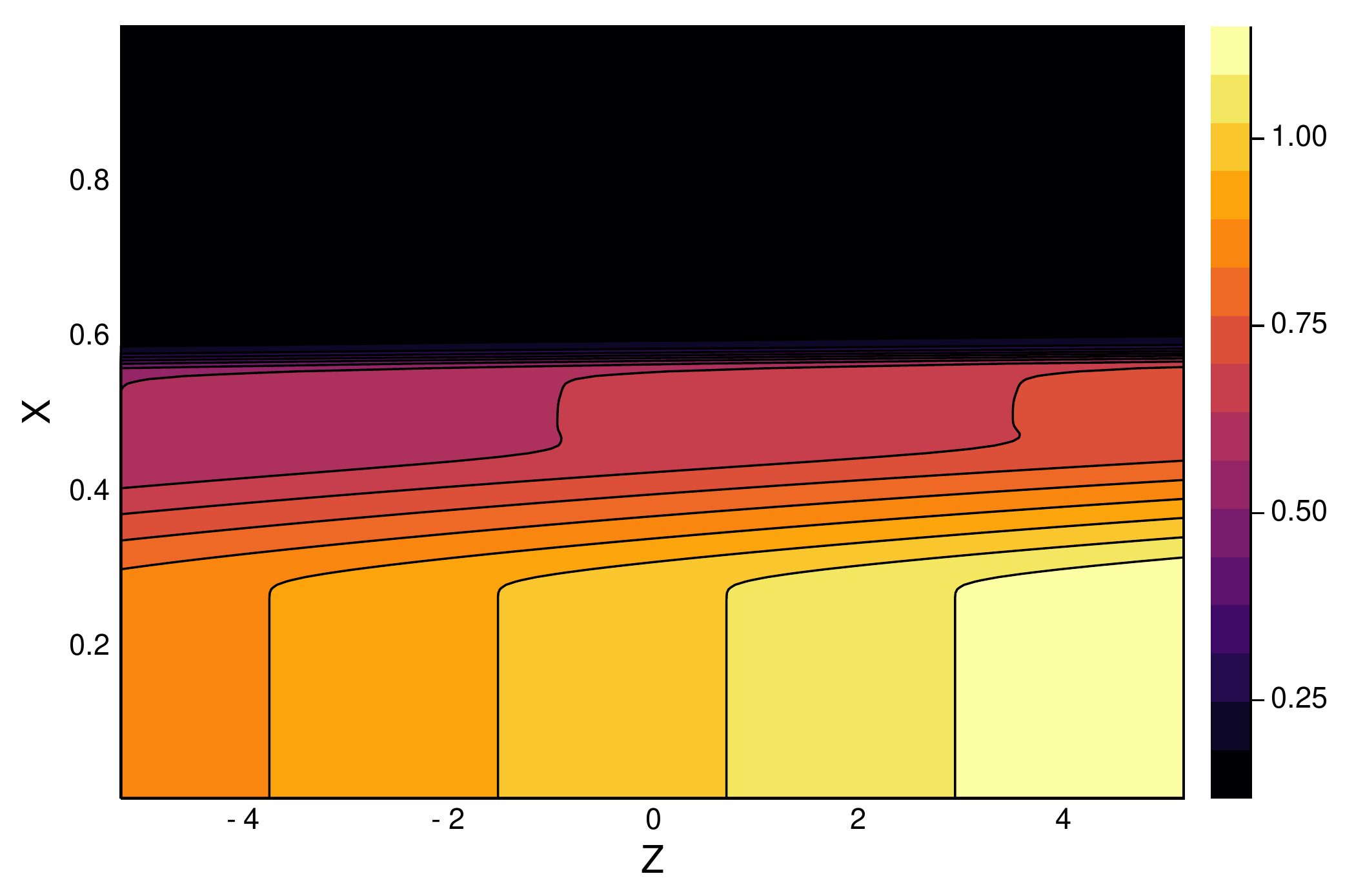}
	}
	\subfigure[Velocity]{
		\includegraphics[width=5cm]{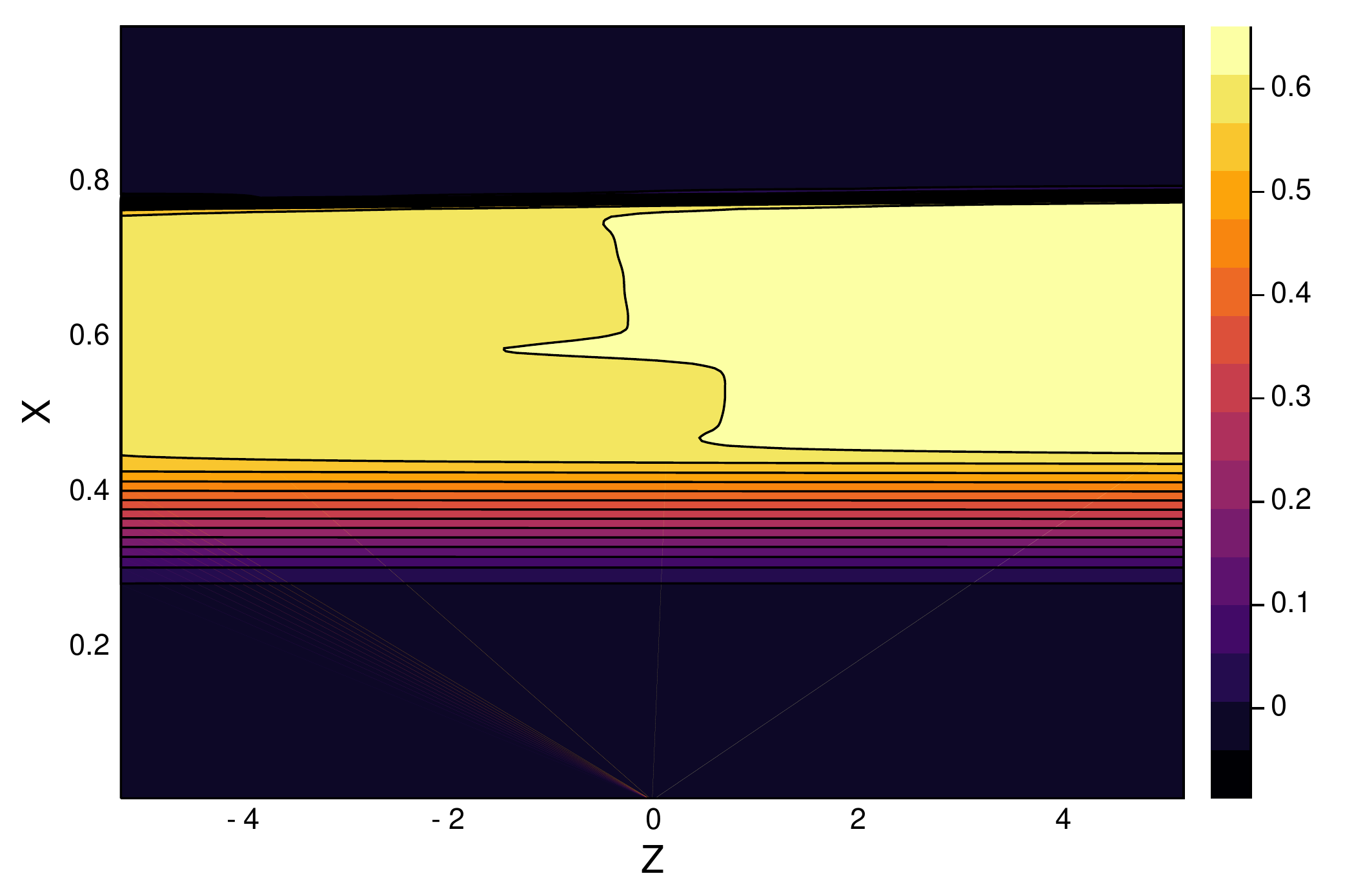}
	}
	\subfigure[Temperature]{
		\includegraphics[width=5cm]{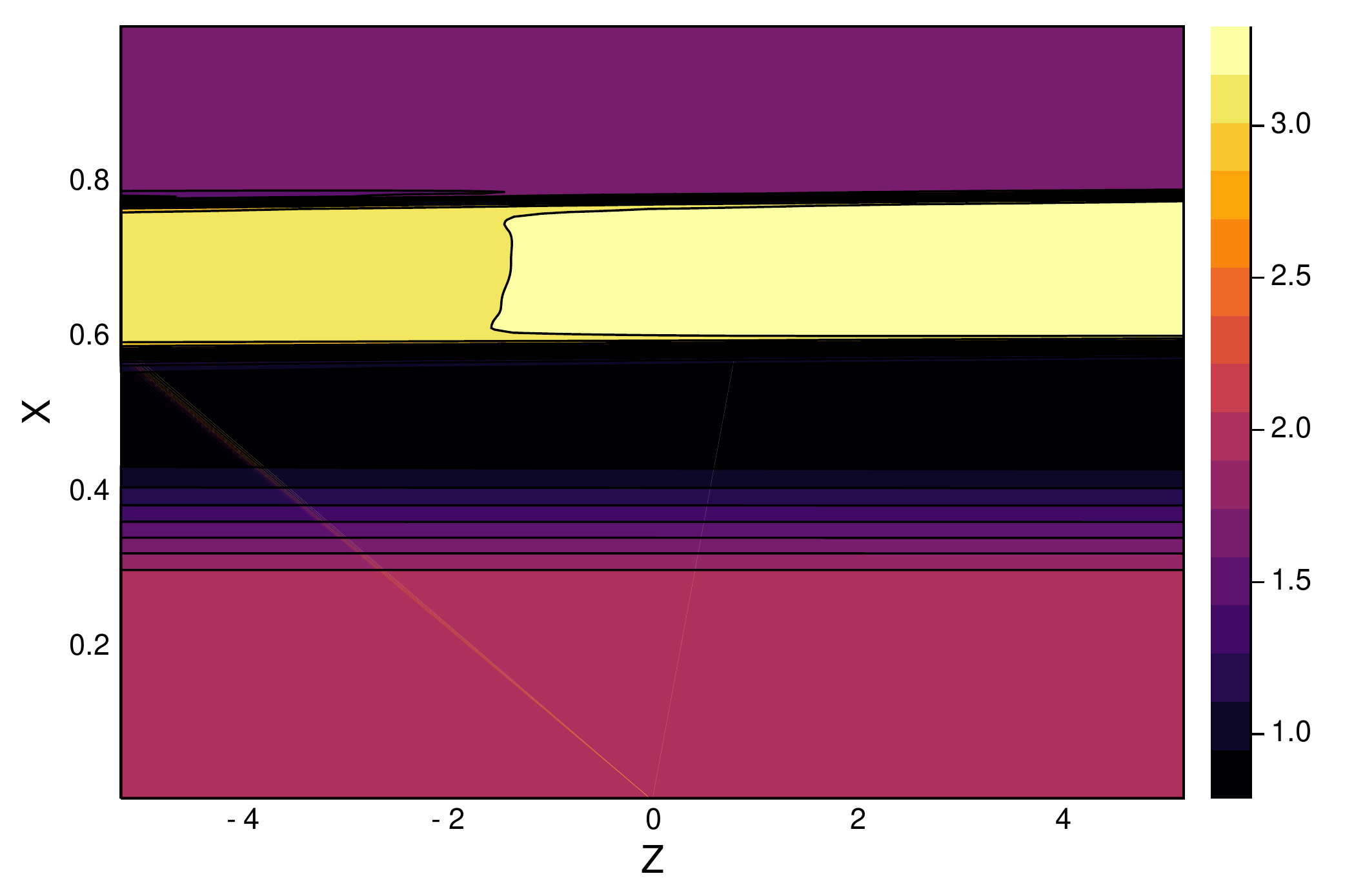}
	}
	\subfigure[Density]{
		\includegraphics[width=5cm]{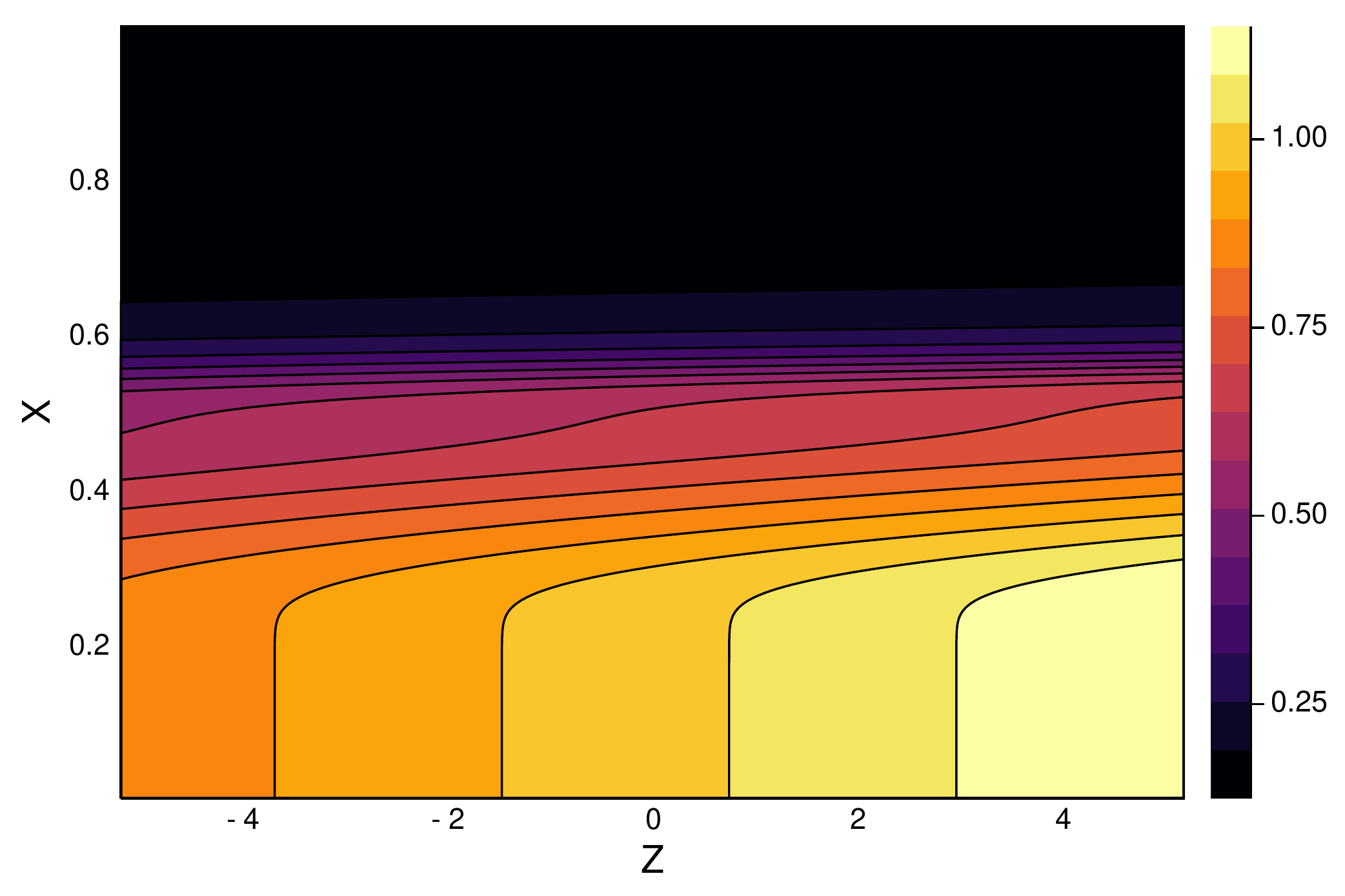}
	}
	\subfigure[Velocity]{
		\includegraphics[width=5cm]{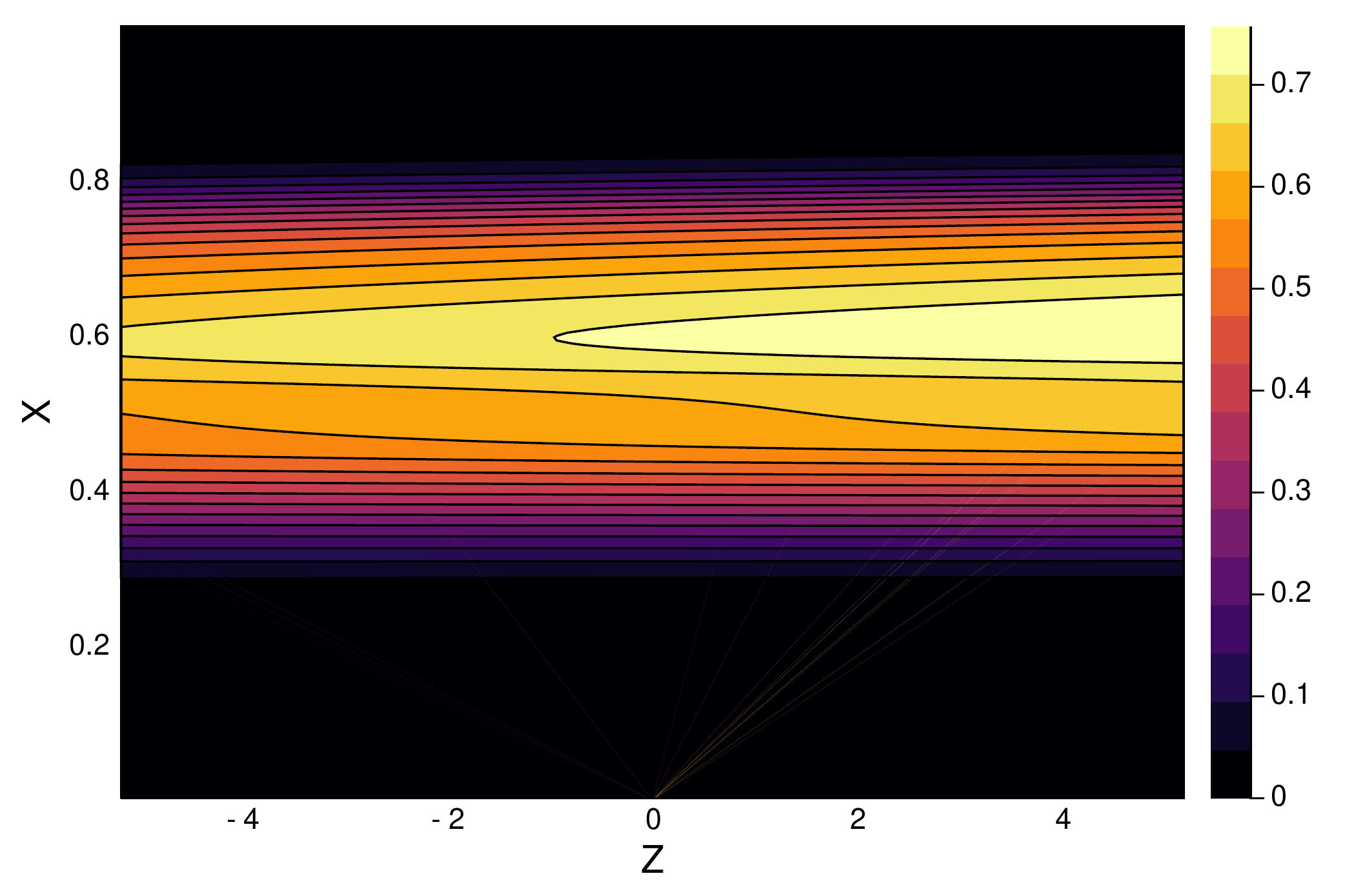}
	}
	\subfigure[Temperature]{
		\includegraphics[width=5cm]{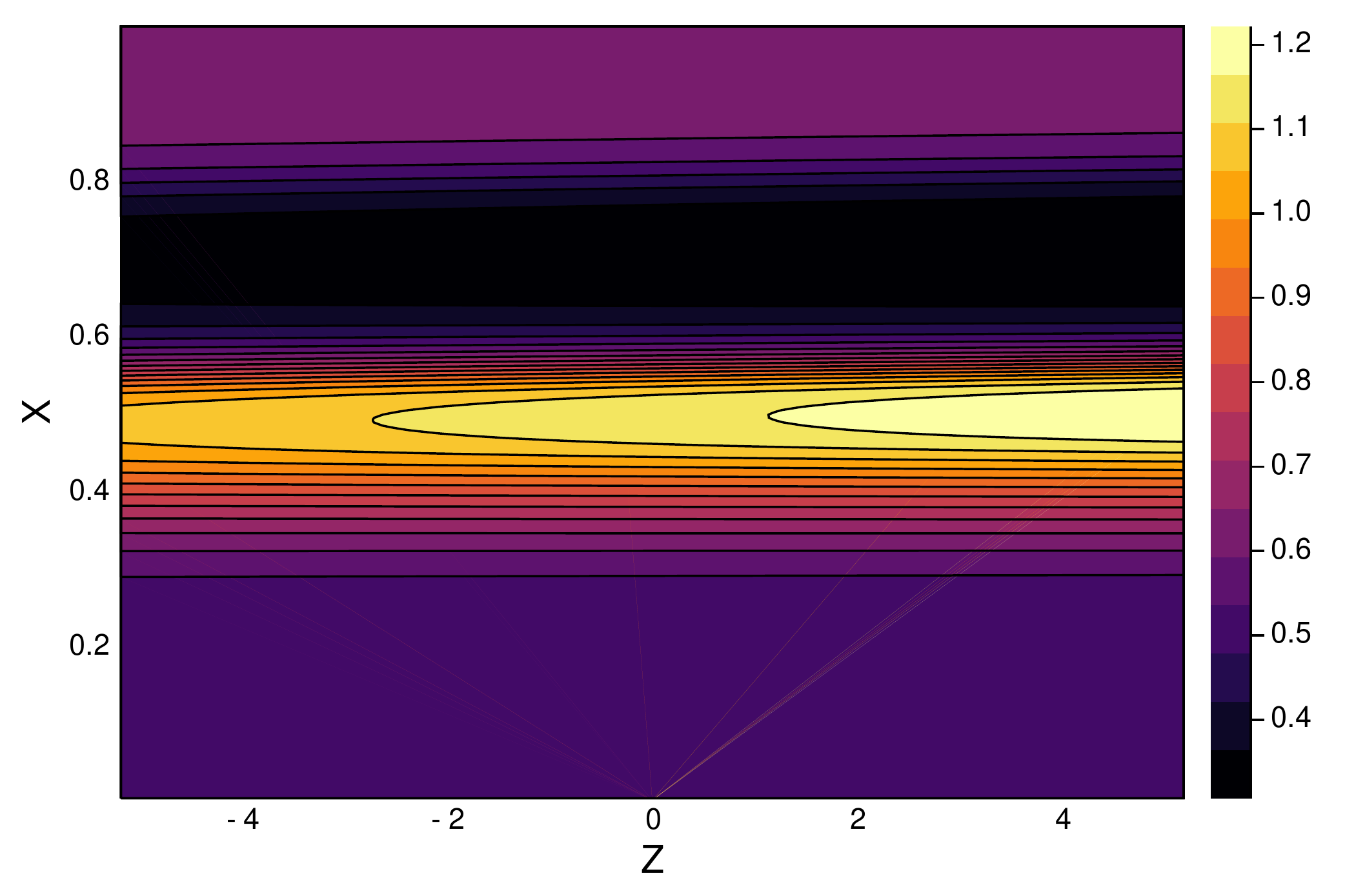}
	}
	\subfigure[Density]{
		\includegraphics[width=5cm]{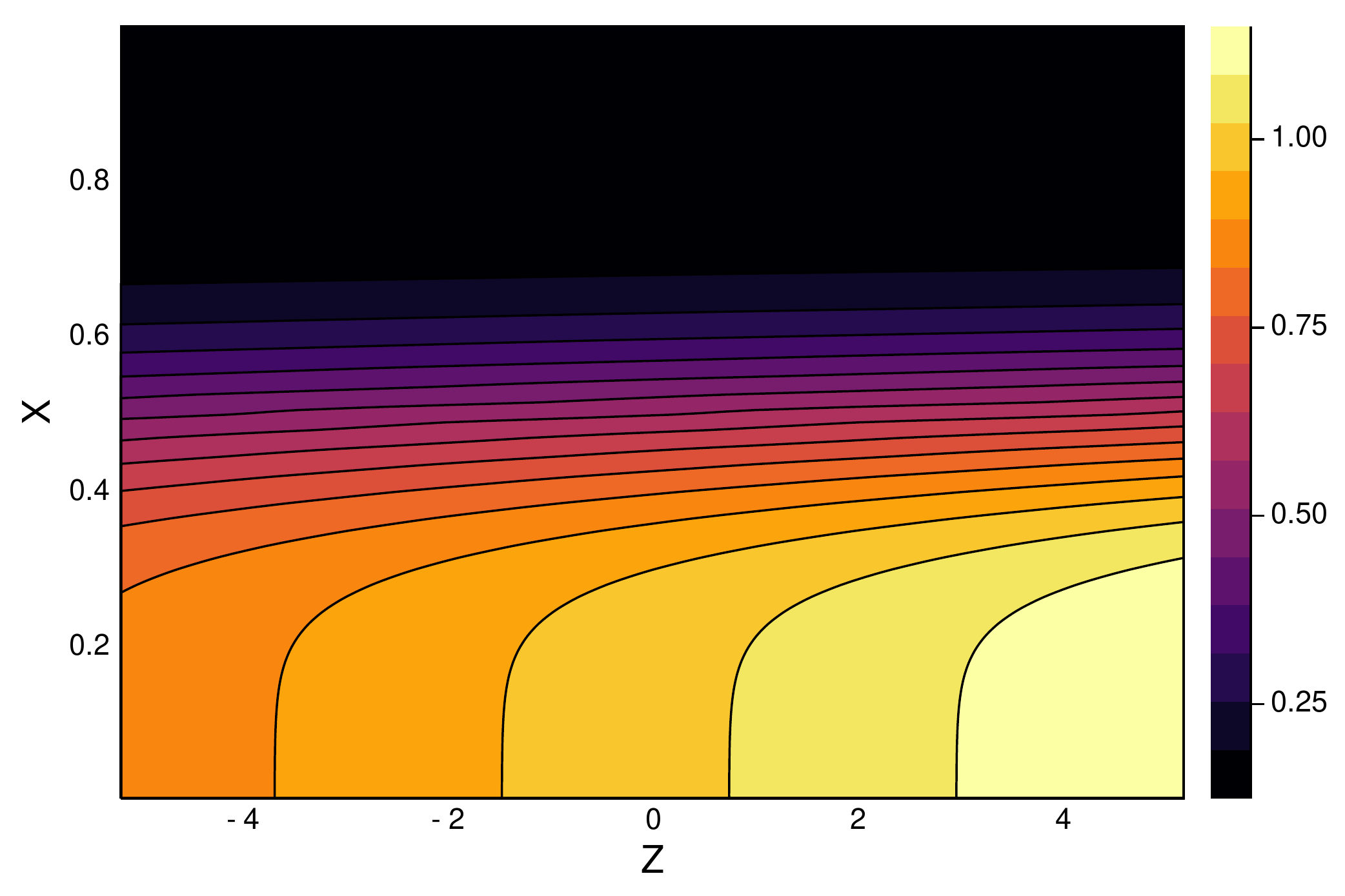}
	}
	\subfigure[Velocity]{
		\includegraphics[width=5cm]{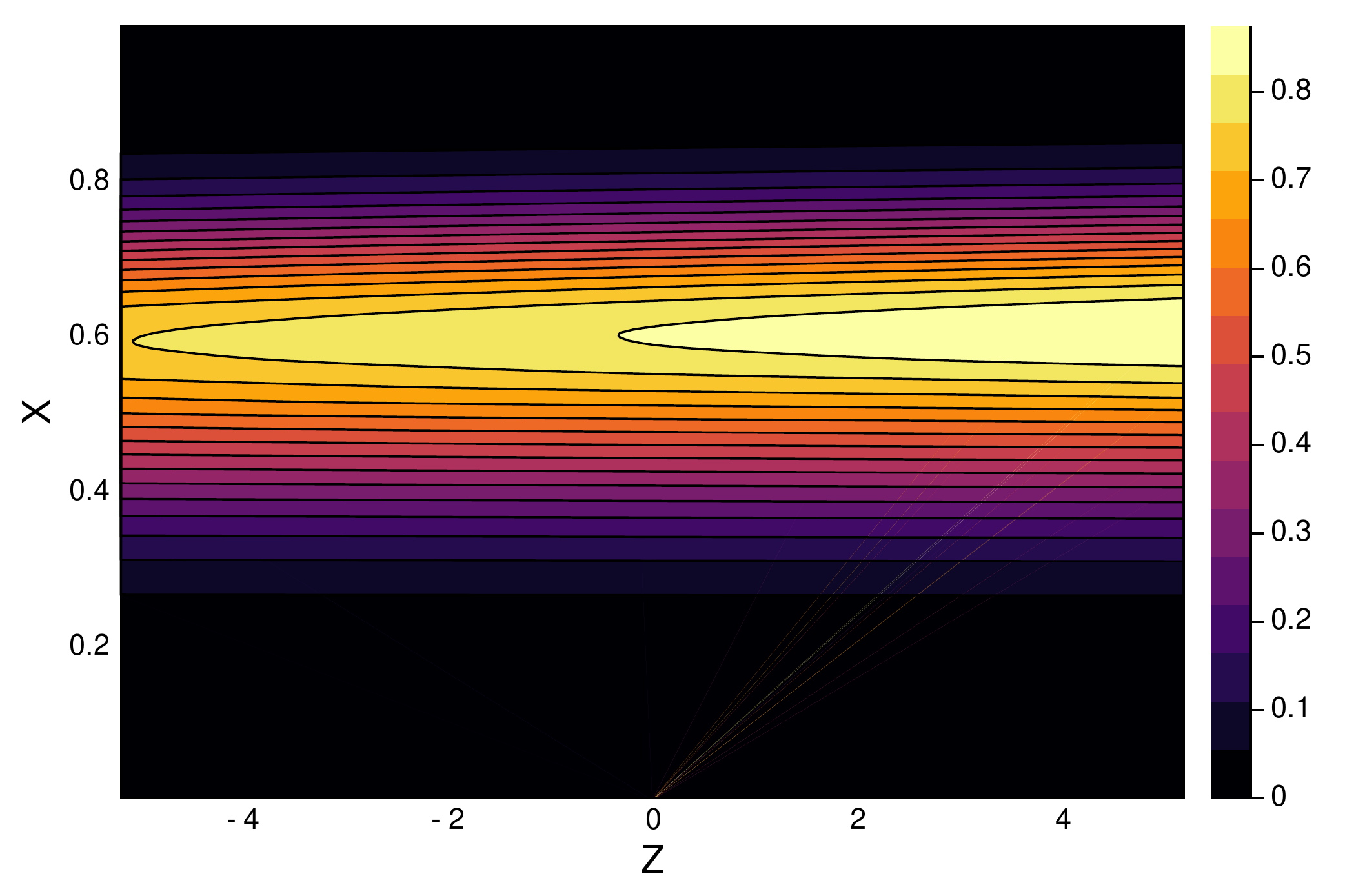}
	}
	\subfigure[Temperature]{
		\includegraphics[width=5cm]{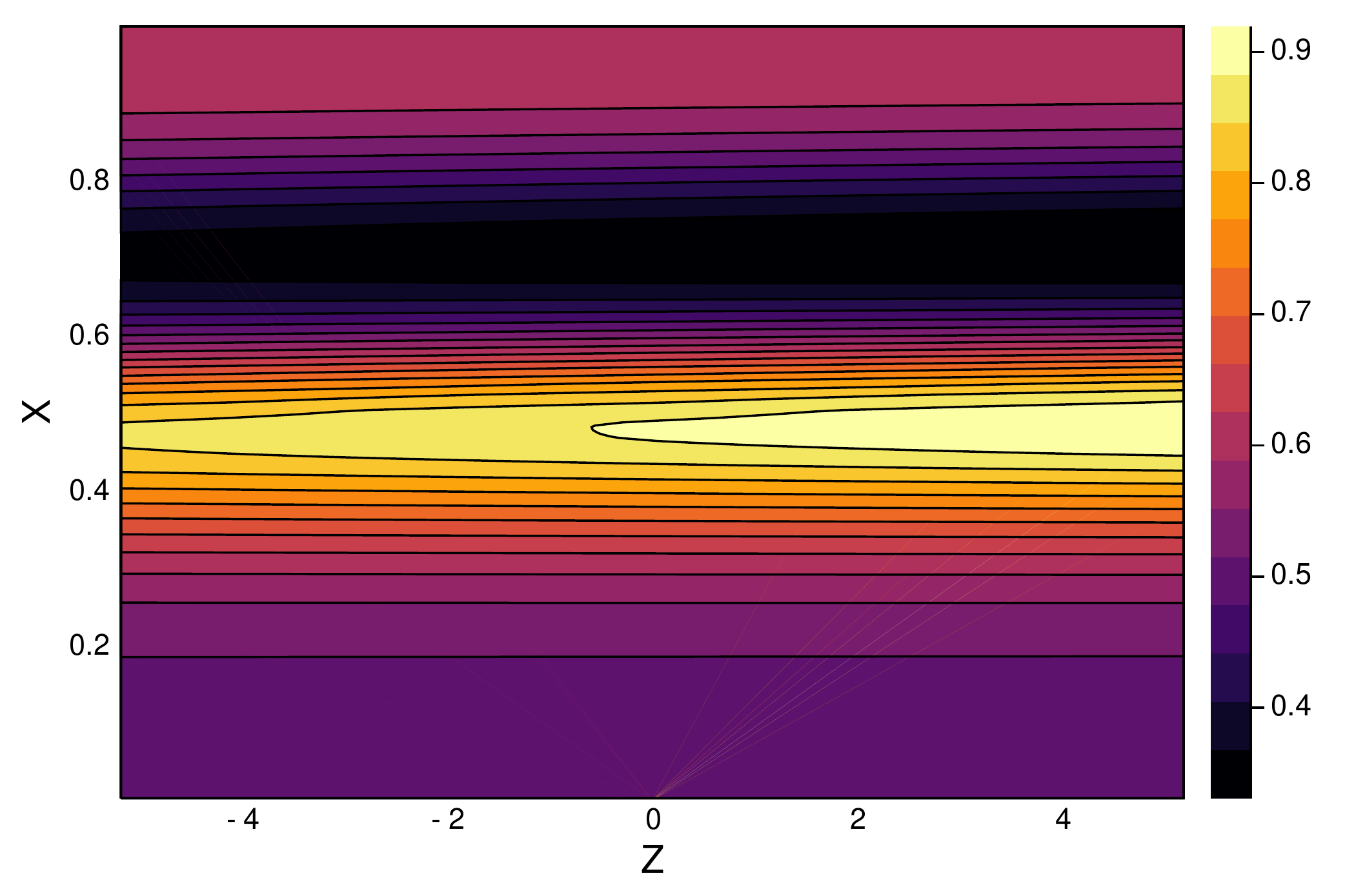}
	}
	\caption{Evaluations of gPC expansions of macroscopic flow variables over random space inside shock tube at $\mathrm{Kn}_{ref}=0.001$ (first row), $\mathrm{Kn}_{ref}=0.01$ (second row) and $\mathrm{Kn}_{ref}=0.1$ (third row).}
	\label{pic:sod ran}
\end{figure}

\begin{figure}[htb!]
	\centering
	\subfigure[Expectation]{
		\includegraphics[width=7.5cm]{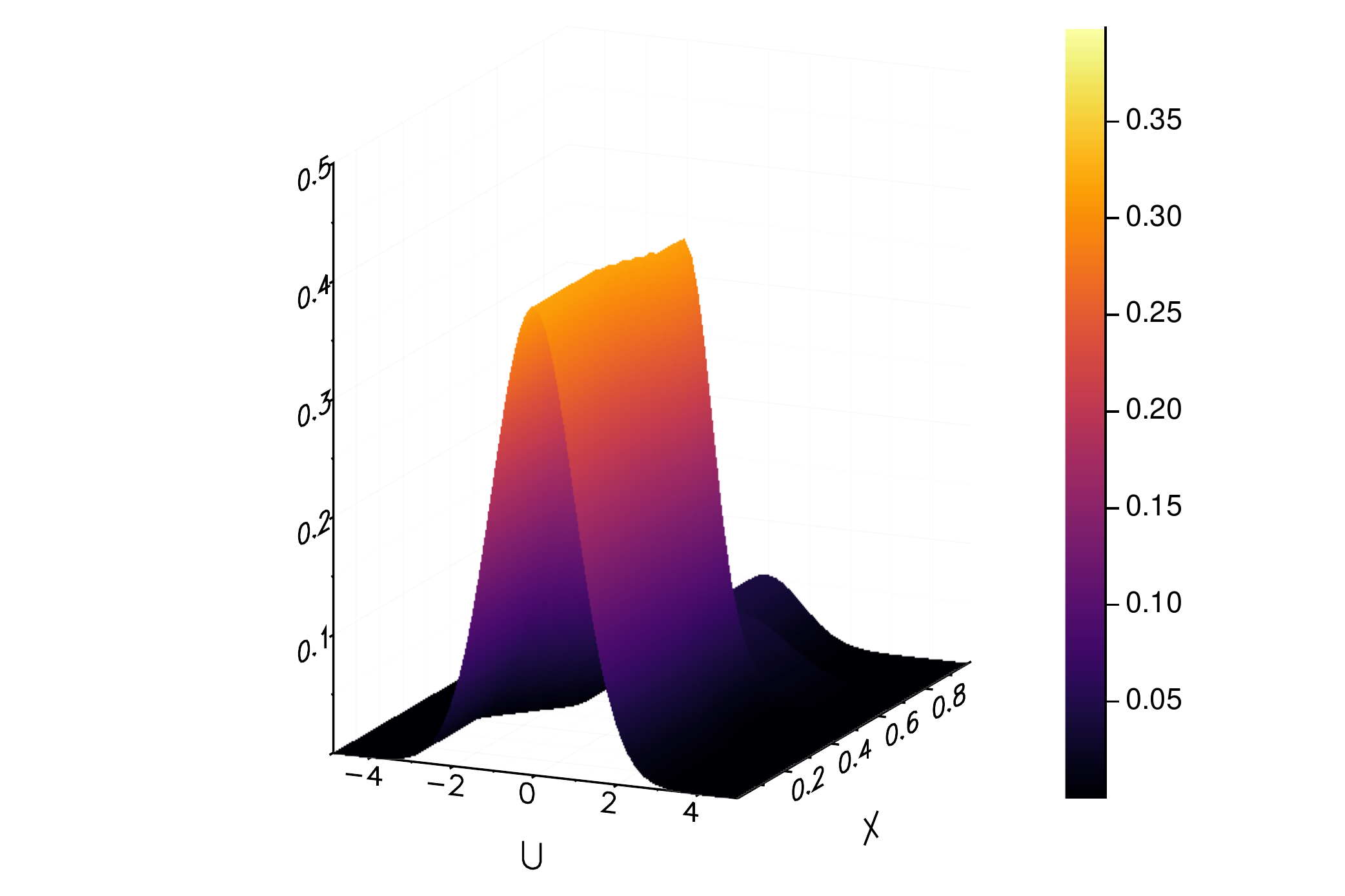}
	}
	\subfigure[Standard deviation]{
		\includegraphics[width=7.5cm]{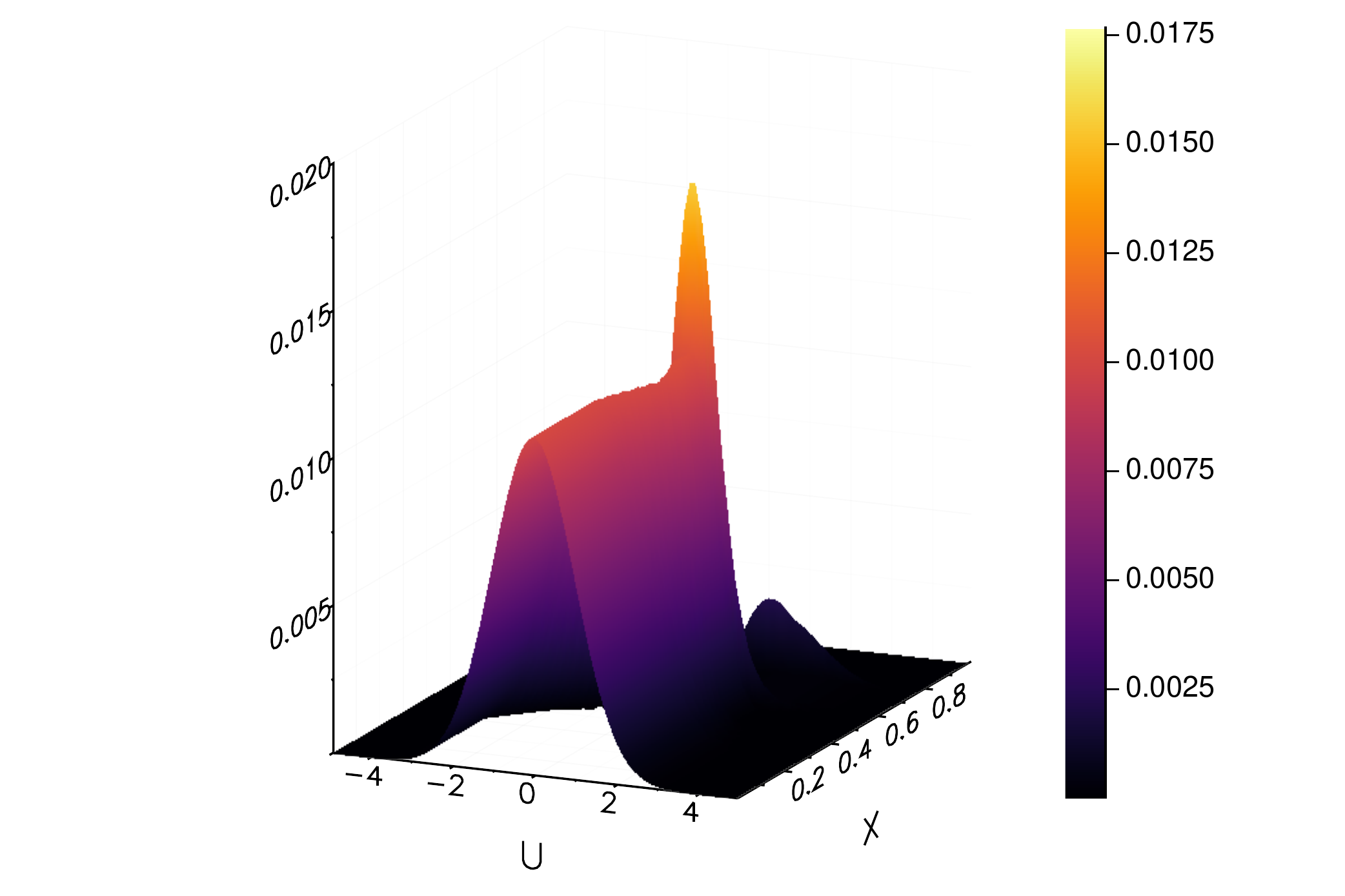}
	}
	\subfigure[Expectation]{
		\includegraphics[width=7.5cm]{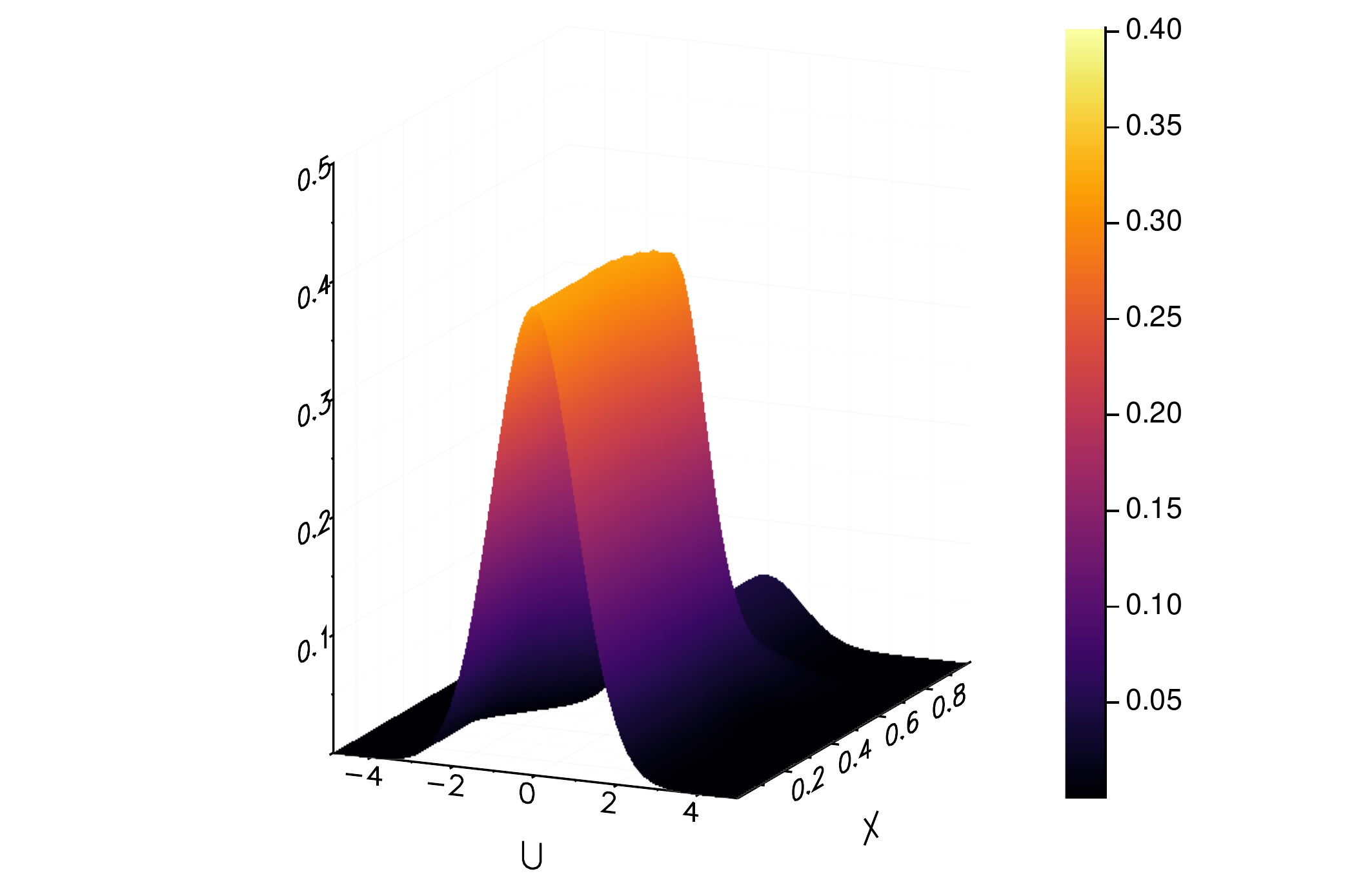}
	}
	\subfigure[Standard deviation]{
		\includegraphics[width=7.5cm]{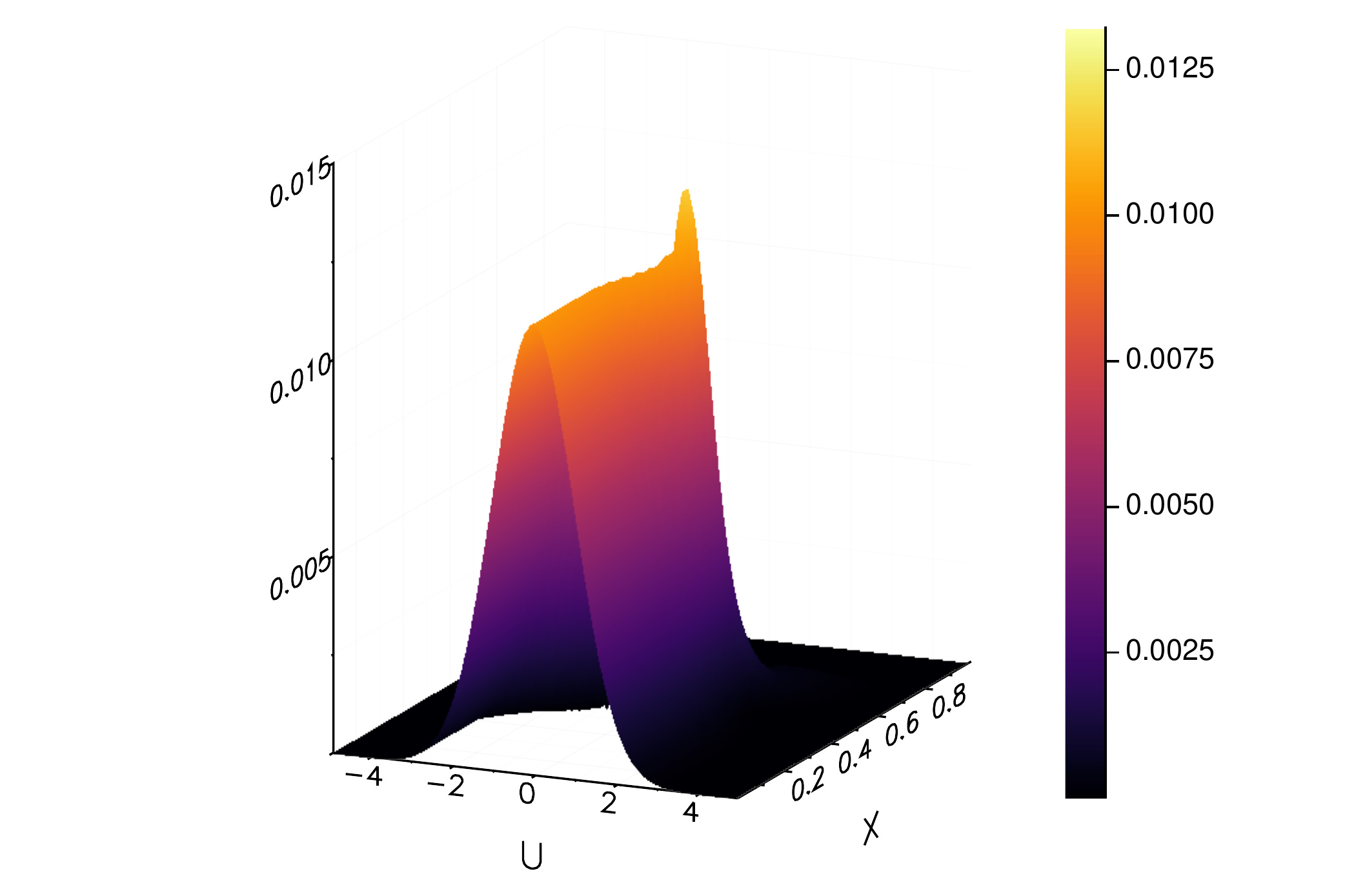}
	}
	\subfigure[Expectation]{
		\includegraphics[width=7.5cm]{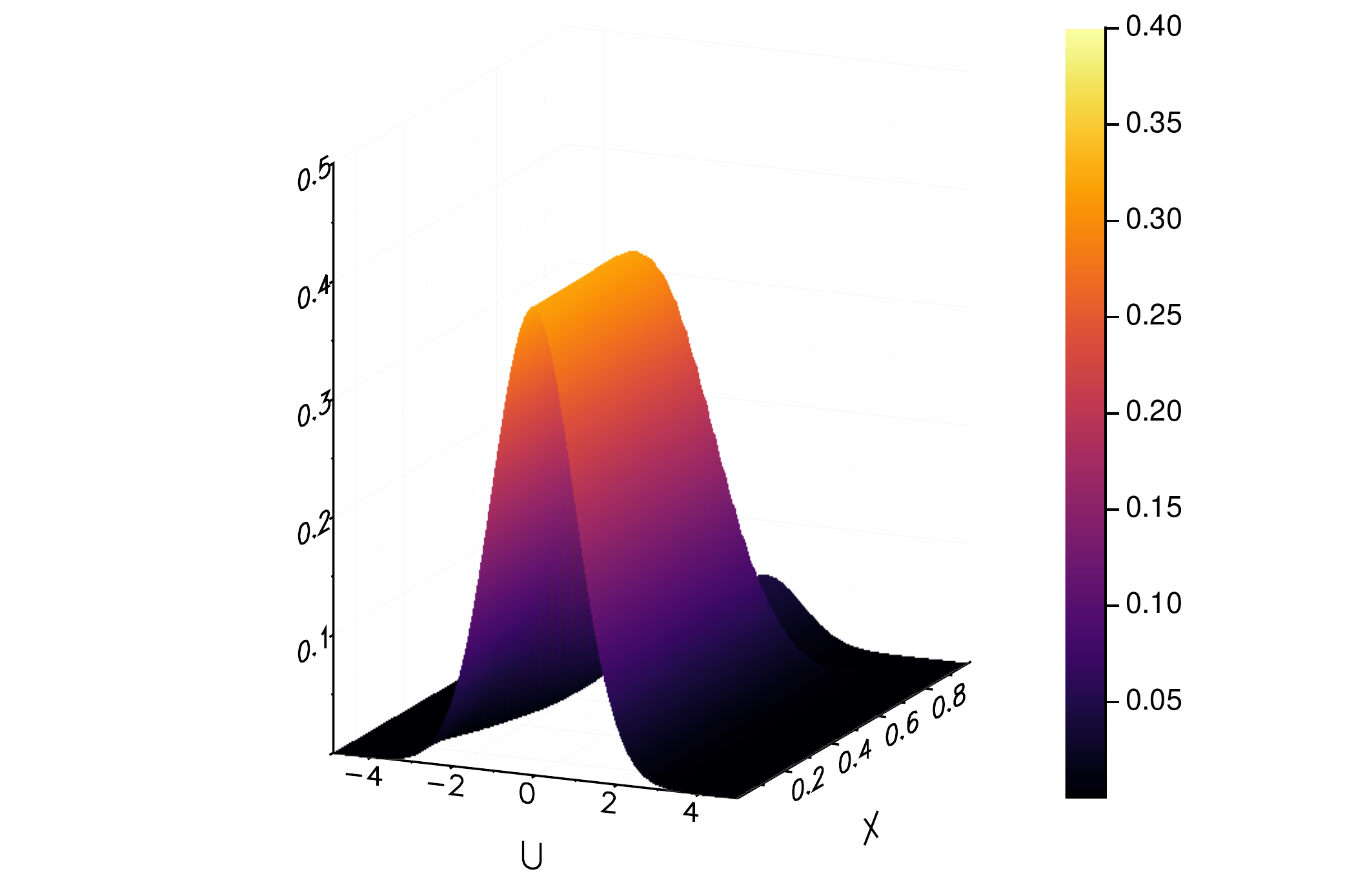}
	}
	\subfigure[Standard deviation]{
		\includegraphics[width=7.5cm]{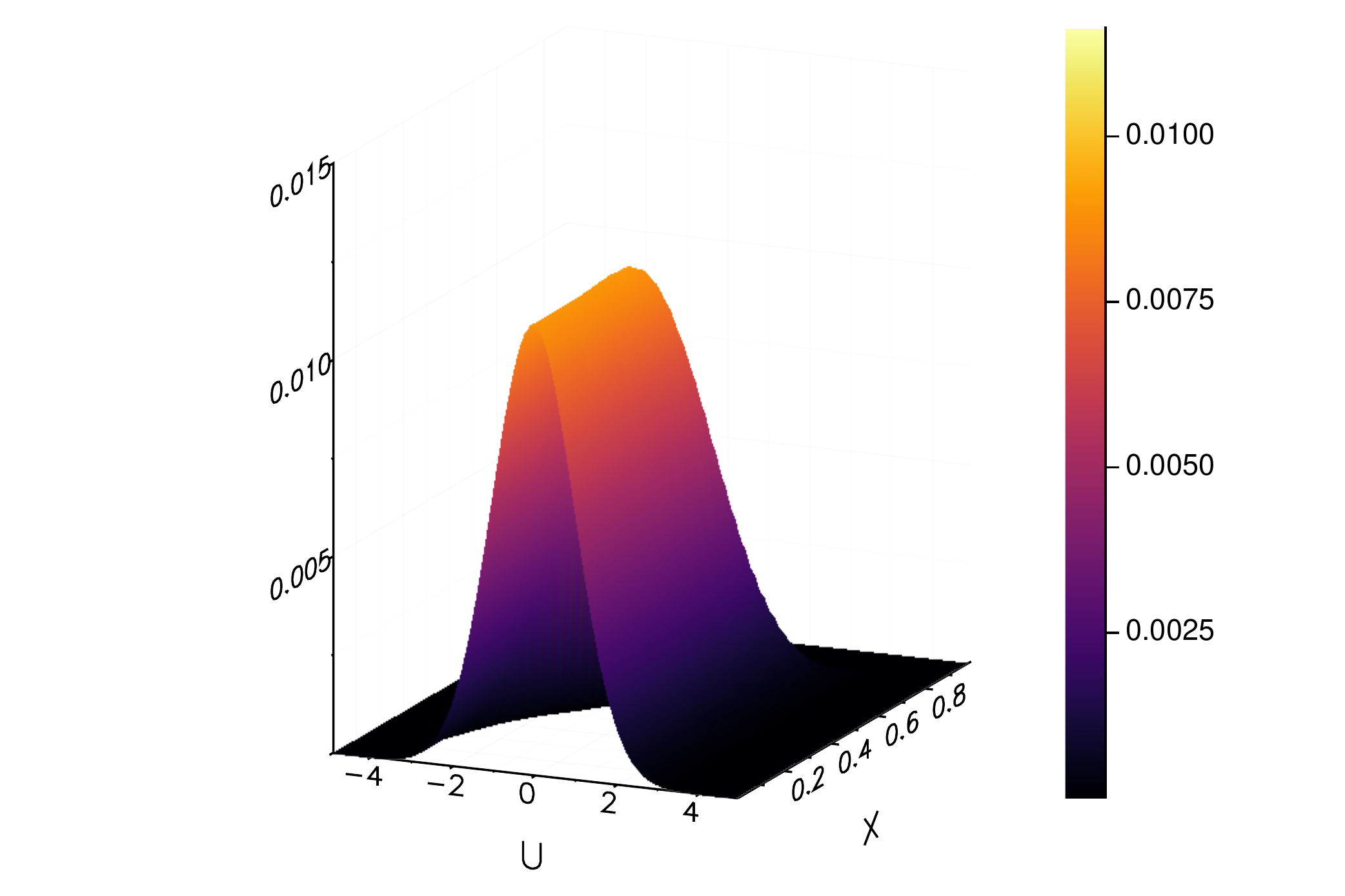}
	}
	\caption{Expectation values and standard deviations of particle distribution function over velocity space inside shock tube at $\mathrm{Kn}_{ref}=0.001$ (first row), $\mathrm{Kn}_{ref}=0.01$ (second row) and $\mathrm{Kn}_{ref}=0.1$ (third row).}
	\label{pic:sod distribution}
\end{figure}

In Fig.\ref{pic:sod ran} we present the evaluation of gPC expansions of macroscopic flow variables over the phase space $\{ x\times z \}$, where the expectation and standard deviation can be determined by integrating the contour value along the $z$-axis along with probability density.
From the contours, we clearly see that it is the horizontal gradients that determine the variances of flow variables.
With increasing Knudsen numbers, although the initial density keeps the same, the dominant physical mechanism in the gas dynamic system turns to particle transport from wave interaction.
As a result, the enhanced transport phenomena lead significant dissipation along the random $z$-axis. 
Therefore, the magnitude of standard deviations reduces correspondingly.
Moreover, with the correspondence between macroscopic and mesoscopic formulations, the stochastic kinetic scheme also provides us the chance to quantify the uncertainties in the evolution process of particle distribution function.
Fig.\ref{pic:sod distribution} presents the expectations and standard deviations of particle distribution function at different reference Knudsen numbers.
As is shown, the overshoots in macroscopic standard deviations at $\mathrm{Kn}_{ref} = 1.0\times 10^{-4}$ come from the uncertainties contained in the particle distribution function near the center of velocity space.
From continuum to rarefied regimes, the randomness on particles get reduced and smoothed, resulting in gentle profiles of macroscopic quantities.

This test case clearly shows the consistency and distinction of propagation modes between expectation value and variance.
It also illustrates the capacity of current scheme to simulate multi-scale flow physics and capture evolution of uncertainties in different regimes.

\subsection{Random boundary condition: suddenly heating wall problem}

The last case comes from \cite{aoki1991numerical, filbet2012deterministic, hu2016stochastic}.
The initial gas is uniformly and deterministically distributed inside the domain,
\begin{equation*}
\rho_0=1,\ U_0=0, \ T_0=1,
\end{equation*} 
with the particle distribution function being Maxwellian everywhere.
From $t>0$, a heating wall is suddenly put on the left boundary of the domain, with the temperature being
\begin{equation*}
T_w=2+0.4z,
\end{equation*} 
where $z \sim \mathcal U[-1,1]$ is a random variable which follows uniform distribution.

In the simulation, the physical domain $x\in[0,0.5]$ is discretized by 200 uniform cells, and the particle velocity space is truncated into $u\in[-5,5]$ with 48 uniform quadrature points.
The variable hard-sphere model is employed with the same parameter setup given in Sec. \ref{sec:sod}, and the Knudsen numbers in the reference state are chosen as $\mathrm{Kn}_{ref}=0.001$, $0.01$ and $0.1$.
The Maxwell’s fully diffusive boundary is adopted at the left wall, and the right boundary is treated with extrapolation.
The hybrid Galerkin-collocation method is employed with 6th-order gPC expansion and 11 Gauss collocation points, while the reference solutions are conducted by the collocation method with 800 uniform cells.

\begin{figure}[htb!]
	\centering
	\subfigure[$\mathbb E(\rho)$]{
		\includegraphics[width=7.5cm]{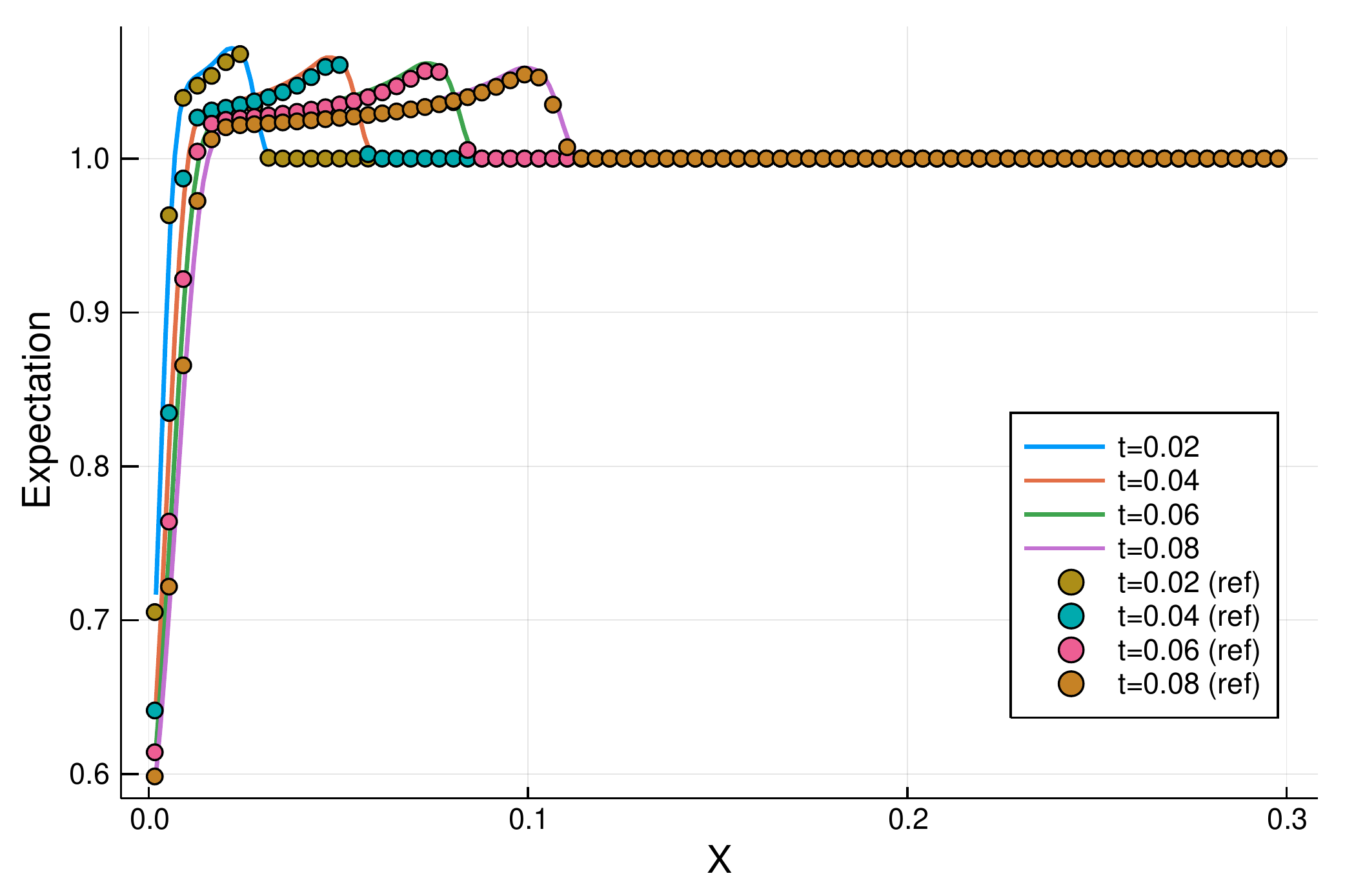}
	}
	\subfigure[$\mathbb S(\rho)$]{
		\includegraphics[width=7.5cm]{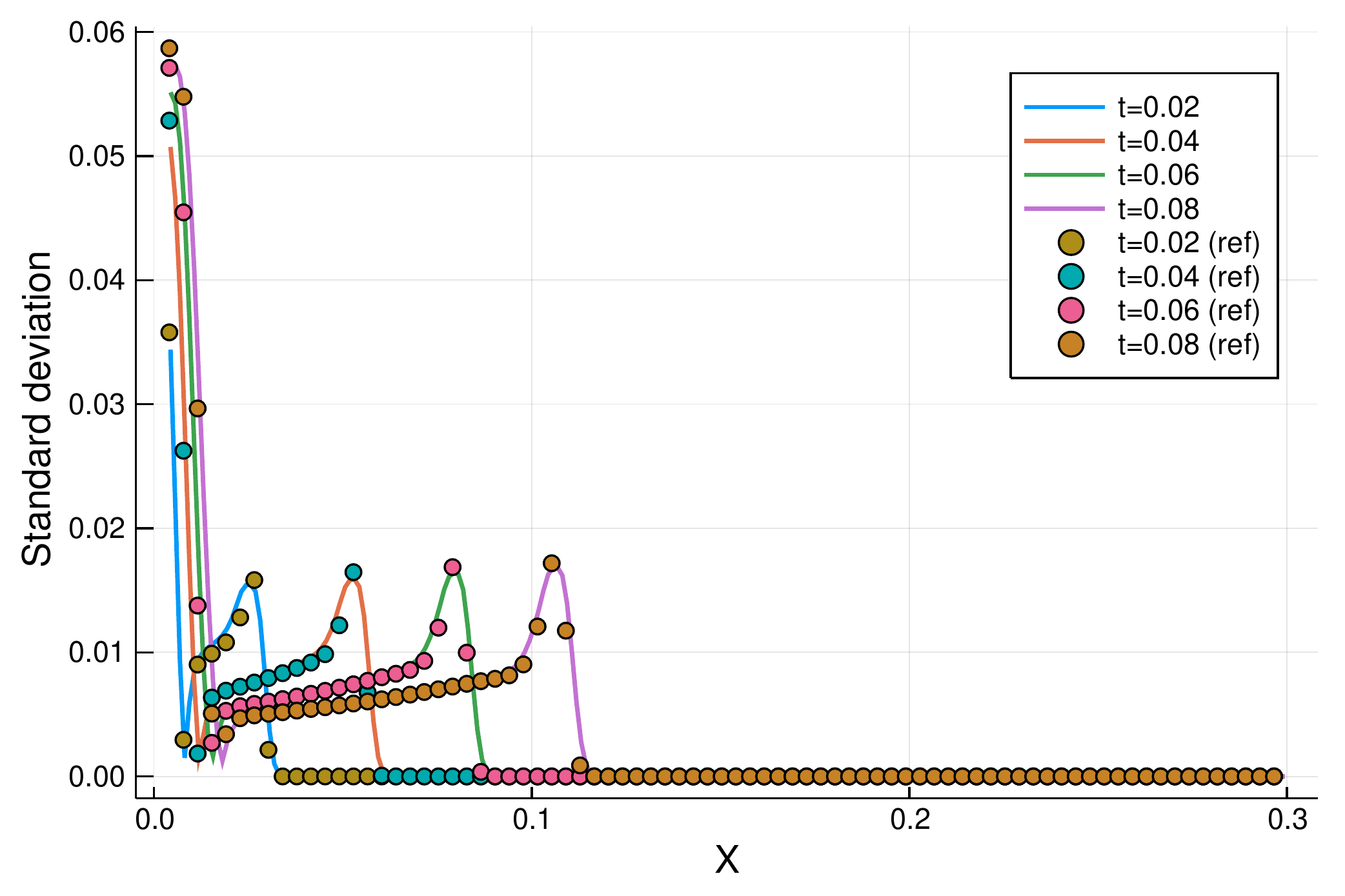}
	}
	\subfigure[$\mathbb E(U)$]{
		\includegraphics[width=7.5cm]{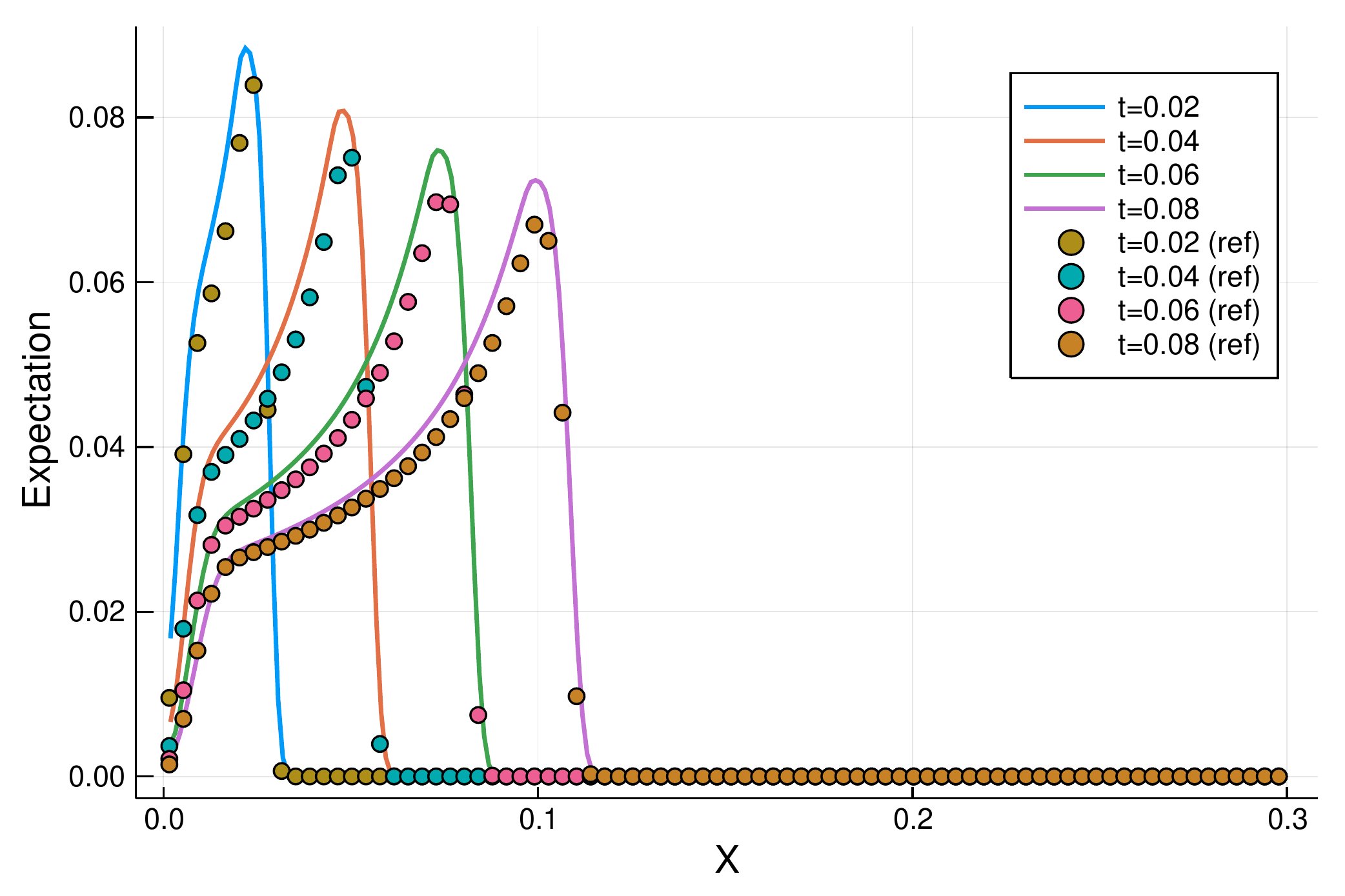}
	}
	\subfigure[$\mathbb S(U)$]{
		\includegraphics[width=7.5cm]{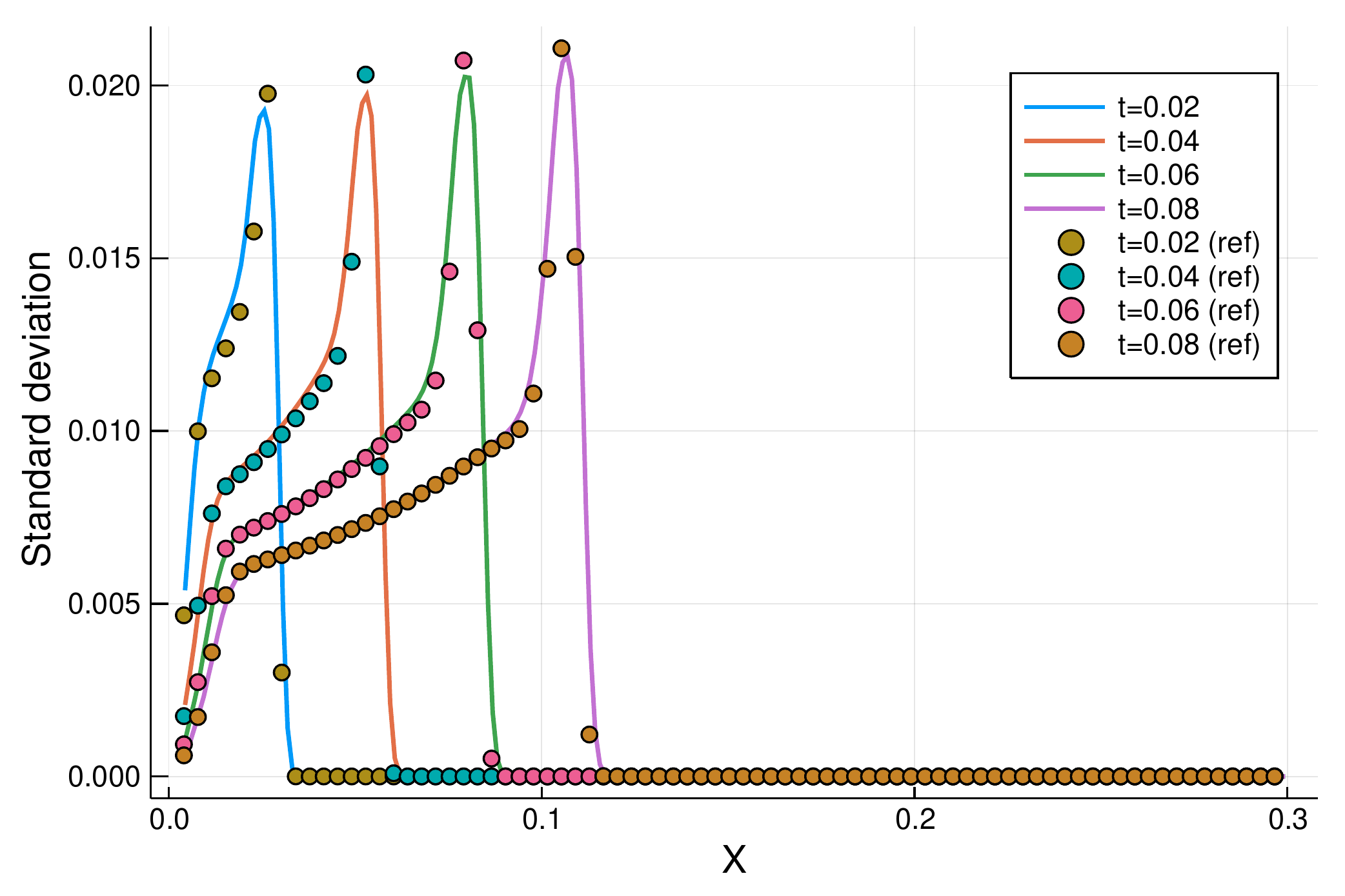}
	}
	\subfigure[$\mathbb E(T)$]{
		\includegraphics[width=7.5cm]{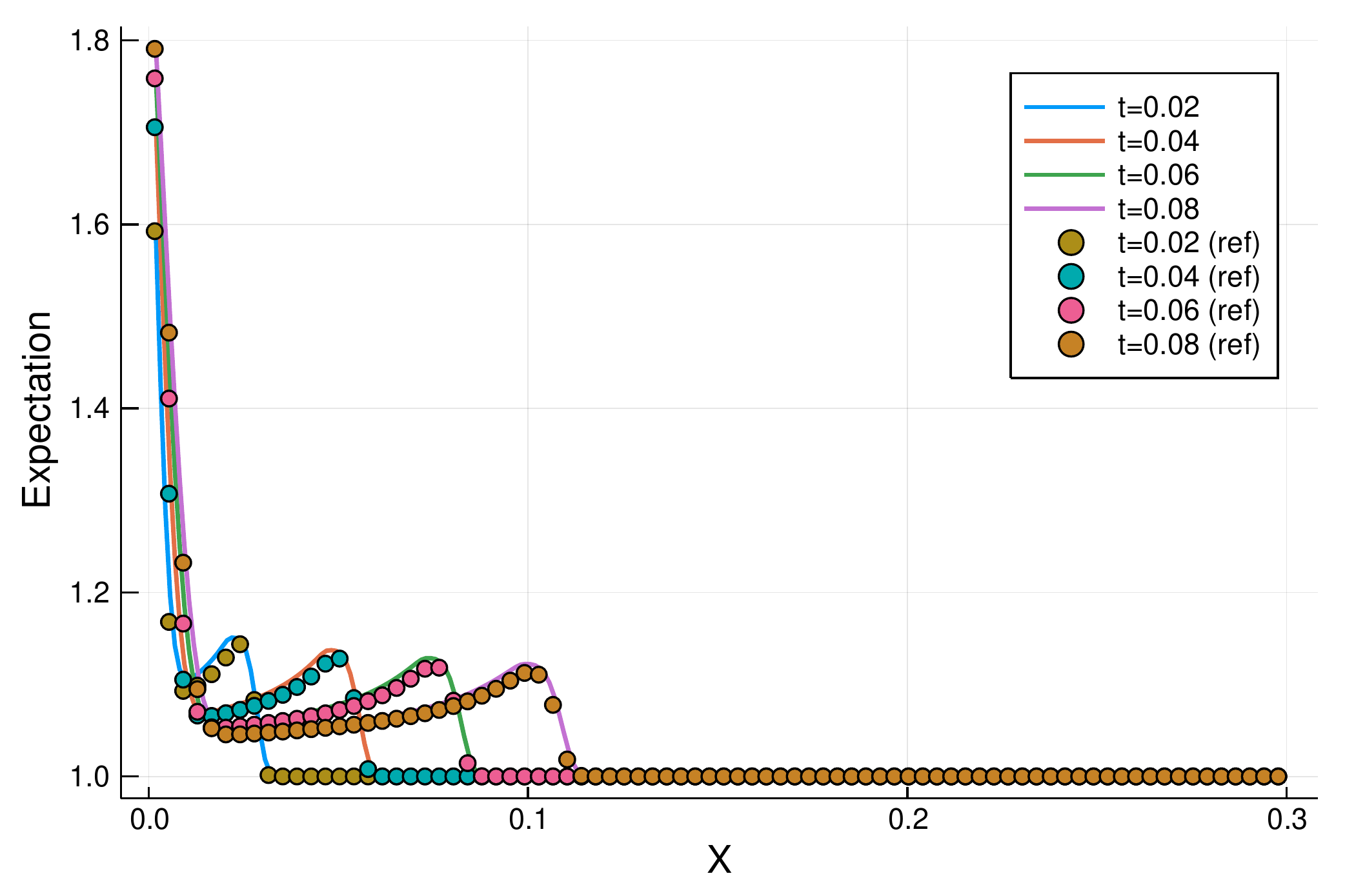}
	}
	\subfigure[$\mathbb S(T)$]{
		\includegraphics[width=7.5cm]{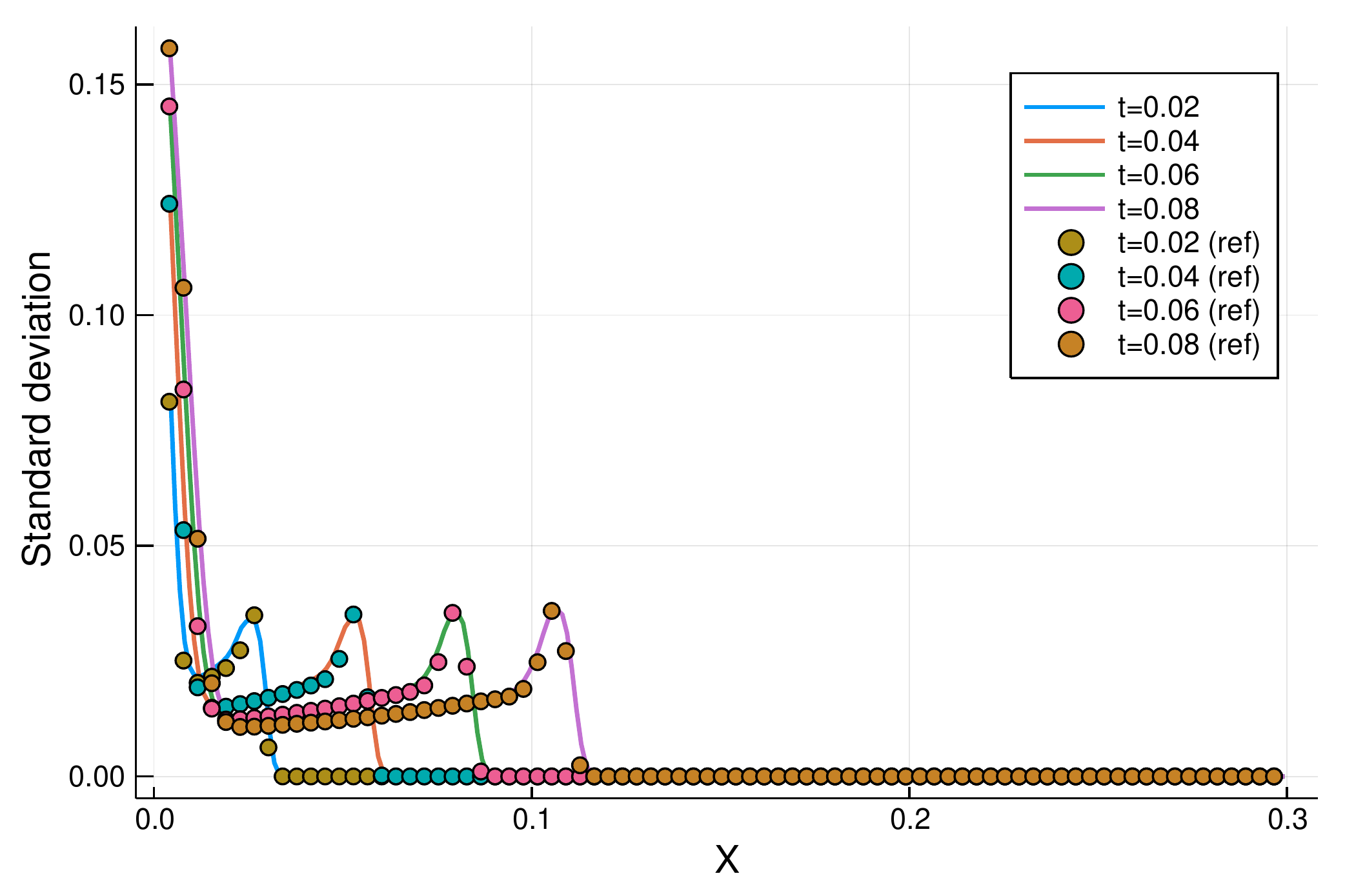}
	}
	\caption{Temporal evolutions of expectations and standard deviations of density (first row), velocity (second row) and temperature (third row) near the heat wall at $\mathrm Kn_{ref}=0.001$.}
	\label{pic:heat macro kn1}
\end{figure}

\begin{figure}[htb!]
	\centering
	\subfigure[$\mathbb E(\rho)$]{
		\includegraphics[width=7.5cm]{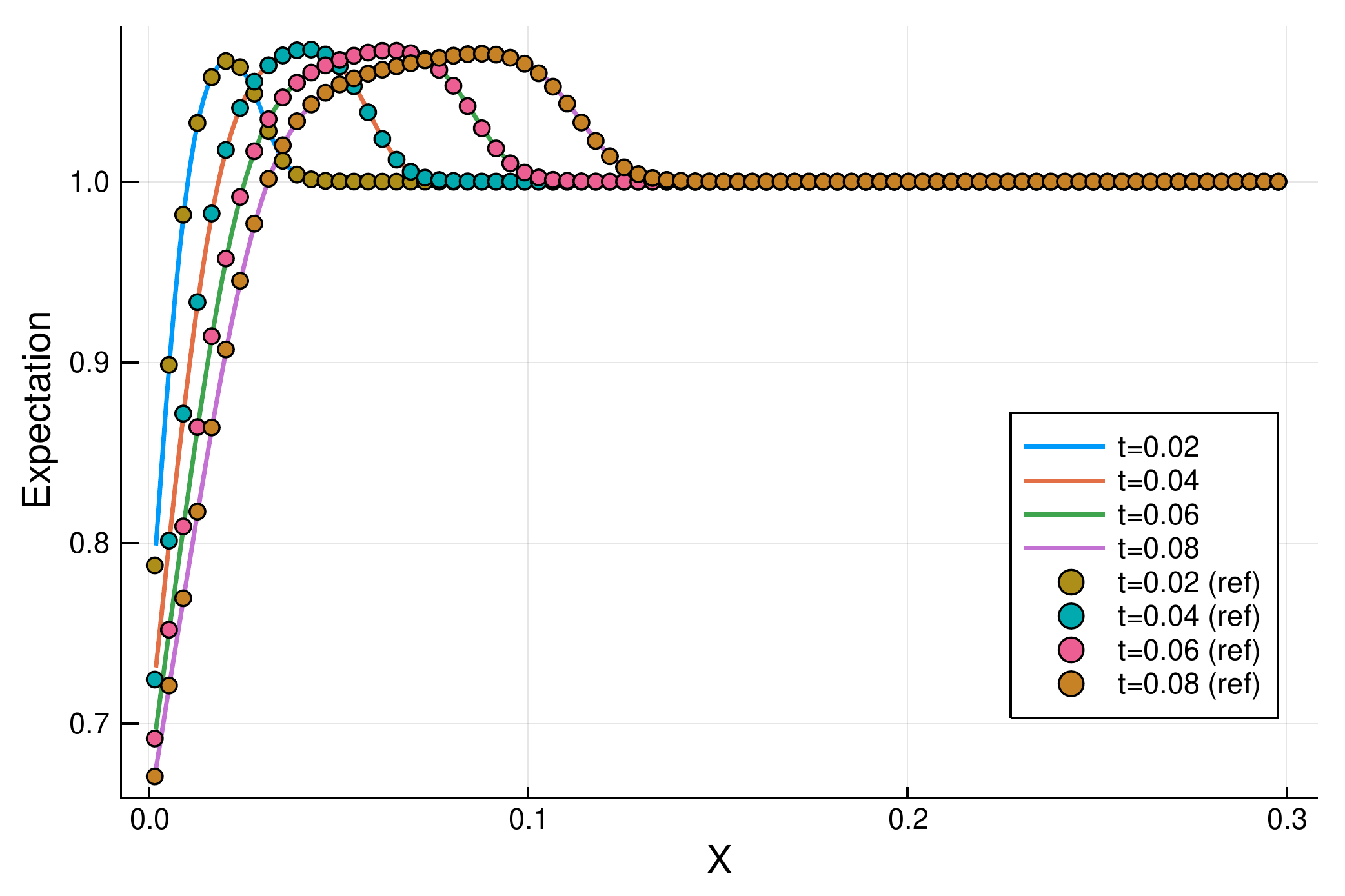}
	}
	\subfigure[$\mathbb S(\rho)$]{
		\includegraphics[width=7.5cm]{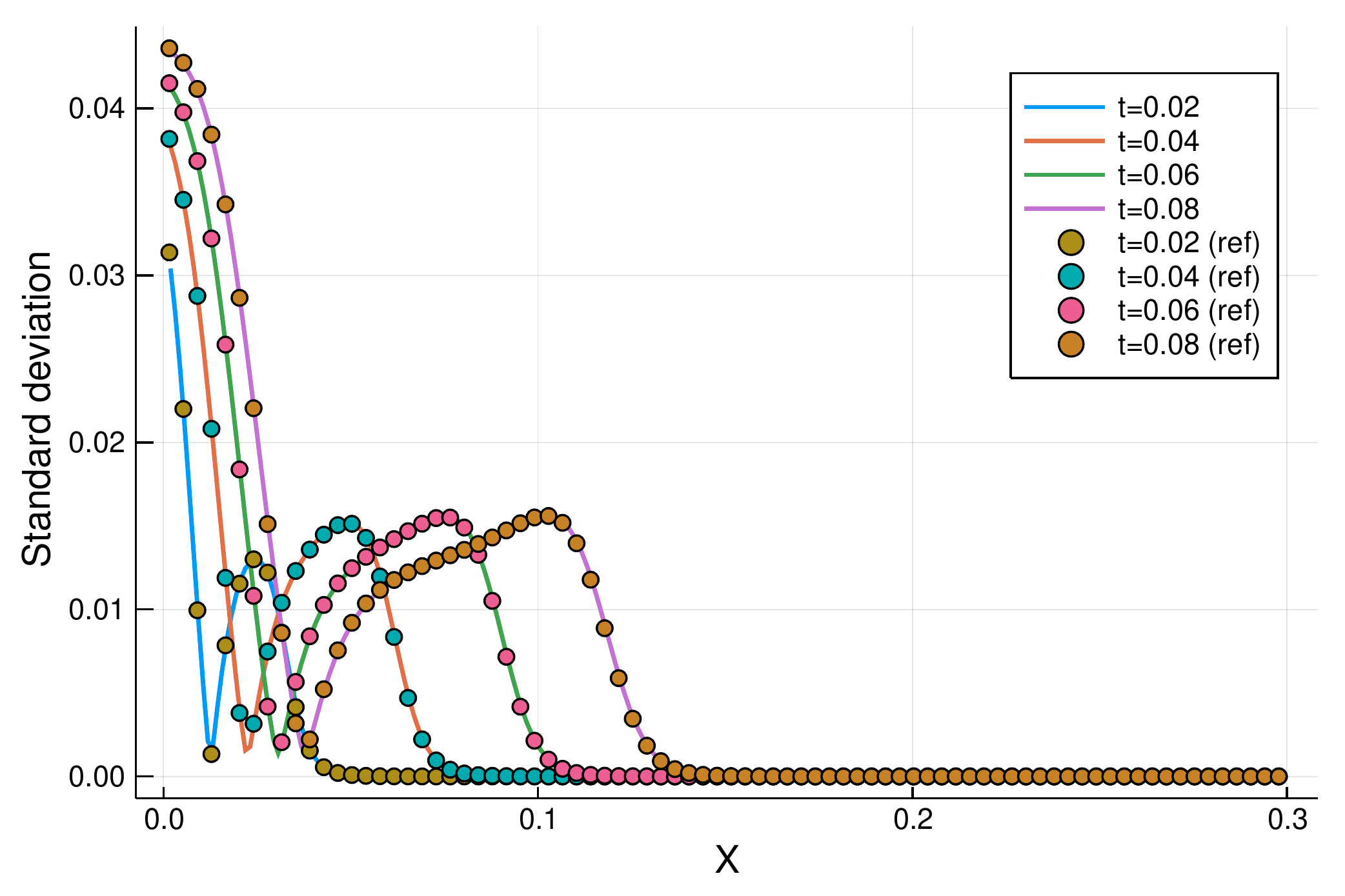}
	}
	\subfigure[$\mathbb E(U)$]{
		\includegraphics[width=7.5cm]{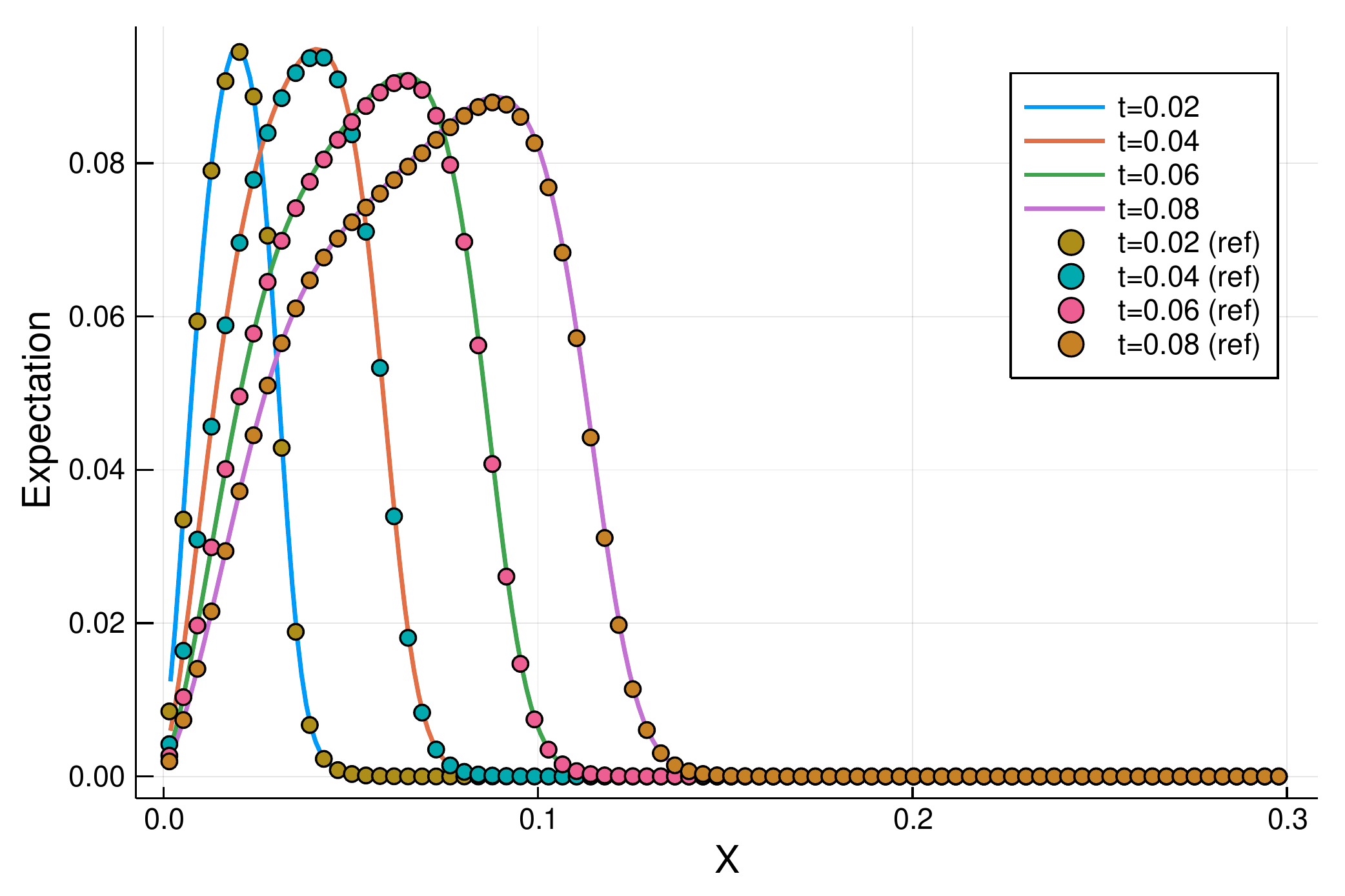}
	}
	\subfigure[$\mathbb S(U)$]{
		\includegraphics[width=7.5cm]{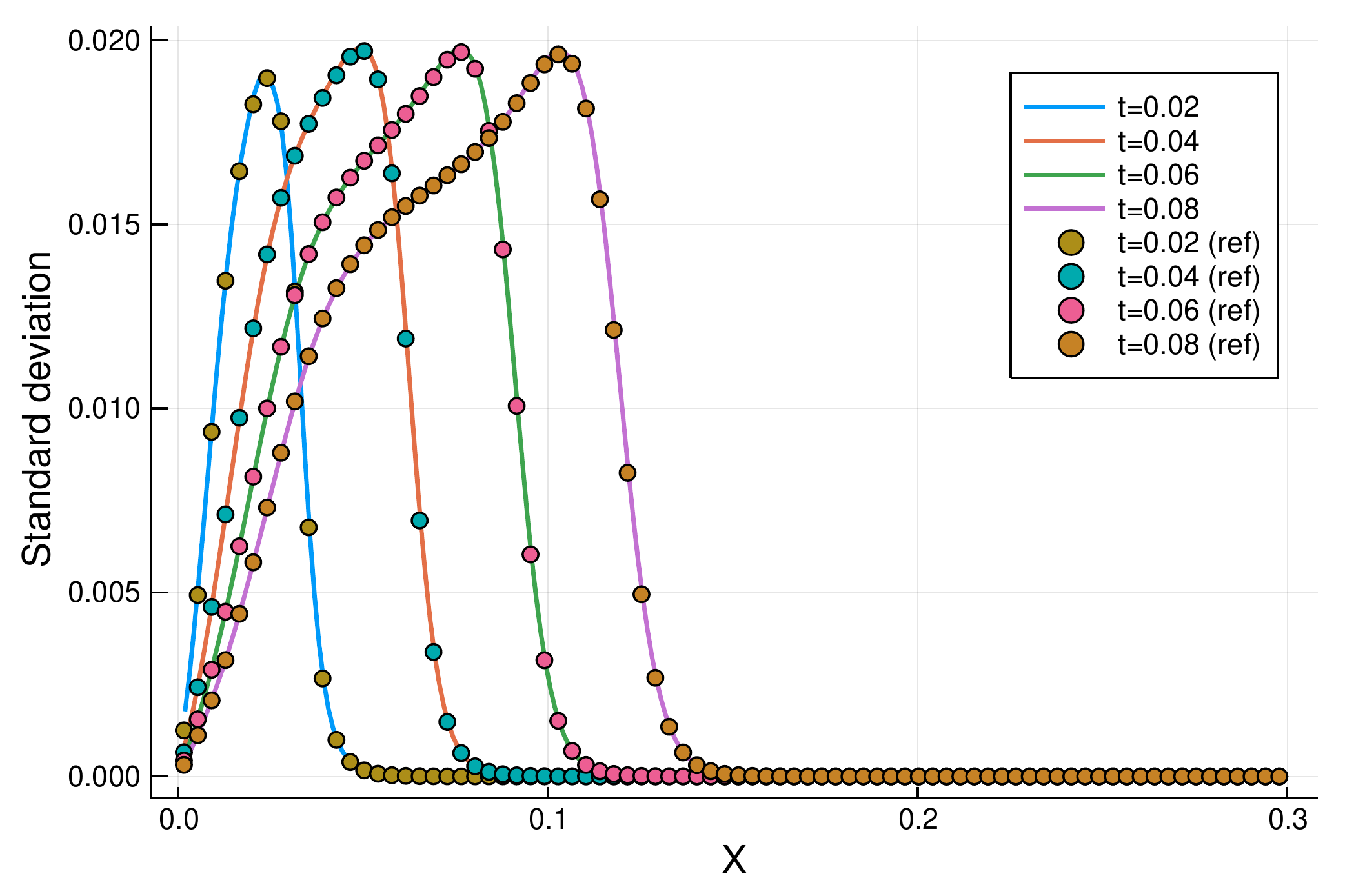}
	}
	\subfigure[$\mathbb E(T)$]{
		\includegraphics[width=7.5cm]{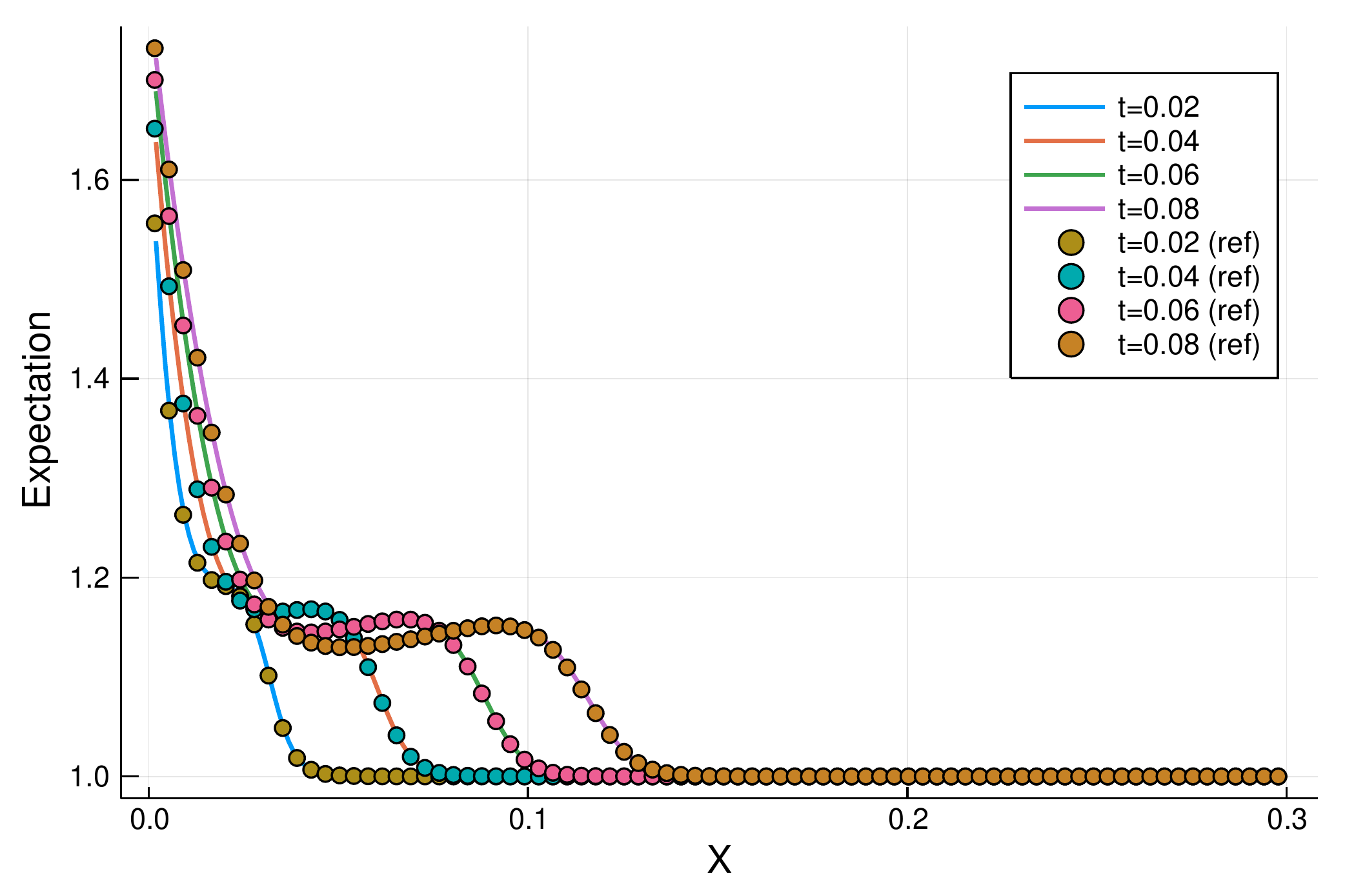}
	}
	\subfigure[$\mathbb S(T)$]{
		\includegraphics[width=7.5cm]{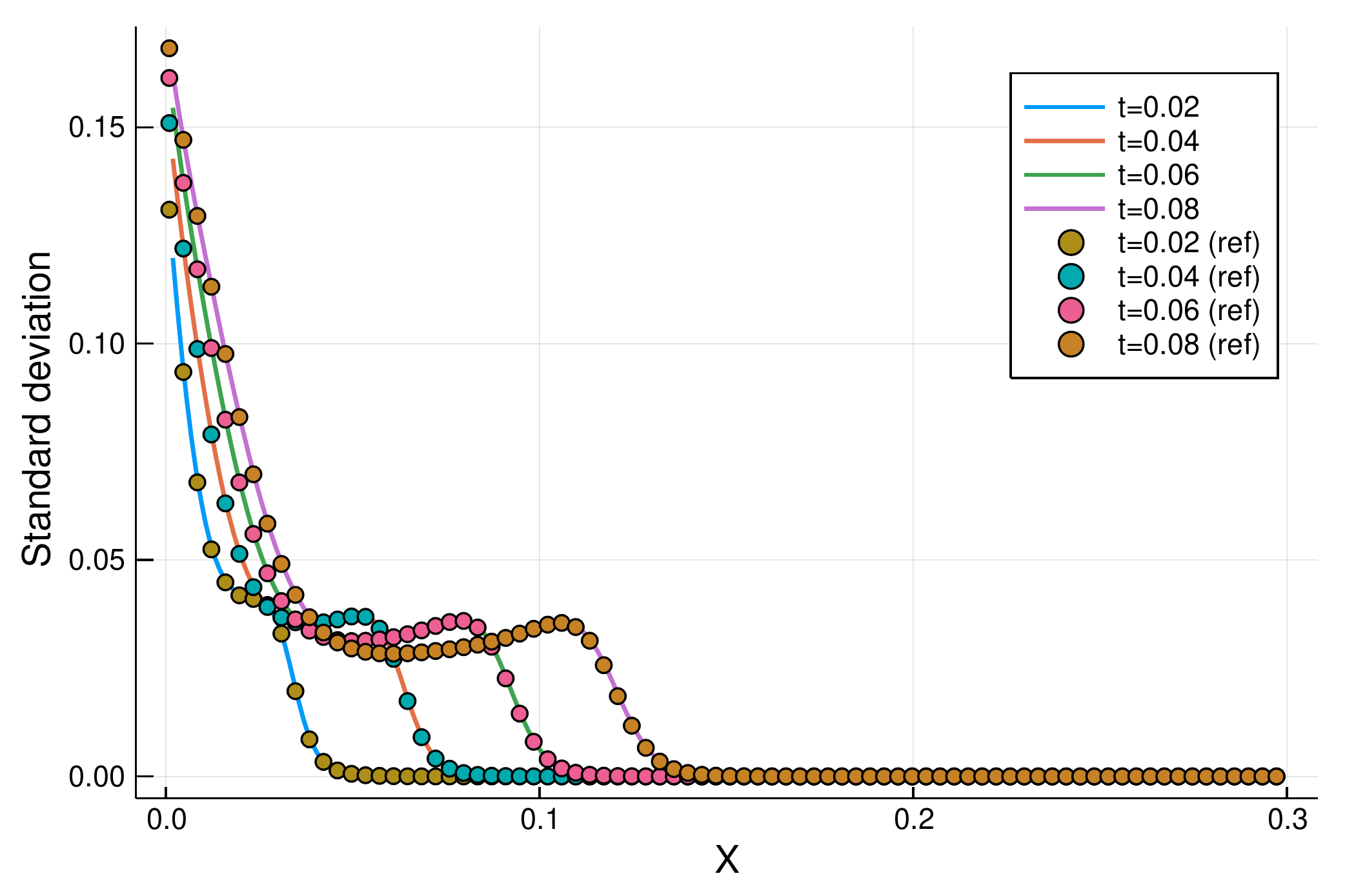}
	}
	\caption{Temporal evolutions of expectations and standard deviations of density (first row), velocity (second row) and temperature (third row) near the heat wall at $\mathrm Kn_{ref}=0.01$.}
	\label{pic:heat macro kn2}
\end{figure}

\begin{figure}[htb!]
	\centering
	\subfigure[$\mathbb E(\rho)$]{
		\includegraphics[width=7.5cm]{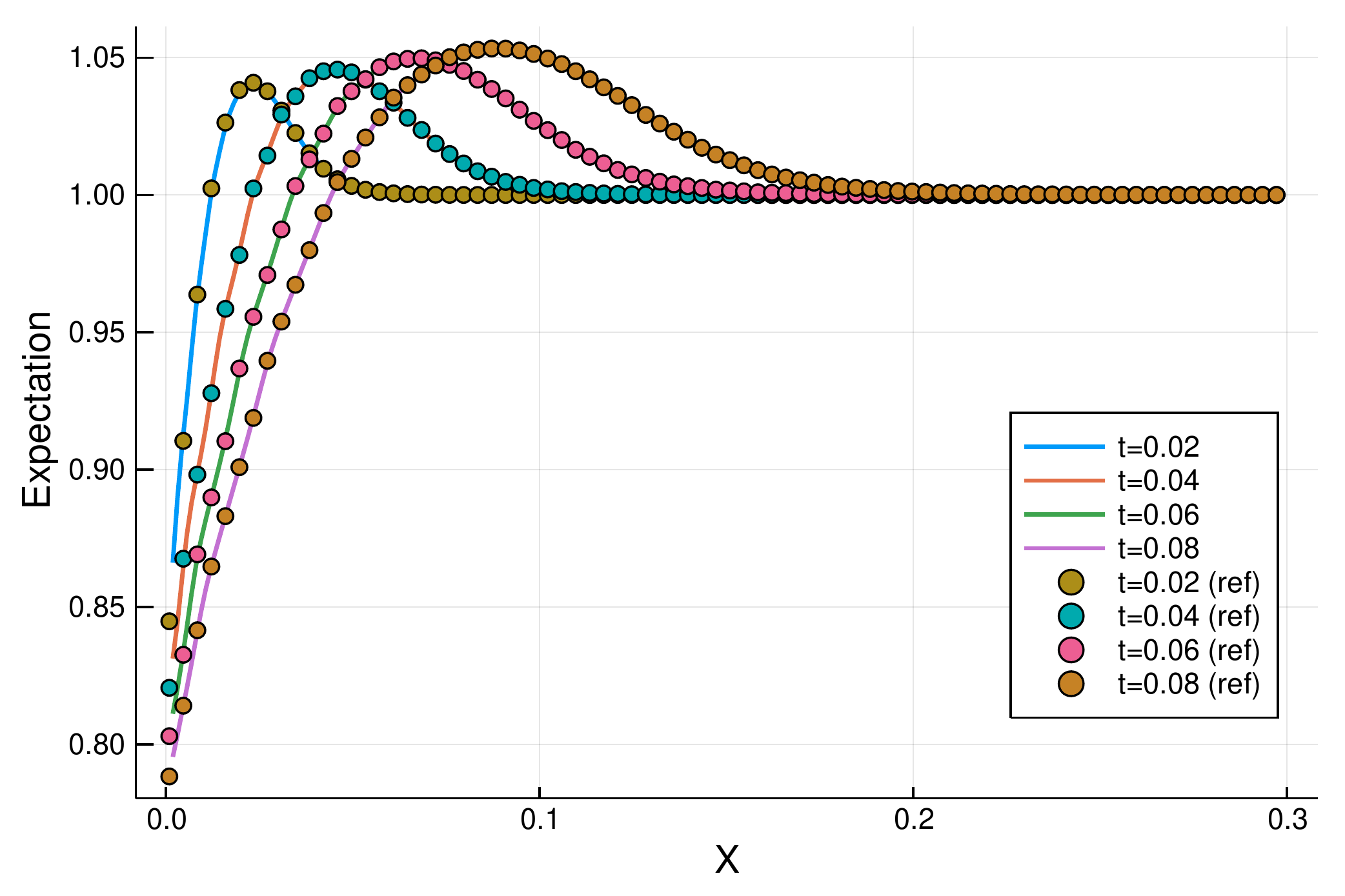}
	}
	\subfigure[$\mathbb S(\rho)$]{
		\includegraphics[width=7.5cm]{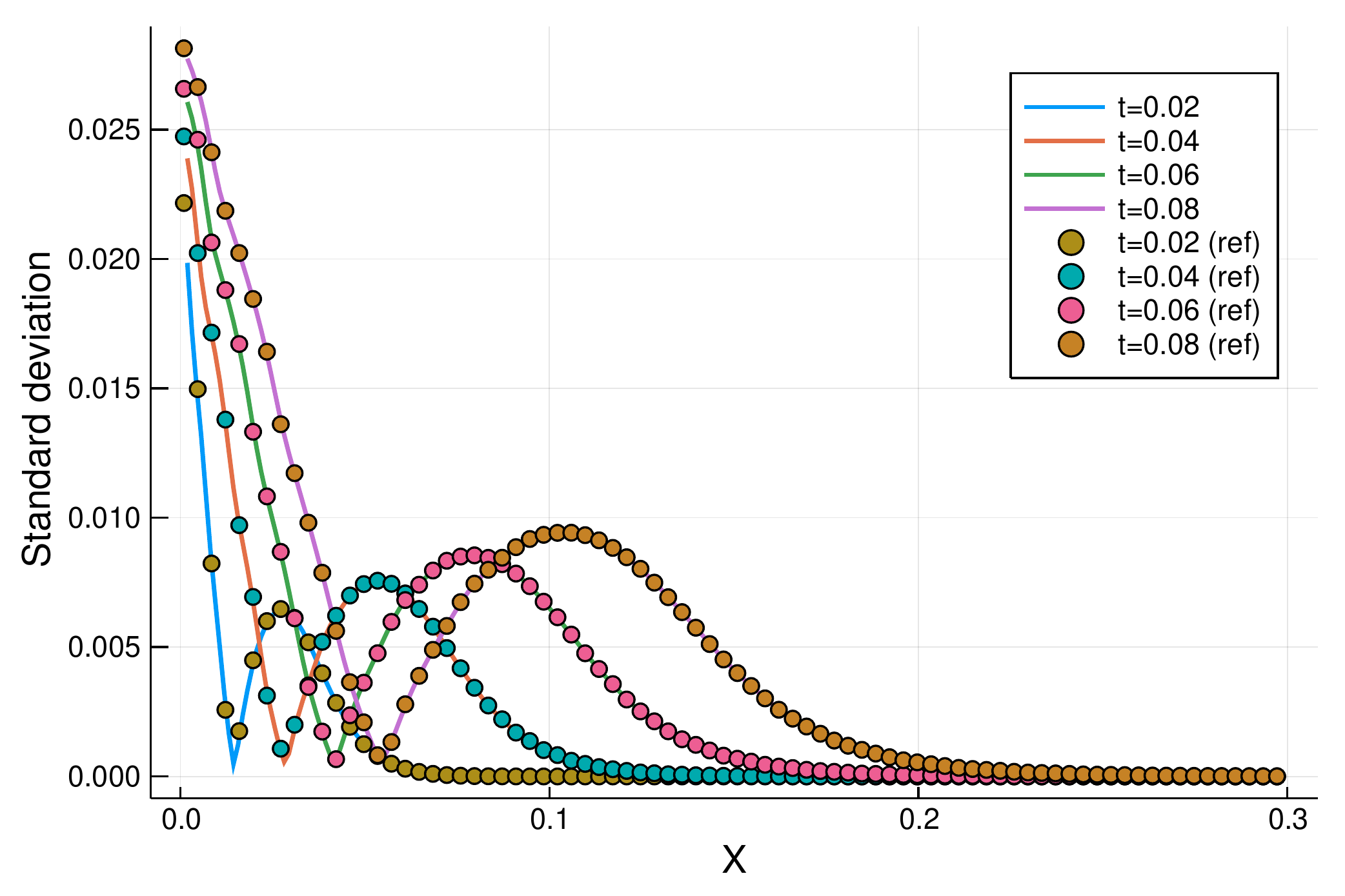}
	}
	\subfigure[$\mathbb E(U)$]{
		\includegraphics[width=7.5cm]{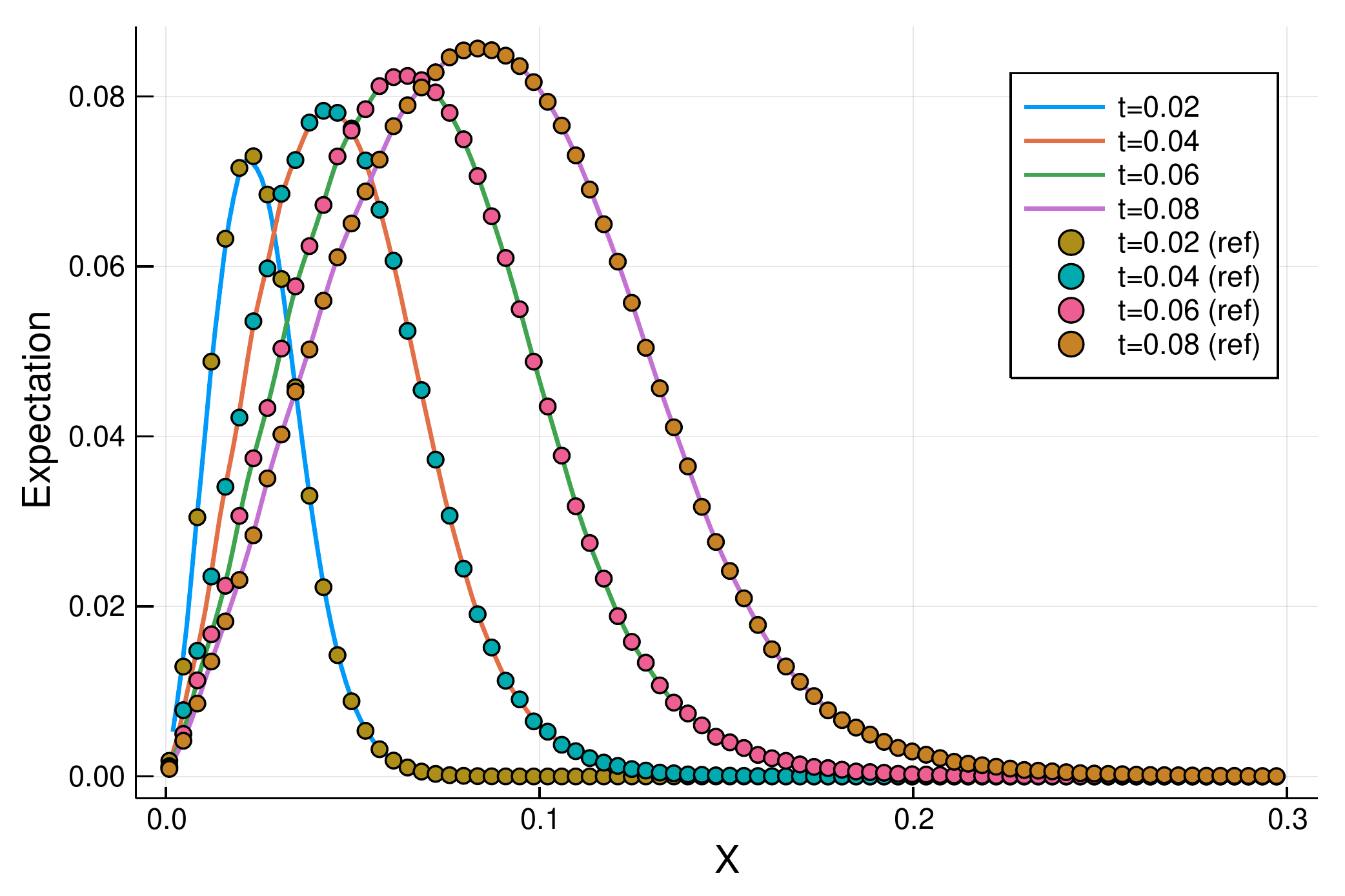}
	}
	\subfigure[$\mathbb S(U)$]{
		\includegraphics[width=7.5cm]{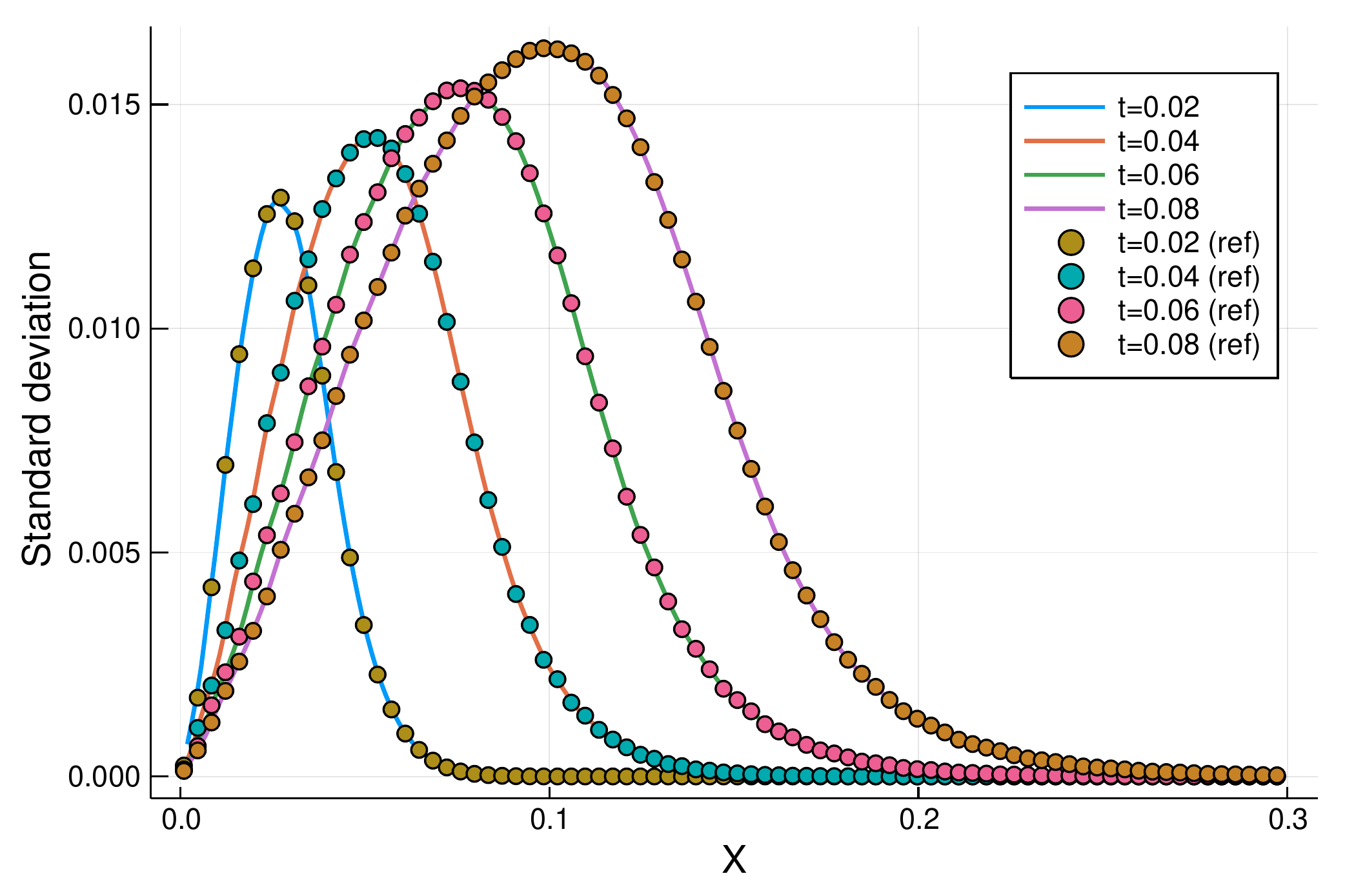}
	}
	\subfigure[$\mathbb E(T)$]{
		\includegraphics[width=7.5cm]{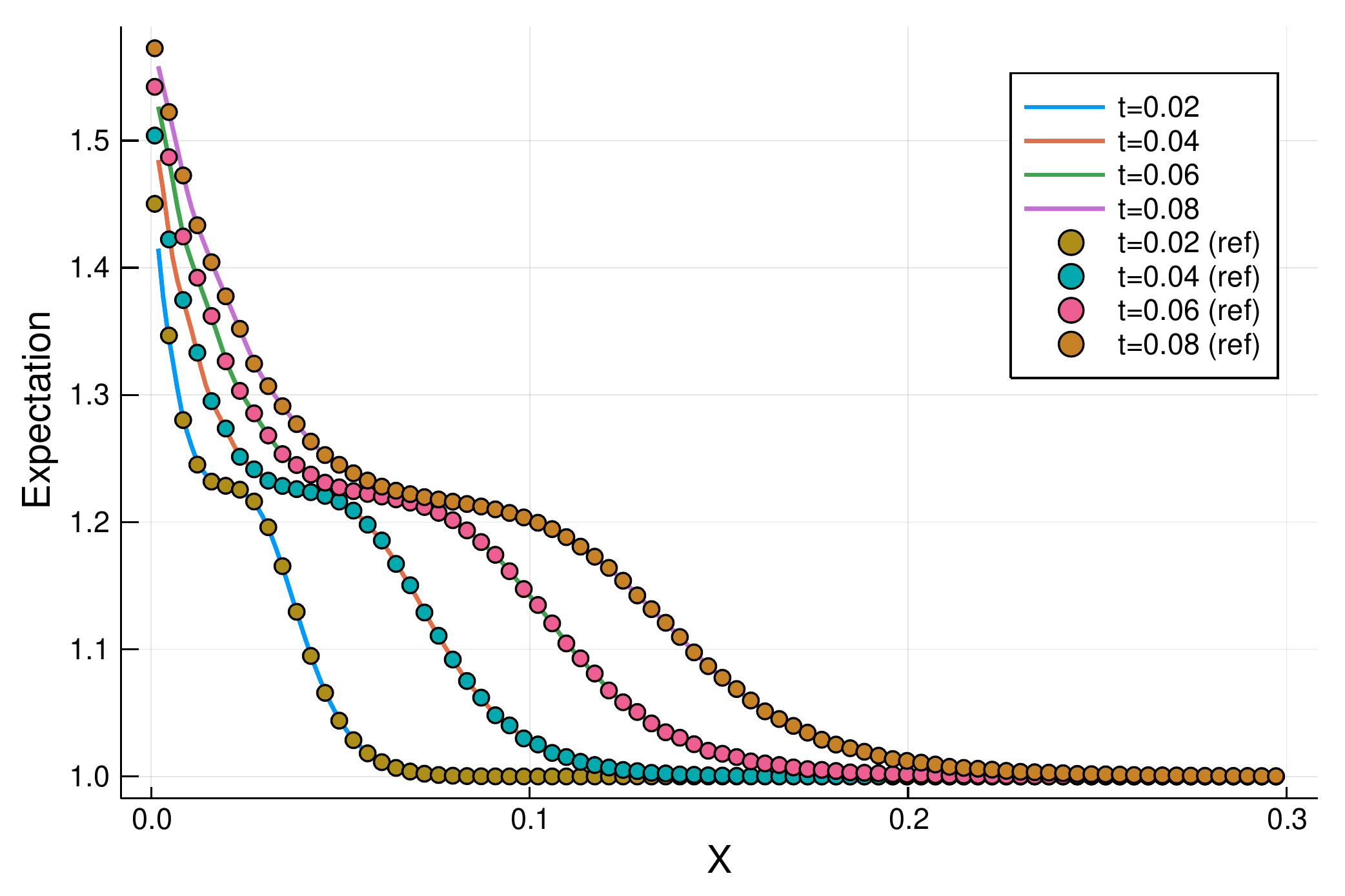}
	}
	\subfigure[$\mathbb S(T)$]{
		\includegraphics[width=7.5cm]{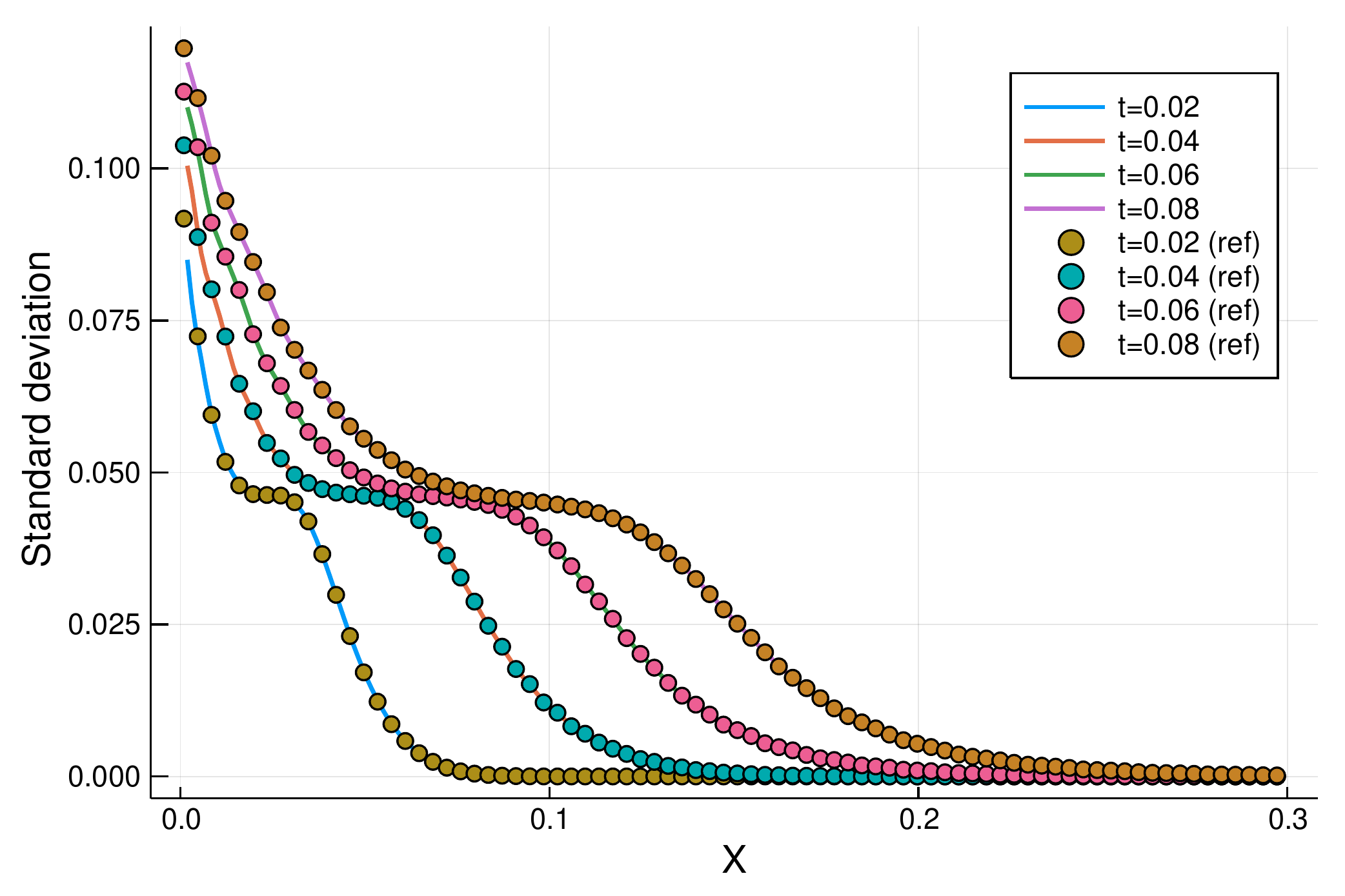}
	}
	\caption{Temporal evolutions of expectations and standard deviations of density (first row), velocity (second row) and temperature (third row) near the heat wall at $\mathrm{Kn}_{ref}=0.1$.}
	\label{pic:heat macro kn3}
\end{figure}

Fig.\ref{pic:heat macro kn1}, \ref{pic:heat macro kn2} and \ref{pic:heat macro kn3} present the expectation values and standard deviations of macroscopic gas density, velocity and temperature at different time instants $t=0.02$, $0.04$, $0.06$, and $0.08$.
As is shown, with the heating wall, the gas temperature and pressure near the wall rise quickly, forming a shock wave propagating towards the bulk region.
In Fig.\ref{pic:heat macro kn1}(a), (c) and (e), when $\mathrm{Kn}_{ref}=0.001$ with moderate viscosity and heat conductivity, the fine-scale structure cannot be resolved by the limited grid points, resulting in sharp discontinuity at the front head of shock wave, and the kinetic scheme becomes a shock capturing method.
In this case, slight deviations exist between the numerical and reference solutions due to the limited resolution, but it is clear seen that all the key wave-interaction structures are preserved.
With the increasing Knudsen number, the loose coupling between particle flight and collision leads to enhanced transport phenomena, and thus the diffusion process is accelerated and the steep shock discontinuity is smoothed into a milder profile.
Compare Fig.\ref{pic:heat macro kn3} with Fig.\ref{pic:heat macro kn1}, we see that the shock wave at $\mathrm{Kn}_{ref}=0.1$ travels twice faster than that at $\mathrm{Kn}_{ref}=0.001$.

Besides the evolution of mean field, the stochastic scheme provides us the opportunity to study the modes of uncertainty propagation from boundary to bulk region quantitatively.
As shown in the second columns of Fig.\ref{pic:heat macro kn1}, \ref{pic:heat macro kn2} and \ref{pic:heat macro kn3}, with randomly distributed boundary temperature, the near-wall gas holds the maximal variances of temperature and density, while the velocity variance is absent due to no-penetration condition across the wall.
As time evolves, another local maximum of variance emerges and propagates rightwards inside the flow field along with the shock wave.
For velocity and temperature, the propagating patterns of variances show clear similarity with expectation values.
However, the standard deviation of density decreases linearly first and arise again towards the front of shock.
The local mininum of variance locates near the starting point of intermediate regions with mitigatory temperature slope.
At $\mathrm{Kn}_{ref}=0.001$ and $0.01$, due to the existence of viscosity, the traveling shock waves are gradually dissipated and the macroscopic flows are 
decelerated with smaller peak velocities.
However, it seems that the strength of standard deviations is preserved and even enhanced as time goes, which indicates the accumulative effect for the propagation of uncertainties.
Moreover, in all cases especially at $\mathrm{Kn}_{ref}=0.1$, it is noticed that the uncertainty travels a little faster than the mean field itself, which demonstrates the particular wave-propagation nature of uncertainty.
It may be explained by the stronger sensitivity of uncertainty over mean field, which means the still gas will feel the existence of uncertainties in front of the shock.

\begin{figure}[htb!]
	\centering
	\subfigure[$\mathbb E(f)$]{
		\includegraphics[width=7.5cm]{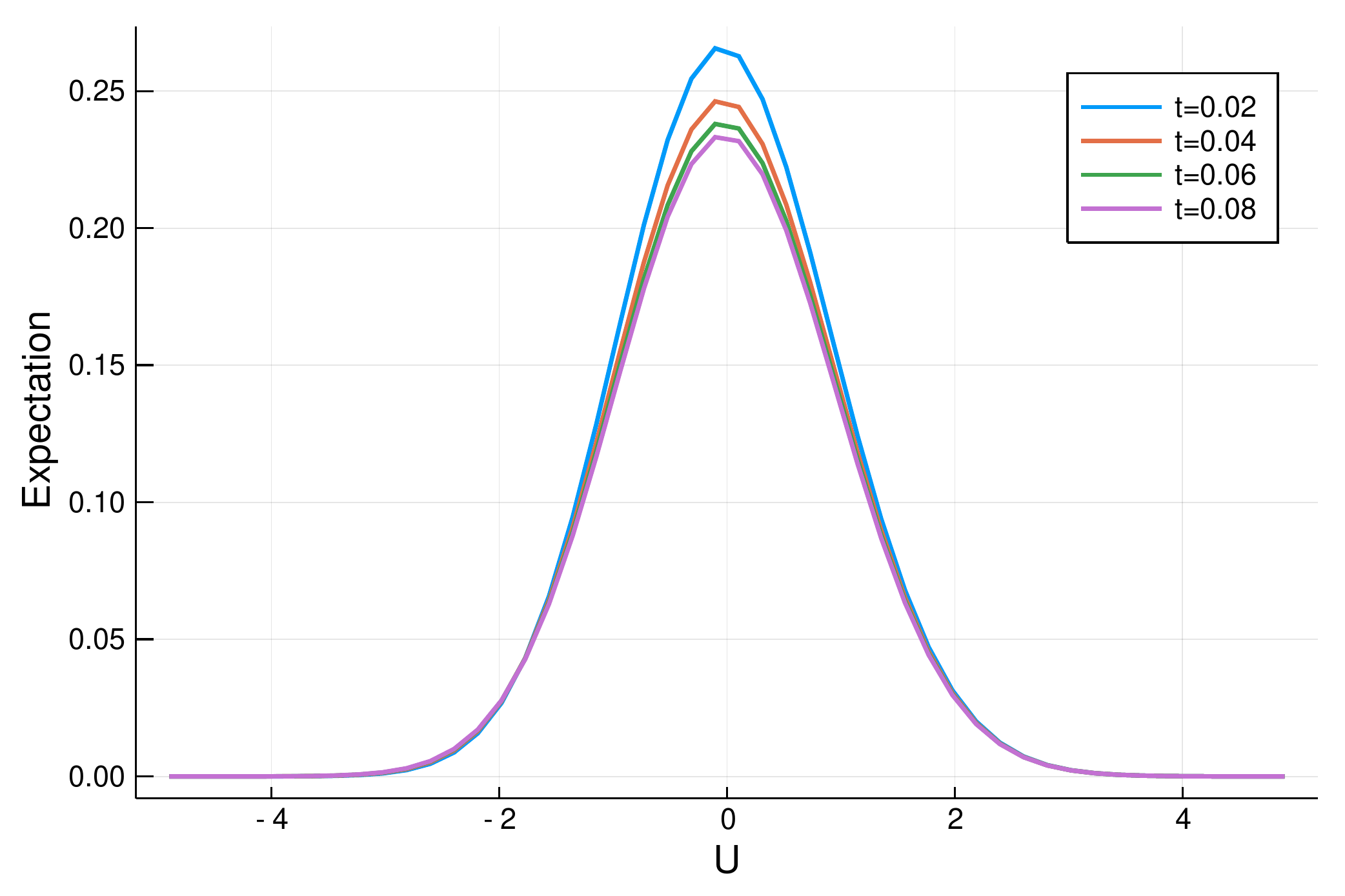}
	}
	\subfigure[$\mathbb S(f)$]{
		\includegraphics[width=7.5cm]{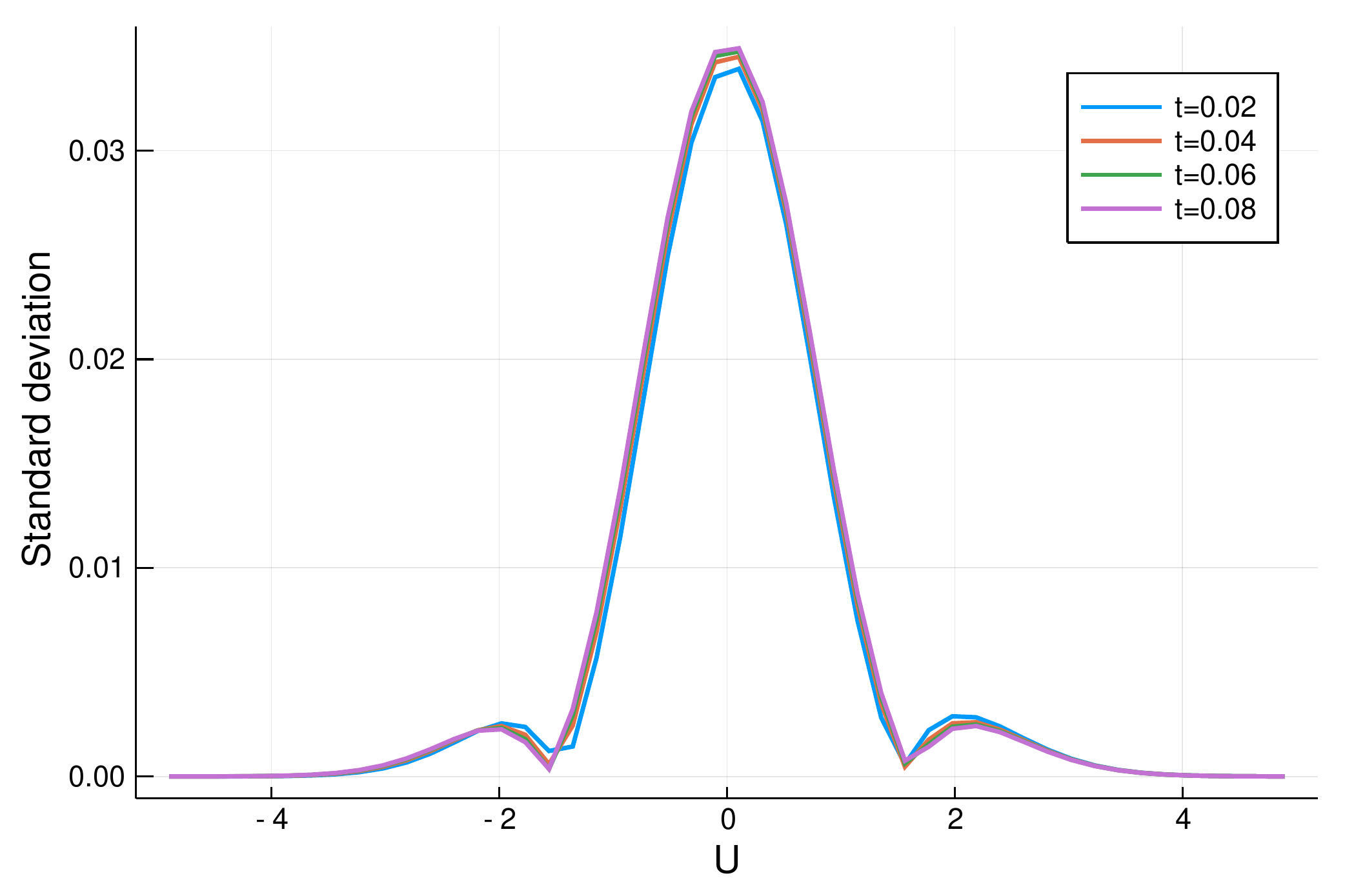}
	}
	\subfigure[$\mathbb E(f)$]{
		\includegraphics[width=7.5cm]{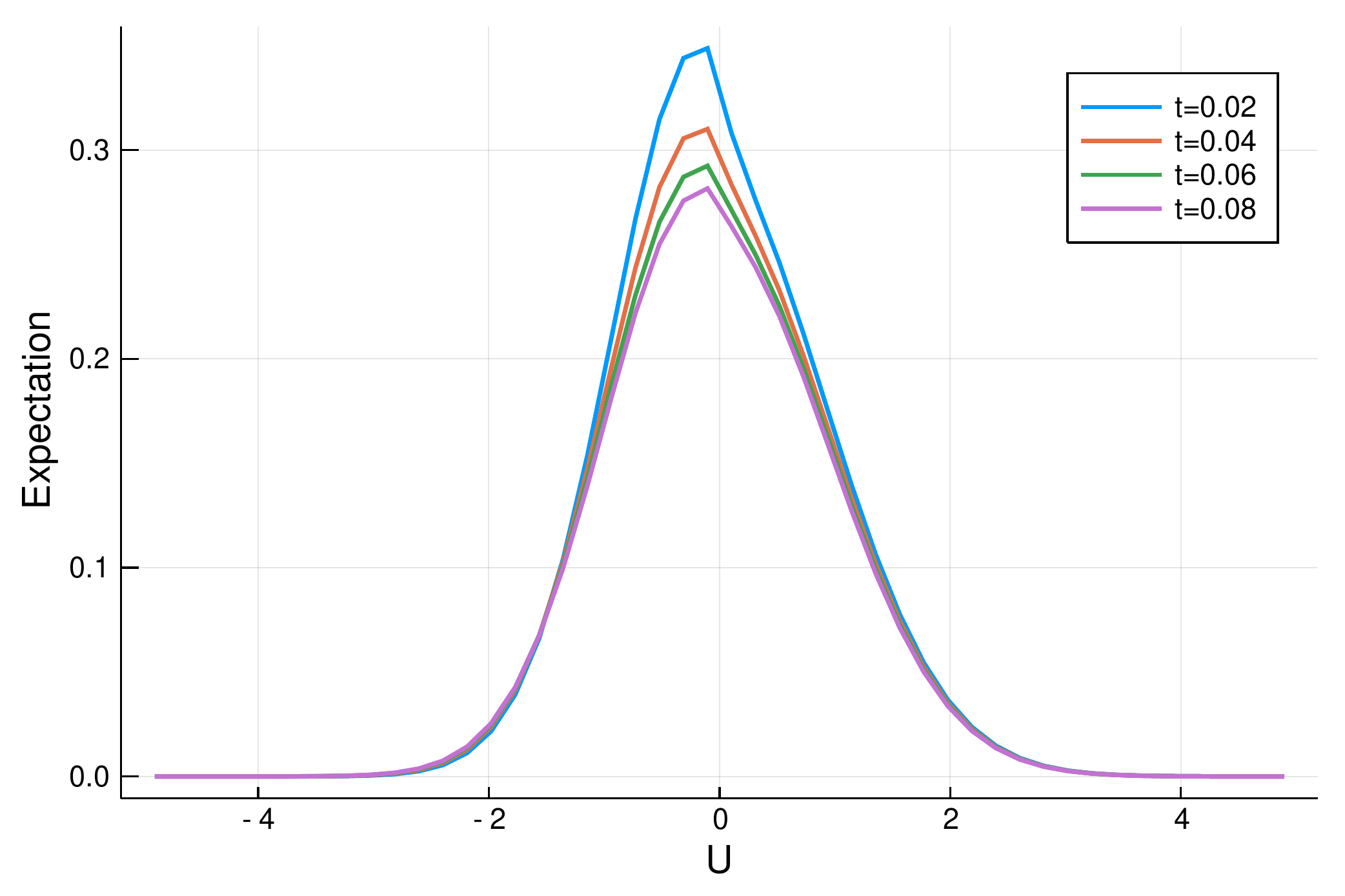}
	}
	\subfigure[$\mathbb S(f)$]{
		\includegraphics[width=7.5cm]{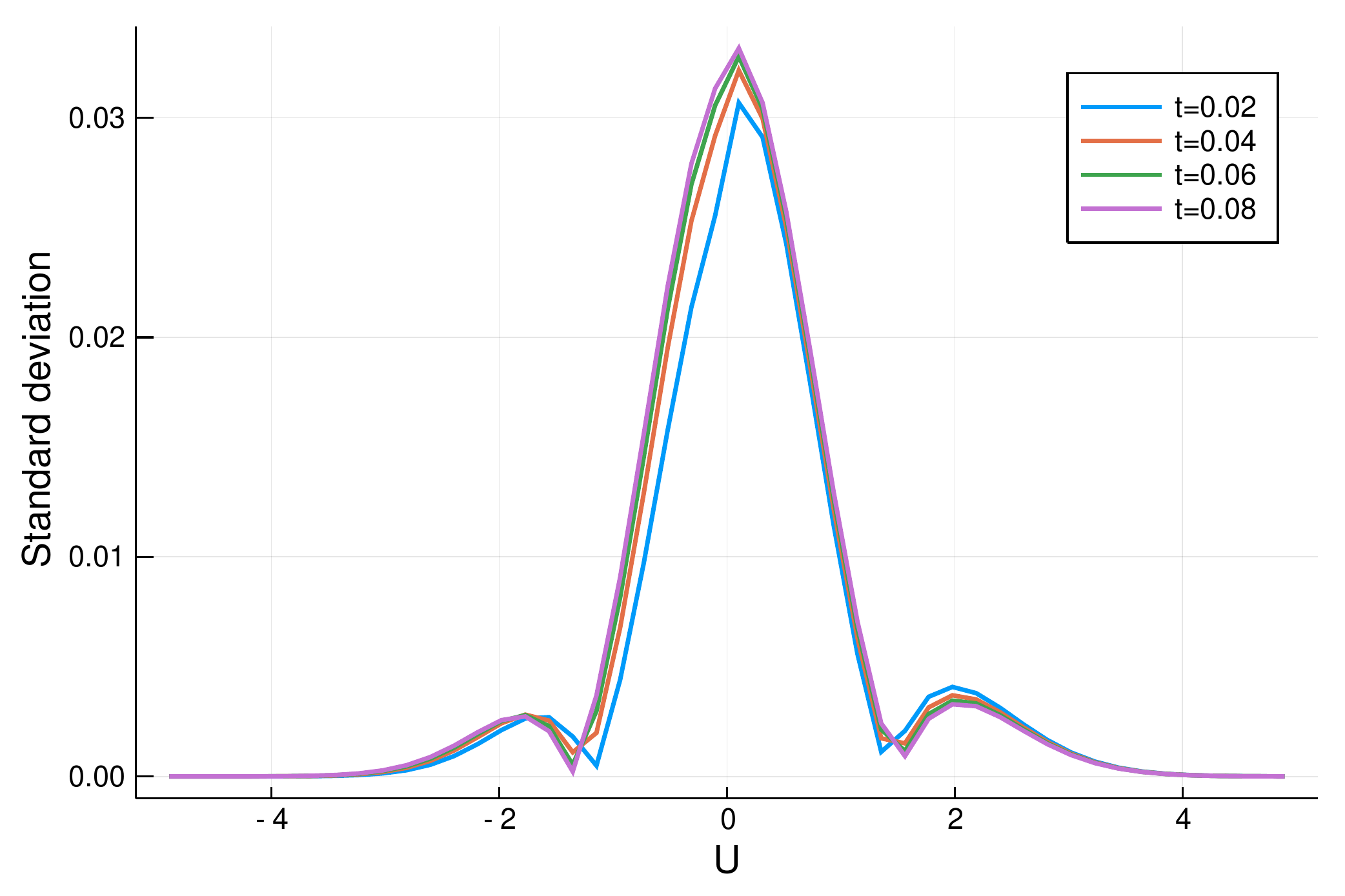}
	}
	\subfigure[$\mathbb E(f)$]{
		\includegraphics[width=7.5cm]{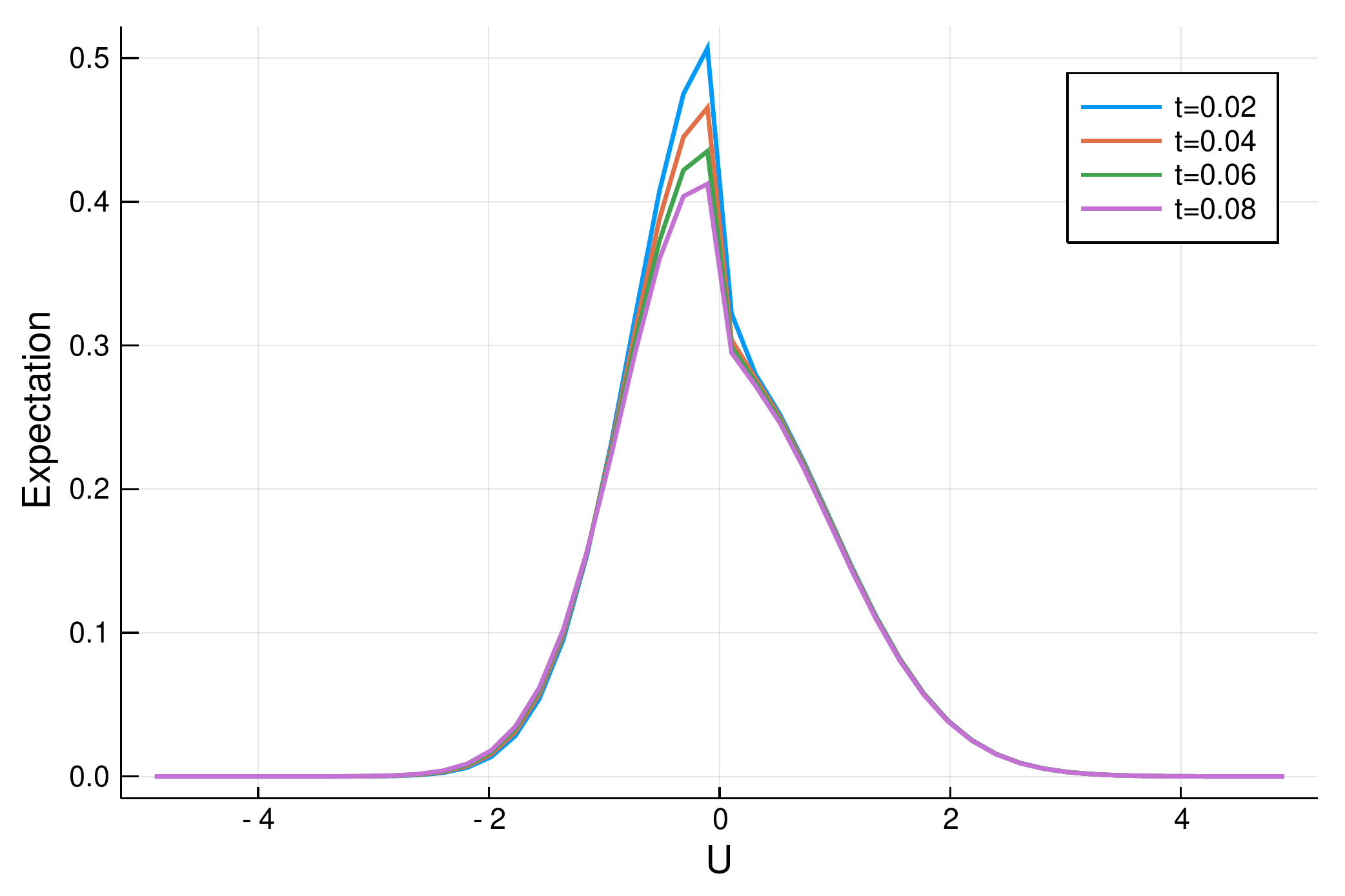}
	}
	\subfigure[$\mathbb S(f)$]{
		\includegraphics[width=7.5cm]{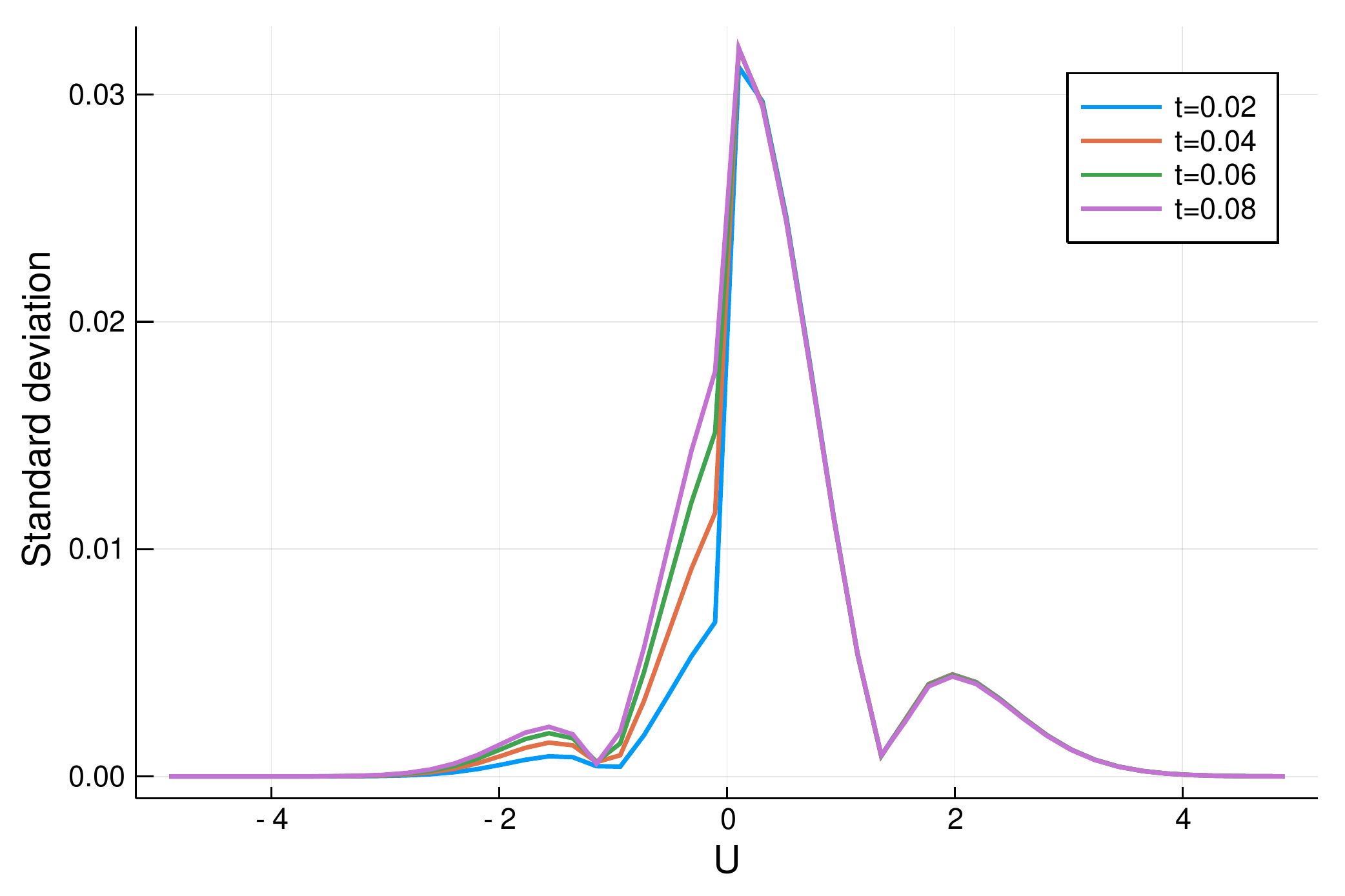}
	}
	\caption{Expectations and standard deviations of boundary particle distribution function at $\mathrm{Kn}_{ref}=0.001$ (first row), $\mathrm{Kn}_{ref}=0.01$ (second row) and $\mathrm{Kn}_{ref}=0.1$ (third row).}
	\label{pic:heat distribution boundary}
\end{figure}

\begin{figure}[htb!]
	\centering
	\subfigure[$t=0.02$]{
		\includegraphics[width=7.5cm]{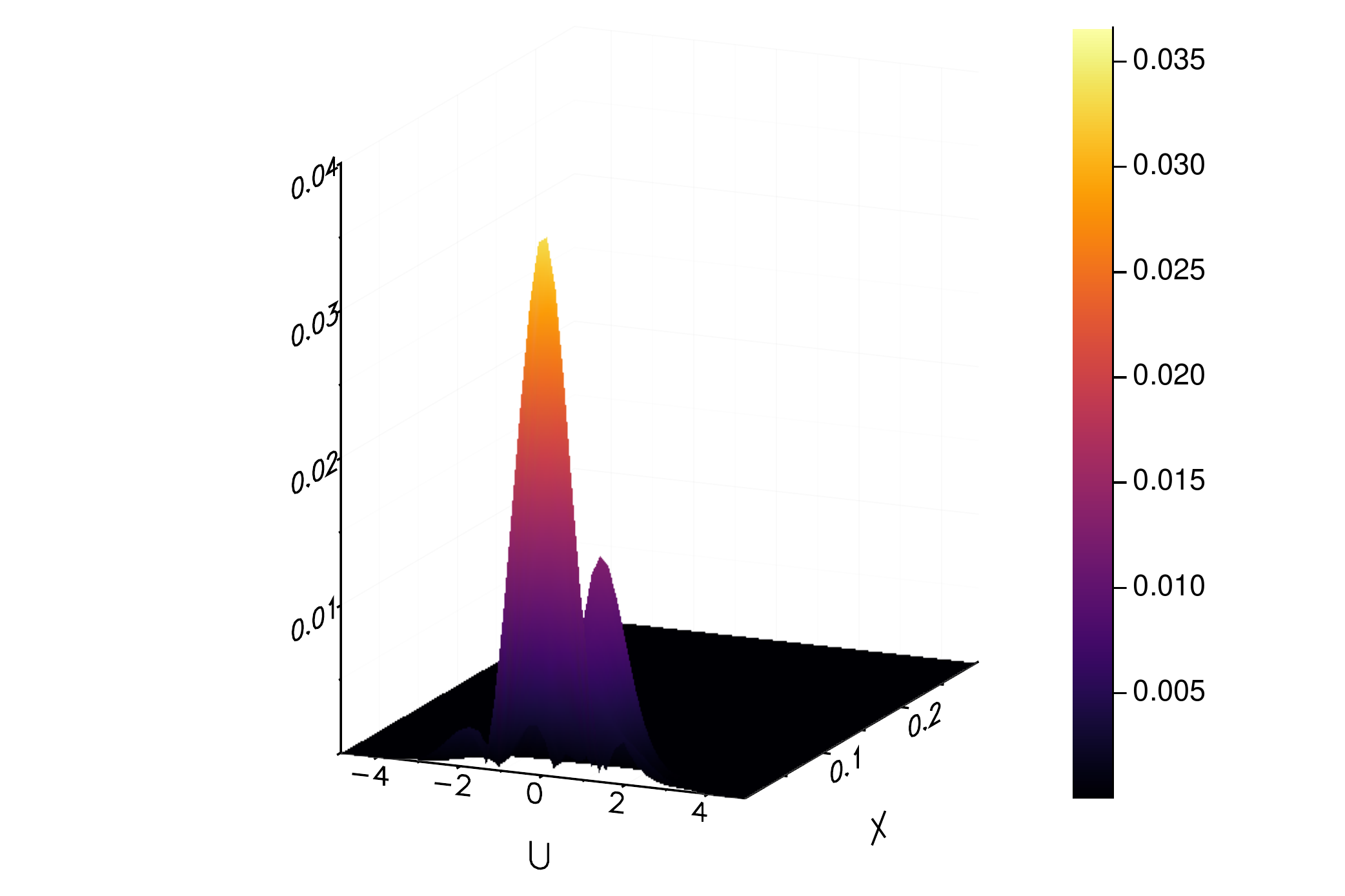}
	}
	\subfigure[$t=0.04$]{
		\includegraphics[width=7.5cm]{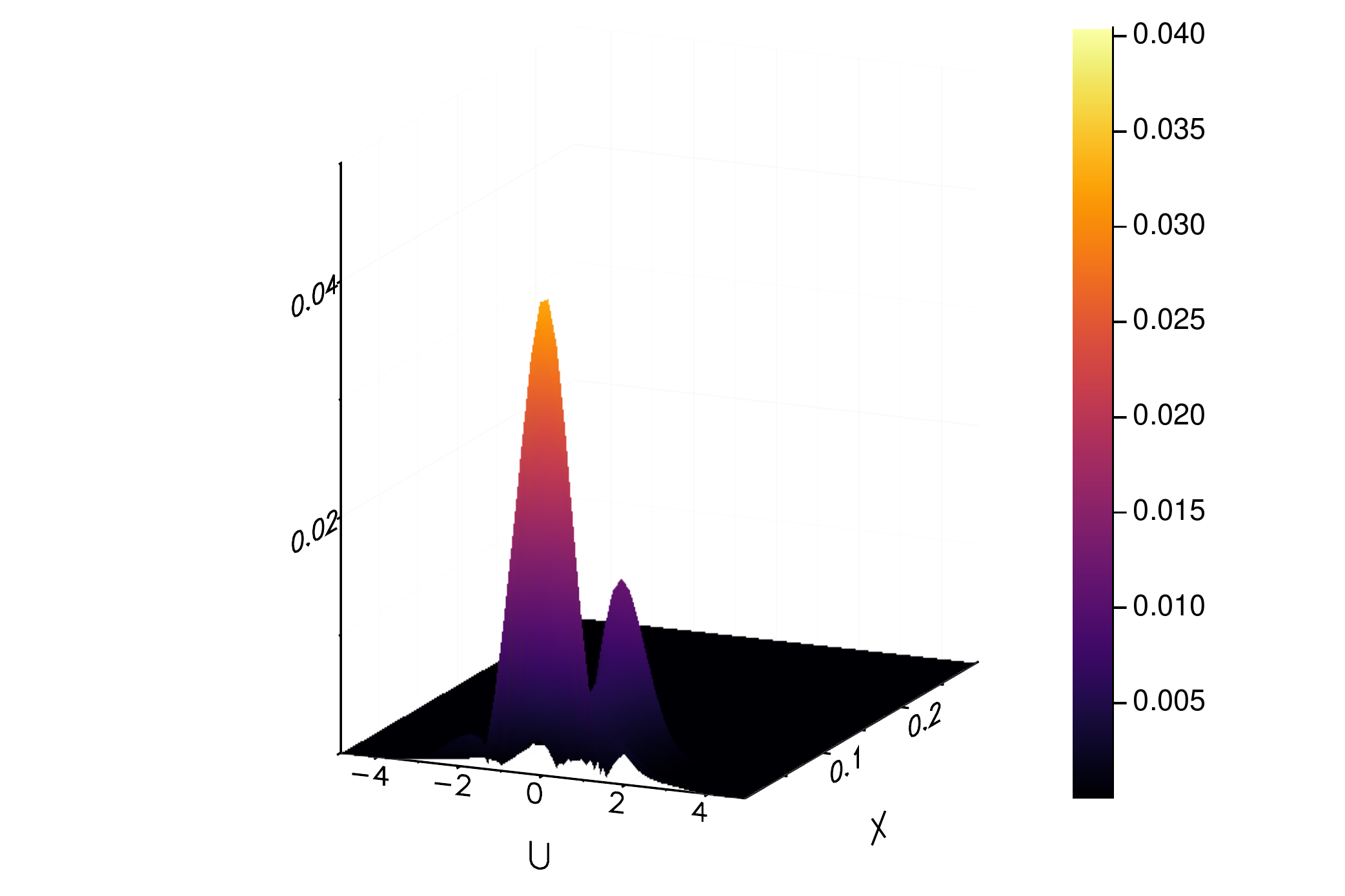}
	}
	\subfigure[$t=0.06$]{
		\includegraphics[width=7.5cm]{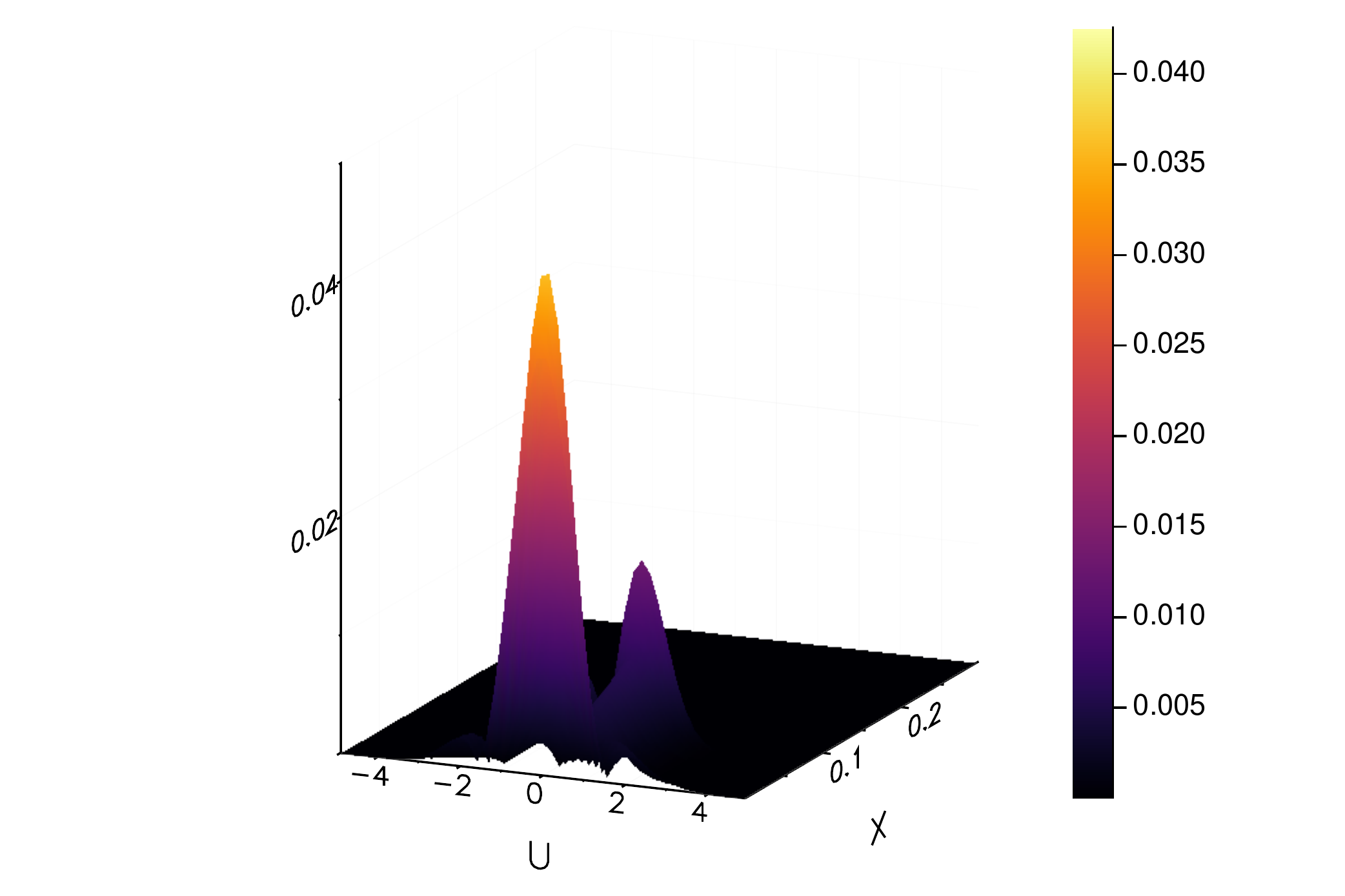}
	}
	\subfigure[$t=0.08$]{
		\includegraphics[width=7.5cm]{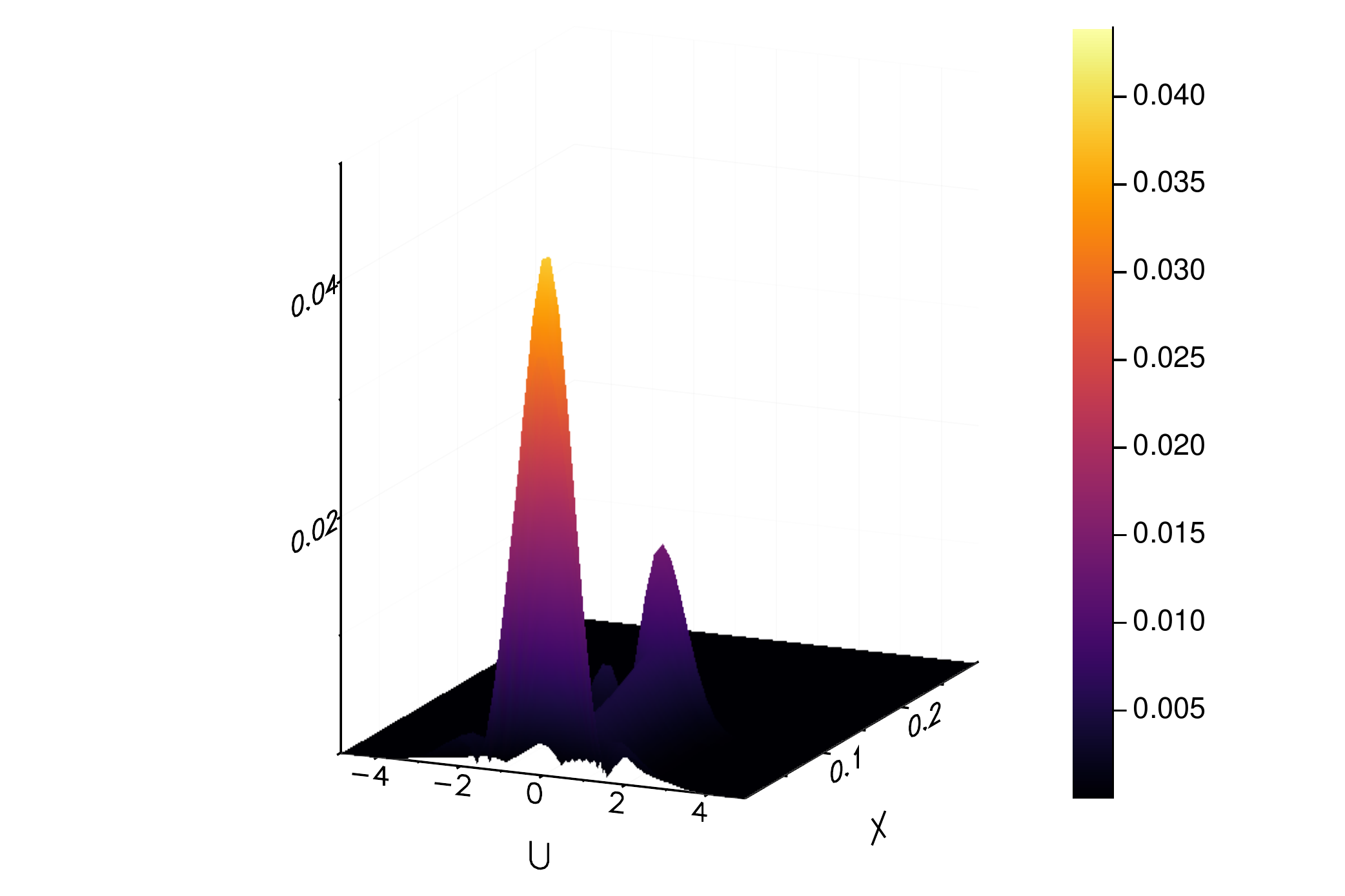}
	}
	\caption{Standard deviations of particle distribution function near the heat wall at $\mathrm{Kn}_{ref}=0.001$.}
	\label{pic:heat distribution std kn1}
\end{figure}

\begin{figure}[htb!]
	\centering
	\subfigure[$t=0.02$]{
		\includegraphics[width=7.5cm]{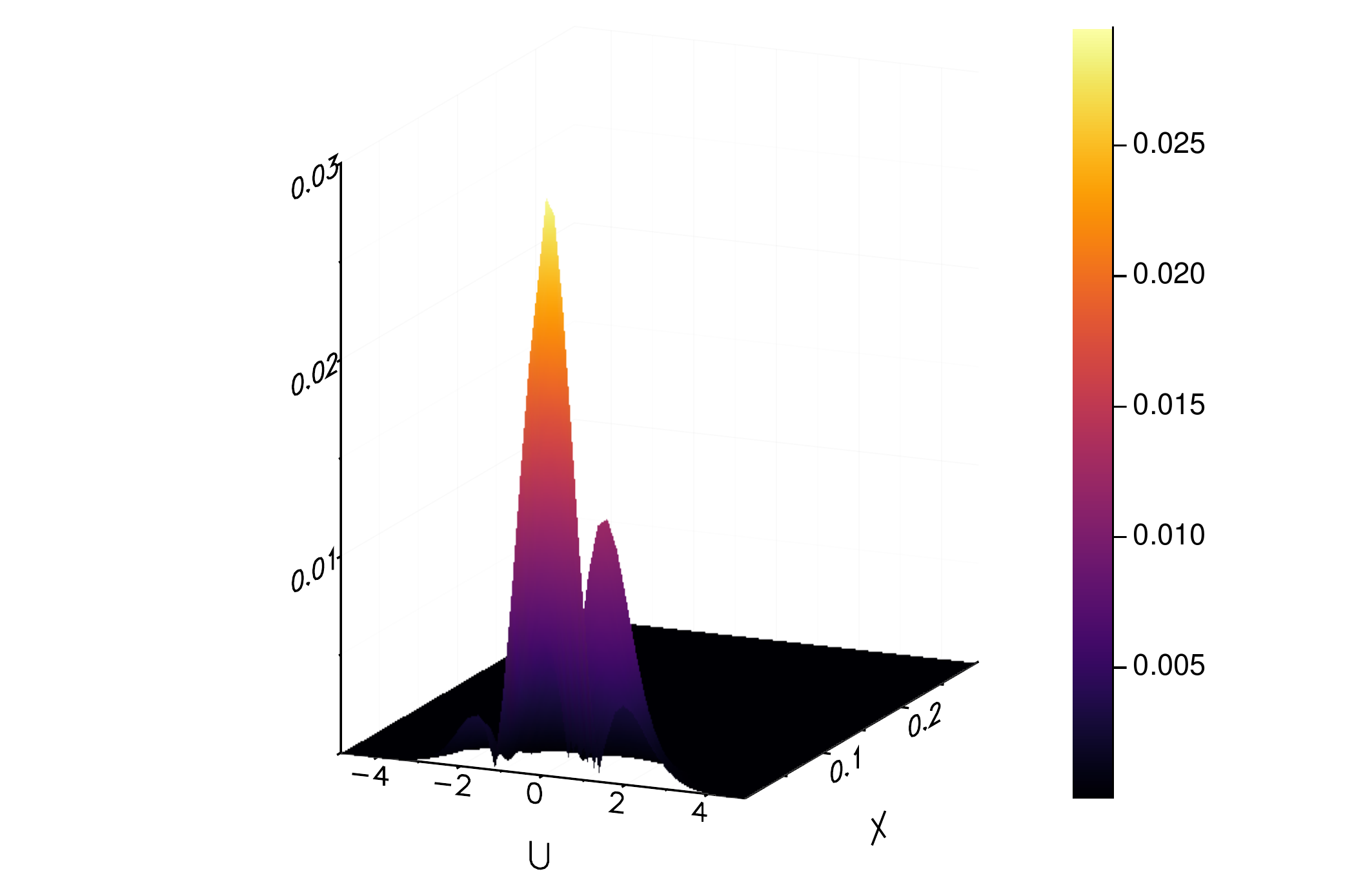}
	}
	\subfigure[$t=0.04$]{
		\includegraphics[width=7.5cm]{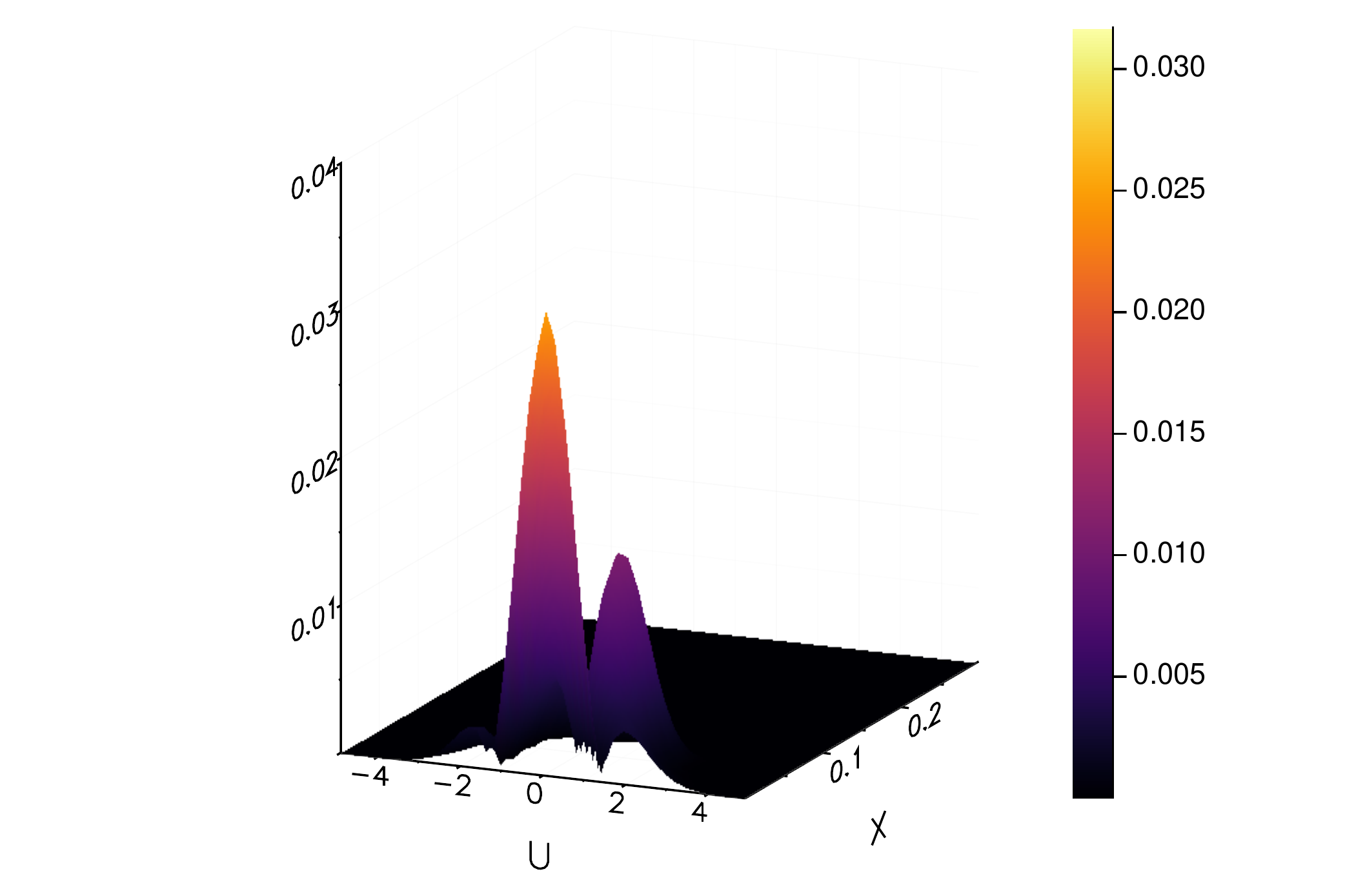}
	}
	\subfigure[$t=0.06$]{
		\includegraphics[width=7.5cm]{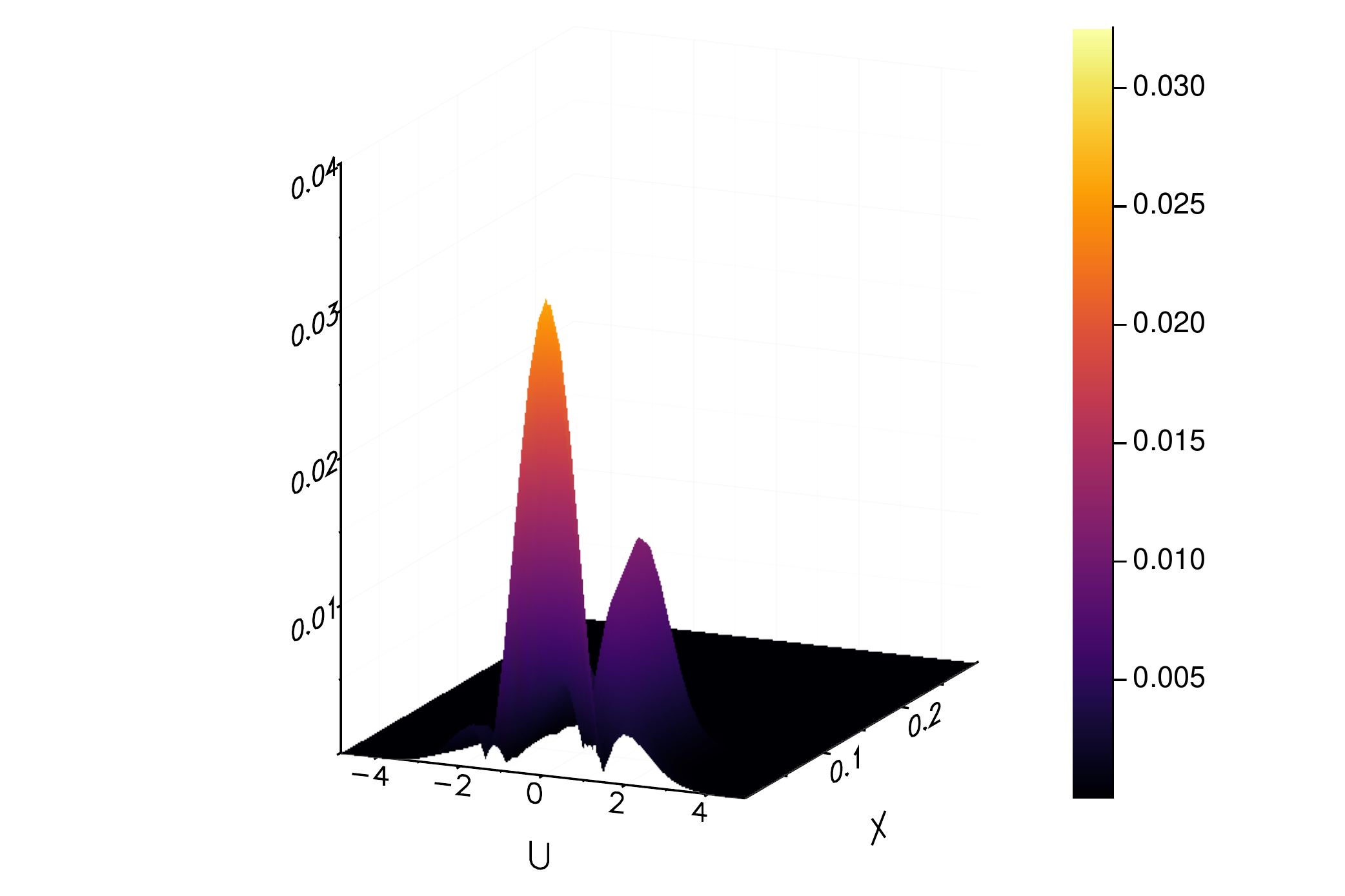}
	}
	\subfigure[$t=0.08$]{
		\includegraphics[width=7.5cm]{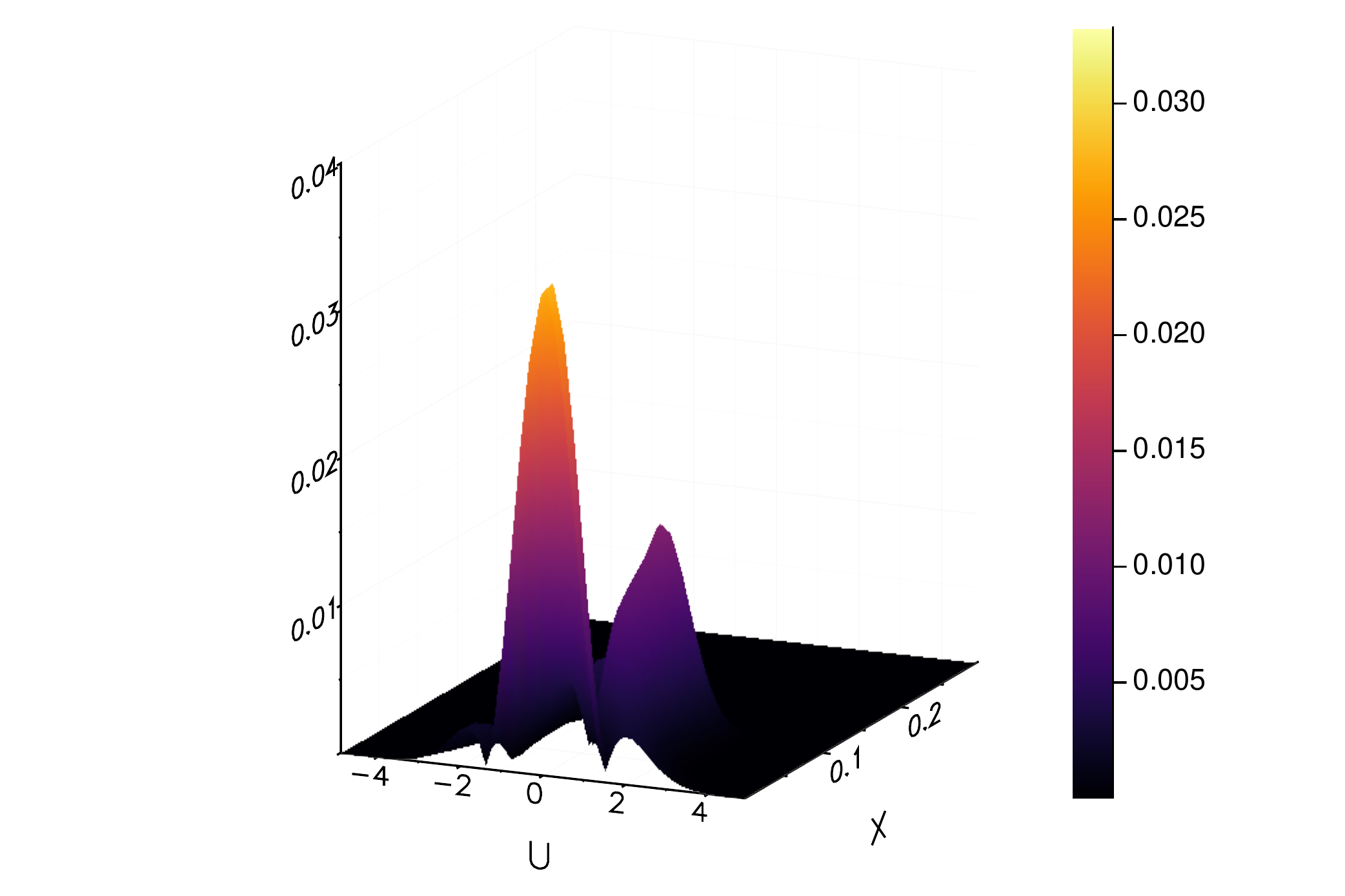}
	}
	\caption{Standard deviations of particle distribution function near the heat wall at $\mathrm{Kn}_{ref}=0.01$.}
	\label{pic:heat distribution std kn2}
\end{figure}

\begin{figure}[htb!]
	\centering
	\subfigure[$t=0.02$]{
		\includegraphics[width=7.5cm]{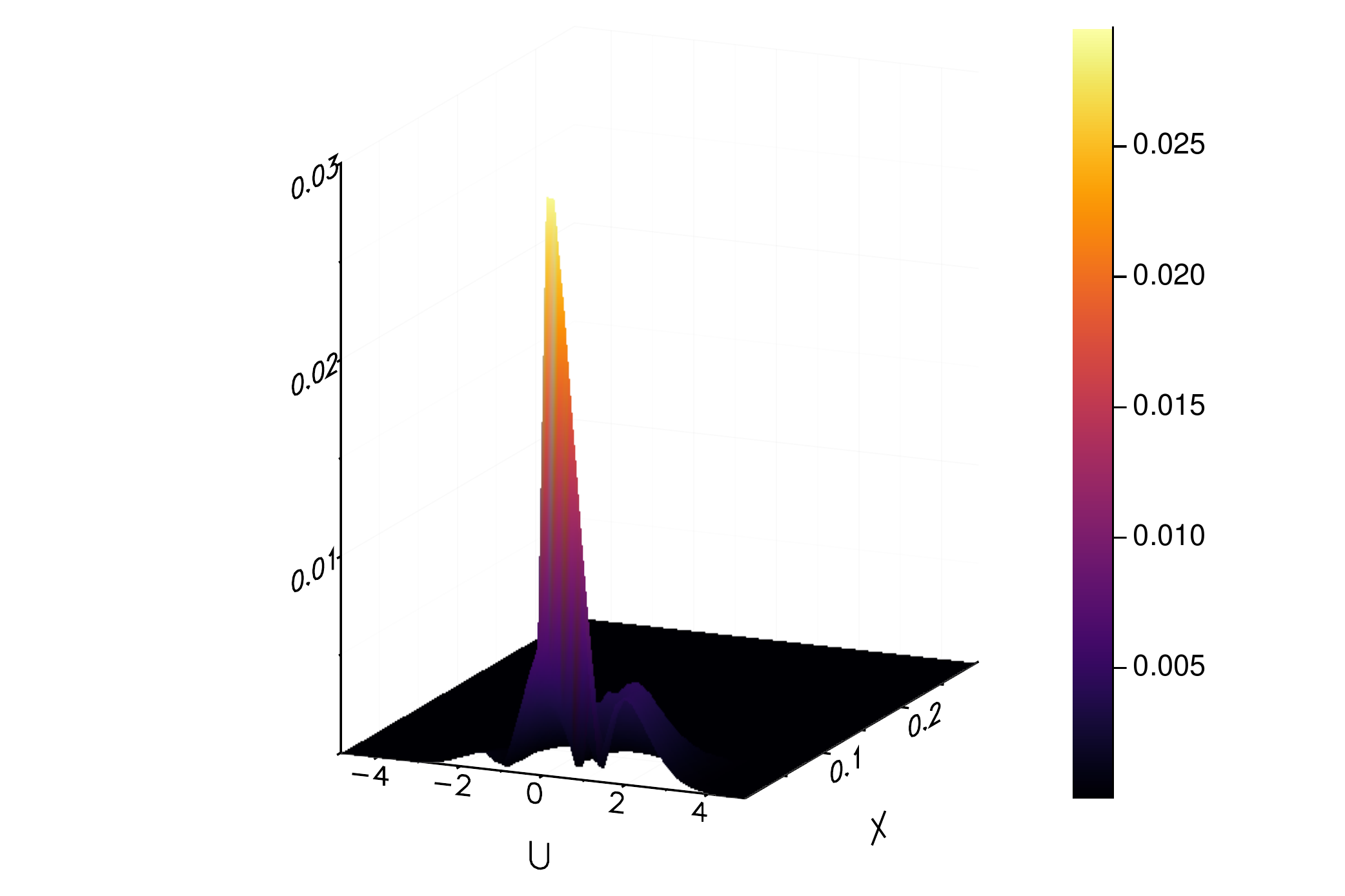}
	}
	\subfigure[$t=0.04$]{
		\includegraphics[width=7.5cm]{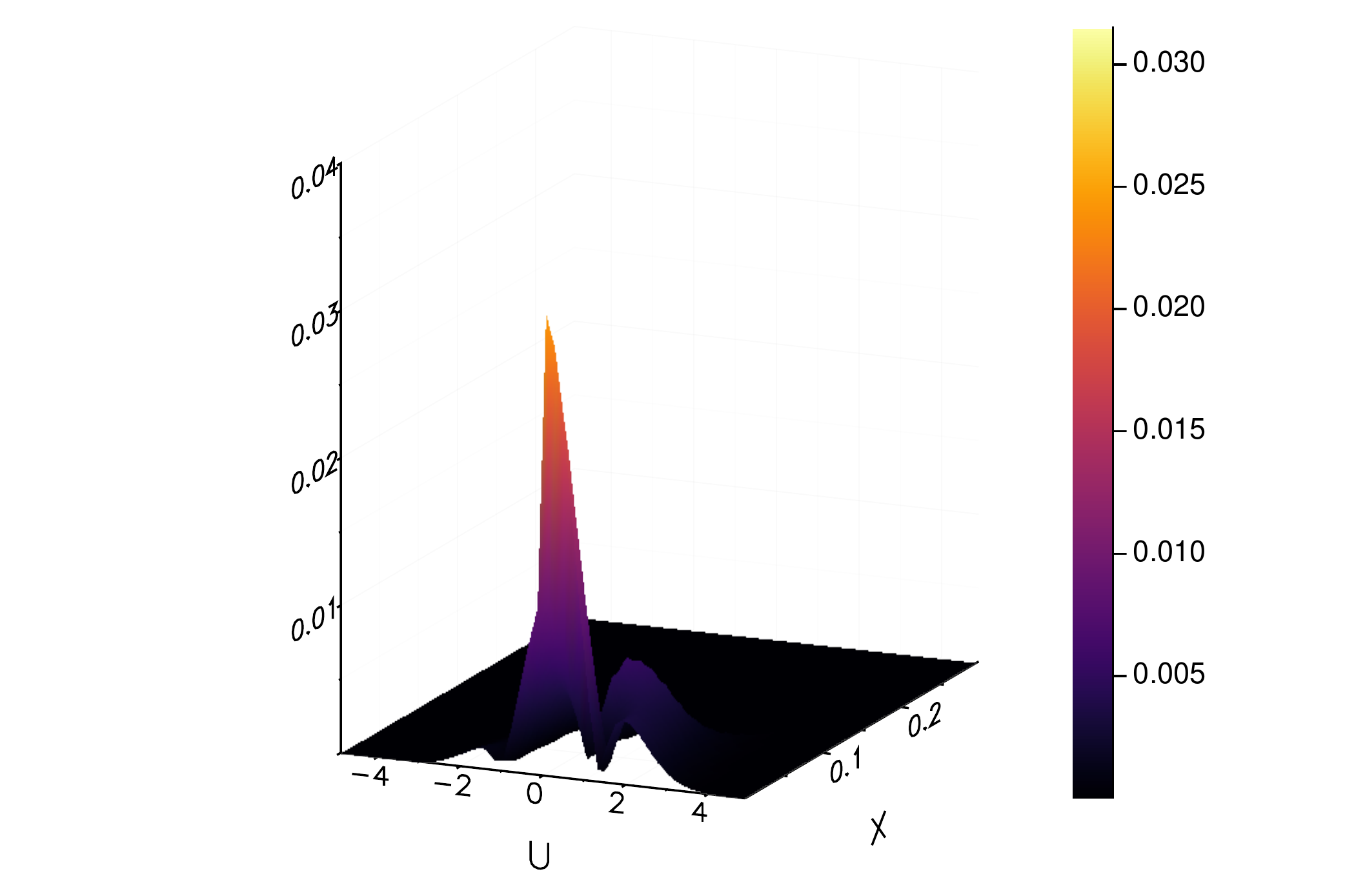}
	}
	\subfigure[$t=0.06$]{
		\includegraphics[width=7.5cm]{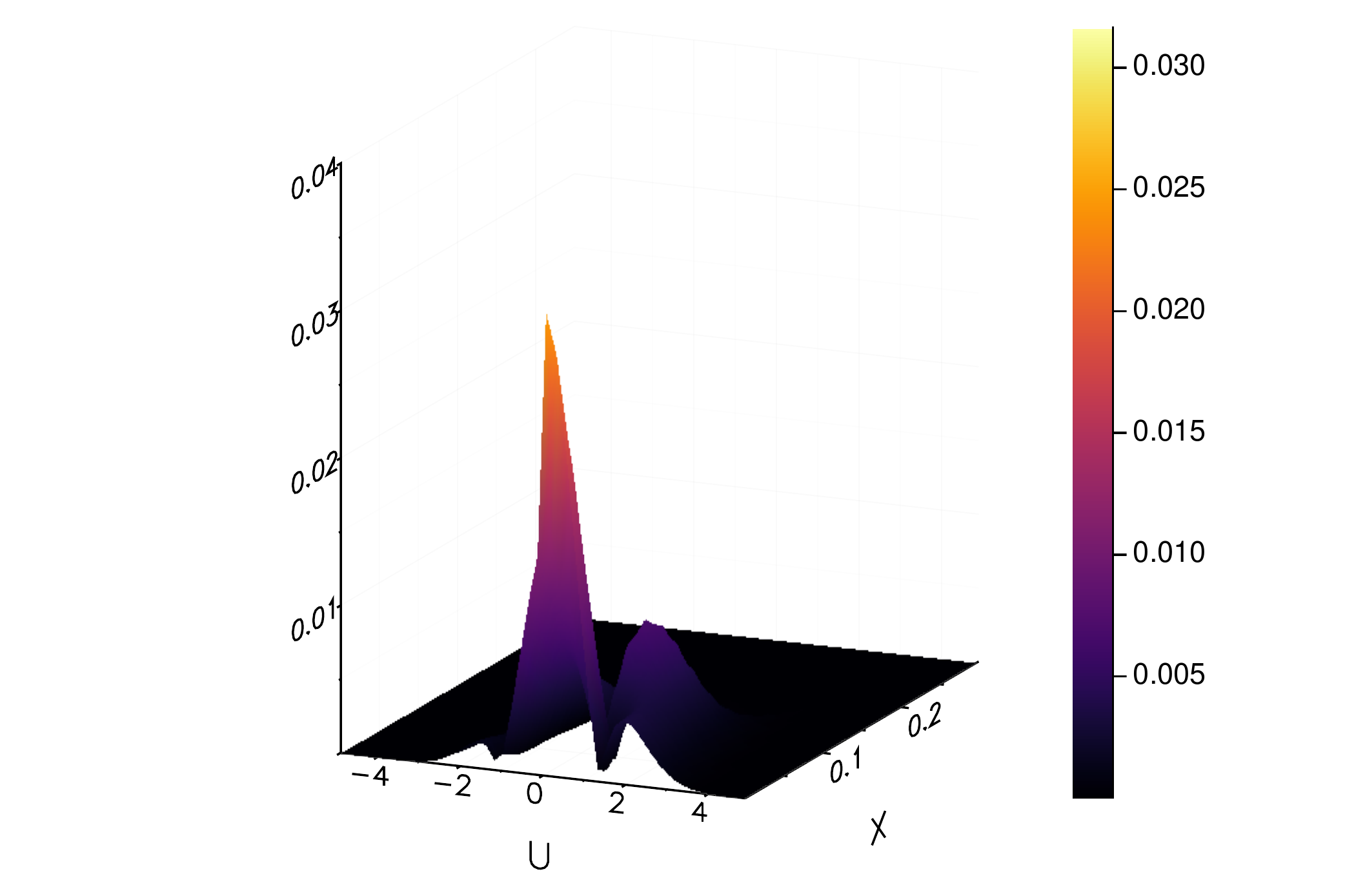}
	}
	\subfigure[$t=0.08$]{
		\includegraphics[width=7.5cm]{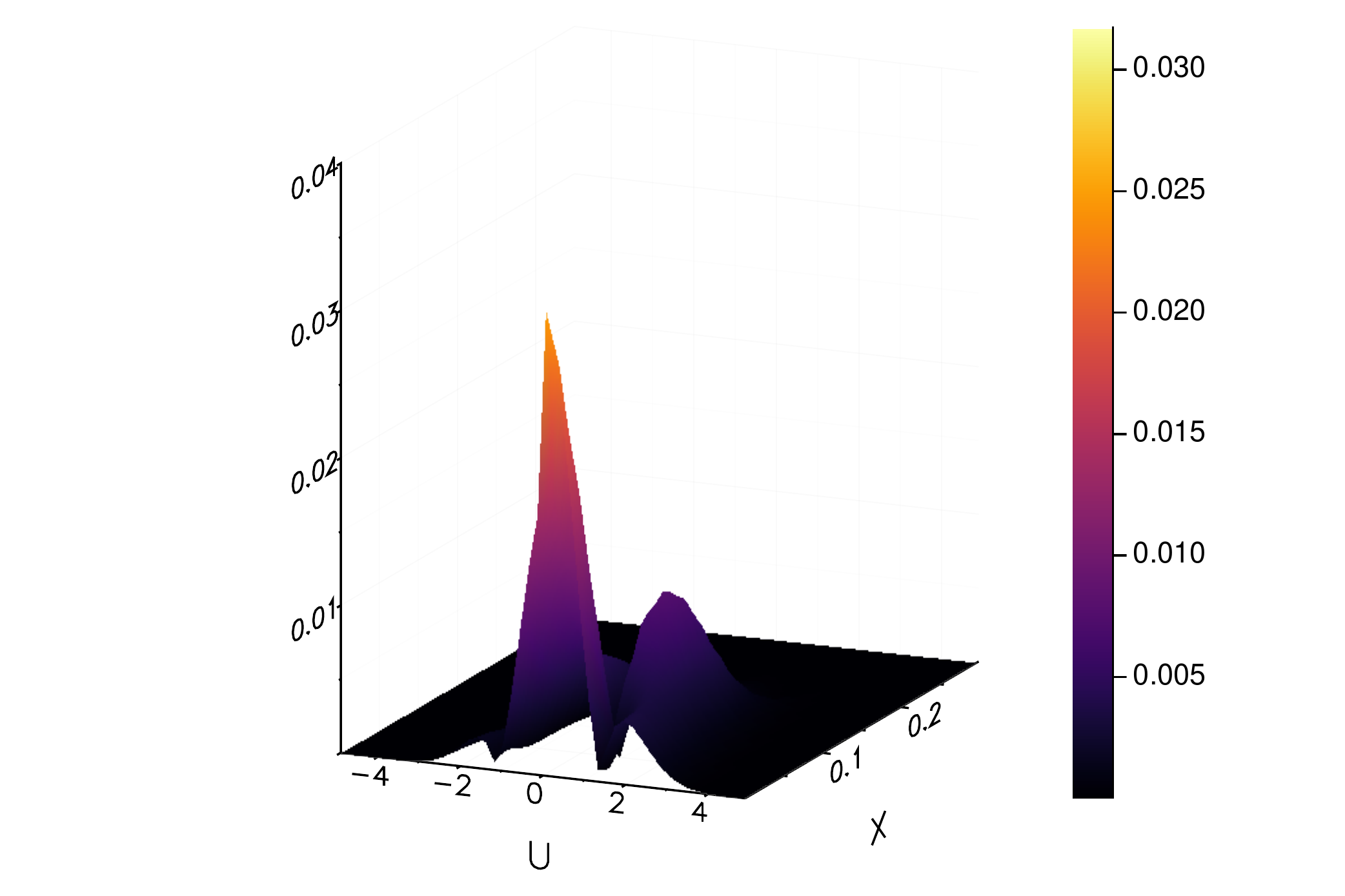}
	}
	\caption{Standard deviations of particle distribution function near the heat wall at $\mathrm{Kn}_{ref}=0.1$.}
	\label{pic:heat distribution std kn3}
\end{figure}

With the one-to-one correspondence between hydrodynamic and mesoscopic formulations, the boundary heating process also evolves particles around the wall and passes uncertainties to the particle distribution function.
Fig.\ref{pic:heat distribution boundary} presents the expectations and standard deviations of particle distribution function on the wall at different reference Knudsen numbers.
Given the Maxwell's diffusive boundary, the wall temperature defines the right half of particle distribution function with positive velocity $u>0$, while the left half is inherited from inner distribution function.
As seen in Fig.\ref{pic:heat distribution boundary}(e) and (f), it leads to a discontinuity in particle distribution at $\mathrm{Kn}_{ref}=0.1$, where the right half possesses much more significant uncertainties correspondingly.
With increasing Knudsen number in Fig.\ref{pic:heat distribution boundary}(a) to (d), frequent intermolecular interactions lead to equipartitions of energy, and thus the inner distribution functions are much closer to Maxwellian.
As a result, the left and right parts of boundary particle distribution coincide with each other and the variances become symmetric with respect to zero velocity point.
Fig.\ref{pic:heat distribution std kn1}, \ref{pic:heat distribution std kn2} and \ref{pic:heat distribution std kn3} show the standard deviations of particle distribution functions at different time instants and reference Knudsen numbers.
Since the macroscopic velocity is very small in this heat diffusion problem, the variances basically keep symmetric along the velocity dimension, and the uncertainty waves propagate inside the phase space by reshaping the heights and widths of particle distribution functions.

\section{Conclusion}

Gas dynamics is a truly multiscale problem due to the large variations of gas density and characteristic length scales of the flow structures. 
Based on a kinetic model equation and its scale-dependent time evolving solution, a stochastic kinetic scheme with both standard stochastic Galerkin and hybrid Galerkin-collocation settings has been constructed in this paper, and both formulations allow for a unified flow simulation in all regimes. Based on multi-scale modeling, the solution algorithm is able to capture both equilibrium and non-equilibrium flow phenomena simultaneously in the flow field, and a continuous spectrum of cross-scale physics can be recovered along with the evolution of randomness. The asymptotic-preserving property of the scheme is validated through theoretical analysis and numerical tests.
In the numerical experiments, for the first time non-equilibrium flow phenomena, such as the wave-propagation patterns of uncertainty from continuum to rarefied gas dynamics, could be clearly identified and quantitatively analyzed. The current scheme provides an efficient and accurate tool for the study of multi-scale non-equilibrium gas dynamics, and may help with the sensitivity analysis in design and applications of fluid machinery with uncertainty quantification. Its extension to multi-dimensional phase space and the analysis of the unified-preserving property \cite{guo2019unified} will be further considered in the future work.

\section*{Acknowledgement}
The authors would like to thank Jonas Kusch and Tillmann Muhlpfordt for the helpful discussion of uncertainty quantification and stochastic methods.

\clearpage
\newpage

\bibliographystyle{unsrt}
\bibliography{tbxiao}
\newpage

\end{document}